\documentclass[prd,aps,onecolumn,floatfix,superscriptaddress,preprintnumbers]{revtex4-2}

\usepackage{amsmath, amssymb, amsfonts, mathtools, cases, slashed, bm, bbm}
\usepackage{array}
\usepackage{verbatim}
\usepackage{epsfig}
\usepackage{graphicx, graphics, color, epsfig, float}
\usepackage[normalem]{ulem}
\usepackage[colorlinks=true,citecolor=blue,linkcolor=blue,urlcolor=blue]{hyperref}
\usepackage[capitalise]{cleveref}
\usepackage[dvipsnames]{xcolor}
\usepackage{multirow}
\usepackage{graphicx} %% include figure files
\usepackage{dcolumn} %% align table columns on decimal point
\usepackage{slashed}
\usepackage{xpatch}
\usepackage[mathlines]{lineno} %% enable numbering of text and display math
\usepackage{comment}
\usepackage{soul}
\usepackage{xcolor}
\usepackage{xfrac}
\usepackage{tabularx}
\usepackage{mathrsfs}
\usepackage{physics}
\usepackage{placeins}
\usepackage{dsfont}
\usepackage{hhline}
\usepackage{amssymb}

\newcommand{\diff}[1]{\mathrm{d}#1}
\newcommand{\xb}{x_{\rm bj}}

\newcommand{\chired}{{\chi^2_{\rm red}}}

\newcommand{\sa}{\sphericalangle}

\newcommand{\Mhb}{\overline{M}_h}

\newcommand{\mean}[1]{\langle {#1} \rangle}

\begin{document}

\preprint{JLAB-THY-23-3901}
%\title{First Simultaneous Global QCD Analysis of Dihadron Fragmentation Functions \\ and Transversity Parton Distribution Functions}
\title{First simultaneous global QCD analysis of dihadron fragmentation functions \\ and transversity parton distribution functions}

\newcommand*{\TU}{Department of Physics, SERC, Temple University, Philadelphia, Pennsylvania 19122, USA}\affiliation{\TU}
\newcommand*{\LVC}{Department of Physics, Lebanon Valley College, Annville, Pennsylvania 17003, USA}\affiliation{\LVC}
\newcommand*{\PSU}{Division of Science, Penn State University Berks, Reading, Pennsylvania 19610, USA}\affiliation{\PSU}
\newcommand*{\JLAB}{Jefferson Lab, Newport News, VA 23606, USA}\affiliation{\JLAB}
\newcommand*{\RHIKEN}{RIKEN BNL Research Center, Upton, New York 11973, USA}
\affiliation{\RHIKEN}

\author{C.~Cocuzza}\affiliation{\TU}
\author{A.~Metz}\affiliation{\TU}
\author{D.~Pitonyak}\affiliation{\LVC}
\author{A.~Prokudin}\affiliation{\PSU}\affiliation{\JLAB}
\author{N.~Sato}\affiliation{\JLAB}
\author{R.~Seidl}\affiliation{\RHIKEN}
\collaboration{{\bf Jefferson Lab Angular Momentum (JAM) Collaboration}}

\begin{abstract}
We perform a comprehensive study within quantum chromodynamics (QCD) of dihadron observables in  electron-positron annihilation, semi-inclusive deep-inelastic scattering, and proton-proton collisions, including recent cross section data from Belle and azimuthal asymmetries from STAR.  We extract simultaneously for the first time $\pi^+\pi^-$ dihadron fragmentation functions (DiFFs) and the nucleon transversity distributions for up and down quarks as well as antiquarks.  For the transversity distributions we impose their small-$x$ asymptotic behavior and the Soffer bound. In addition, we utilize a new definition of DiFFs that has a number density interpretation to then calculate expectation values for the dihadron invariant mass and momentum fraction.  Furthermore, we investigate the compatibility of our transversity  results with those from single-hadron fragmentation (from a transverse momentum dependent/collinear twist-3 framework) and the nucleon tensor charges computed in lattice QCD.  We find a universal nature to all of this available information. 
Future measurements of dihadron production can significantly further this research, especially, as we show, those that are sensitive to the region of large parton momentum fractions.
\end{abstract}

\date{\today}
\maketitle

%%%%%%%%%%%%%%%%%%%%%%%%%%%%%%%%%%%%%%%%%%
\section{Introduction}
\label{s.intro}
%%%%%%%%%%%%%%%%%%%%%%%%%%%%%%%%%%%%%%%%%%

Phenomenological analyses of hadrons, at the level of quarks and gluons (partons), based on high-energy collision measurements rely on two key ingredients:~parton distribution functions (PDFs) and fragmentation functions (FFs).  The former provides insight into the momentum-space partonic structure of the incoming nucleon while the latter encodes how a parton forms multiple (colorless) particles in the final state.  With regard to FFs, there are two common situations:~experiments tagging on one hadron or on a hadron pair (dihadron) within a parton-initiated jet.  The single-hadron unpolarized collinear FF $D_1^{h/i}(z)$ ($h$ denoting the hadron type, $i$ the parton, and $z$ the momentum fraction carried by the hadron) has been  studied in observables described by leading-twist collinear factorization and extracted by numerous groups over the last nearly four decades~\cite{EuropeanMuon:1985nbu, EuropeanMuon:1989srv, Chiappetta:1992uh, Binnewies:1994ju, Binnewies:1995pt, Binnewies:1995kg, deFlorian:1997zj, Kretzer:2000yf, Bourhis:2000gs, Kniehl:2000fe, Kretzer:2001pz, Bourrely:2003wi, Hirai:2007cx, deFlorian:2007aj, deFlorian:2007ekg, Albino:2008fy, Aidala:2010bn, deFlorian:2014xna, Leader:2015hna, Anderle:2015lqa, COMPASS:2016xvm, Sato:2016wqj, deFlorian:2017lwf, Ethier:2017zbq, Bertone:2017tyb, Soleymaninia:2018uiv, Bertone:2018ecm, Soleymaninia:2020bsq, Moffat:2021dji, Khalek:2021gxf, Abdolmaleki:2021yjf, Borsa:2021ran, Borsa:2022vvp, Soleymaninia:2022alt, AbdulKhalek:2022laj, Cocuzza:2022jye}.  When transverse momentum dependent (TMD) observables are considered, then, for unpolarized hadrons, not only is $D_1^{h/i}(z,z^2\vec{k}_T^2)$ relevant ($\vec{k}_T$ giving the transverse momentum of the parton relative to the hadron), but also the chiral-odd Collins TMD FF $H_1^{\perp\,h/i}(z,z^2\vec{k}_T^2)$~\cite{Collins:1992kk}.  Both $D_1^{h/i}(z,z^2\vec{k}_T^2)$~\cite{Anselmino:2005nn, Anselmino:2013lza, Signori:2013mda, Bacchetta:2017gcc, Scimemi:2019cmh, Bacchetta:2022awv, Boglione:2022nzq, Boglione:2023duo} and $H_1^{\perp\,h/i}(z,z^2\vec{k}_T^2)$~\cite{Anselmino:2007fs, Anselmino:2008jk, Anselmino:2013vqa, Anselmino:2015sxa, Kang:2015msa, Lin:2017stx, DAlesio:2020vtw, Cammarota:2020qcw, Gamberg:2022kdb} have been extracted using processes described by leading-power TMD factorization. 
We refer to Ref.~\cite{Metz:2016swz} for a review on FFs.

On the dihadron side, the FFs depend on more variables~\cite{Bianconi:1999cd,Bacchetta:2002ux}, which can be chosen to be the dihadron total momentum fraction $z$, the relative momentum fraction $\zeta$, the relative transverse momentum $\vec{R}_T$ (whose magnitude can be related to the invariant mass $M_h$ of the dihadron~\cite{Bianconi:1999cd, Bacchetta:2002ux}), along with the parton transverse momentum $\vec{k}_T$.  
Consequently, even after integrating over $\vec{k}_T$, two dihadron FFs (DiFFs) remain that can be studied within leading-twist collinear factorization~\cite{Collins:1993kq, Collins:1994ax, Jaffe:1997hf, Jaffe:1997pv, Bianconi:1999cd, Bacchetta:2002ux}:~$D_1^{h_1h_2/i}(z,M_h)$ and $H_1^{\sphericalangle\,h_1h_2/i}(z,M_h)$, with the latter being chiral-odd and sometimes referred to as the interference FF (IFF).
(Although both $D_1$ and $H_1^{\sa}$ are DiFFs, we will often use the term IFF to distinguish $H_1^{\sa}$ from $D_1$.)  
Unlike the single-hadron unpolarized (collinear and TMD) FFs, only one group has extracted these DiFFs~\cite{Courtoy:2012ry,Radici:2015mwa}, and only one measurement exists ($\diff \sigma/ \diff z\, \diff M_h$ cross section from Belle for $e^+e^-\to (h_1h_2)\,X$~\cite{Belle:2017rwm}, which postdates Refs.~\cite{Courtoy:2012ry,Radici:2015mwa}) that is directly sensitive to $D_1^{h_1h_2/i}(z,M_h)$.  We note that a new definition of DiFFs has recently been introduced~\cite{Pitonyak:2023gjx} that, in particular, allows $D_1^{h_1h_2/i}(z,M_h)$ to retain a number density interpretation, from which one can meaningfully calculate expectation values. 

With regard to PDFs, at leading twist the longitudinal momentum structure of the nucleon is characterized by the spin-averaged PDF $f_1(x)$, the helicity PDF $g_1(x)$, and the transversity PDF $h_1(x)$, with $x$ being the momentum fraction carried by the parton.  The transversity PDF quantifies the degree of transverse polarization of quarks within a transversely polarized nucleon. It is also used to calculate the tensor charges of the nucleon:
\begin{align}
\delta u = \int_0^1 \!\!\diff x ~h_1^{u_v}(x;\mu)\,, ~~~~ 
\delta d = \int_0^1 \!\!\diff x ~h_1^{d_v}(x;\mu)\,,
\label{e.tensorcharge}
\end{align}
where $h_1^{q_v} \equiv h_1^q - h_1^{\bar{q}}$ are the valence distributions, and $\mu$ is the renormalization scale.  In addition to phenomenological analyses of experimental data~\cite{Anselmino:2007fs, Anselmino:2008jk, Bacchetta:2012ty, Anselmino:2013vqa, Goldstein:2014aja, Anselmino:2015sxa, Radici:2015mwa, Kang:2015msa, Lin:2017stx, Radici:2018iag, Benel:2019mcq, DAlesio:2020vtw, Cammarota:2020qcw, Gamberg:2022kdb}, the tensor charges can be found through ab initio computations in lattice QCD (LQCD)~\cite{Gupta:2018qil, Gupta:2018lvp, Yamanaka:2018uud, Hasan:2019noy, Alexandrou:2019brg, Harris:2019bih, Horkel:2020hpi, Alexandrou:2021oih, Park:2021ypf, Tsuji:2022ric, QCDSFUKQCDCSSM:2023qlx, Smail:2023eyk} and model calculations~\cite{He:1994gz, Barone:1996un, Schweitzer:2001sr, Gamberg:2001qc, Pasquini:2005dk, Wakamatsu:2007nc, Lorce:2007fa, Yamanaka:2013zoa, Pitschmann:2014jxa, Xu:2015kta, Wang:2018kto, Liu:2019wzj}.

The chiral-odd nature of transversity makes it difficult to extract (relative to the spin-averaged and helicity PDFs) since it must couple to another chiral-odd function.  Three possible candidates are the aforementioned Collins TMD FF $H_1^{\perp\,h/i}(z,z^2\vec{k}_T^2)$ and IFF $H_1^{\sphericalangle\,h_1h_2/i}(z,M_h)$ as well as chiral-odd higher-twist (twist-3) collinear FFs~\cite{Metz:2012ct, Kanazawa:2014dca,Gamberg:2017gle, Cammarota:2020qcw, Gamberg:2022kdb}.  Analyses that extract $h_1(x)$, or its TMD version $h_1(x,\vec{k}_T^2)$, have been performed using single-hadron TMD processes~\cite{Anselmino:2007fs, Anselmino:2008jk, Anselmino:2013vqa, Anselmino:2015sxa, Kang:2015msa, Lin:2017stx, DAlesio:2020vtw}, both single-hadron TMD and collinear twist-3 observables~\cite{Cammarota:2020qcw, Gamberg:2022kdb}, or dihadron reactions~\cite{Bacchetta:2012ty,Radici:2015mwa,Radici:2018iag,Benel:2019mcq}.  Specifically, transverse single-spin asymmetries can be investigated in electron-positron ($e^+ e^-$) annihilation, semi-inclusive deep-inelastic scattering (SIDIS), and proton-proton ($pp$) collisions with transversely polarized nucleons and/or partons involved where either a single hadron or dihadron is detected in the final state.  For the single-hadron case, simultaneous extractions of $h_1(x,\vec{k}_T^2)$ and $H_1^{\perp \,h/i}(z,z^2\vec{k}_T^2)$ in $e^+ e^-$ and SIDIS have been carried out~\cite{Anselmino:2007fs, Anselmino:2008jk, Anselmino:2013vqa, Anselmino:2015sxa, Kang:2015msa, Lin:2017stx, DAlesio:2020vtw}.  Moreover, global analyses that also included $pp$ data and further extracted the chiral-odd collinear twist-3 FF $\tilde{H}^{h/i}(z)$ and Sivers TMD PDF $f_{1T}^\perp(x,\vec{k}_T^2)$ have been performed (JAM3D)~\cite{Cammarota:2020qcw, Gamberg:2022kdb}.  For the dihadron case, $h_1(x)$, $H_1^{\sphericalangle\,h_1h_2/i}(z,M_h)$, and $D_1^{h_1h_2/i}(z,M_h)$ have all been extracted separately~\cite{Courtoy:2012ry,Bacchetta:2012ty,Radici:2015mwa,Radici:2018iag,Benel:2019mcq}, but no group has considered a simultaneous extraction of the three functions using all available dihadron measurements/kinematic binnings.

Given the above discussion, the motivation for this work is the following:
\begin{enumerate}
    \item Perform the first simultaneous global QCD analysis (JAMDiFF) of the $\pi^+ \pi^-$ DiFFs and transversity PDFs (for up and down quarks and antiquarks) from $e^+ e^-$ annihilation, SIDIS, and $pp$ data.
    \item Include, for the first time, the Belle cross section data~\cite{Belle:2017rwm}, the latest $pp$ dihadron measurements from STAR~\cite{STAR:2017wsi}, and all kinematic variable binnings for the relevant processes under consideration, making this the most comprehensive study of dihadron observables to date.
    \item Examine the compatibility between phenomenological results for $h_1(x)$ and/or the tensor charges  based on the dihadron approach, those in the TMD/collinear twist-3 framework, and LQCD computations by implementing theoretical constraints at small $x$~\cite{Kovchegov:2018zeq} and large~$x$~\cite{Soffer:1994ww} (where experimental data is absent) in order to meaningfully calculate the integrals in Eq.~(\ref{e.tensorcharge}) and include LQCD data into the analysis.
\end{enumerate}

Connected to point (3), the ability to probe certain low-energy beyond the Standard Model (BSM) physics relies on knowledge of the isovector tensor charge $g_T = \delta u - \delta d$ as well as the individual quark tensor charges $\delta u$ and $\delta d$ (see, e.g., Refs.~\cite{Herczeg:2001vk, Erler:2004cx, Severijns:2006dr, Cirigliano:2013xha, Courtoy:2015haa, Gonzalez-Alonso:2018omy, Erler:2004cx, Pospelov:2005pr, Yamanaka:2017mef, Liu:2017olr}).  Therefore, it is crucial to have precise values for $\delta u$, $\delta d$, and $g_T$ and to test how compatible they are between different approaches used for their determination.
To highlight the importance of this issue, we have assembled essential results and discussion regarding our extraction of transversity and calculation of the tensor charges, as well as a comparison to other phenomenological work and lattice QCD, in Ref.~\cite{Cocuzza:2023oam}, with some supplemental details on that aspect of the analysis provided here alongside points (1) and (2).

We organize the paper as follows.  
In Sec.~\ref{s.processes} we summarize the factorization formulas for all relevant observables, while in Sec.~\ref{s.pheno} we discuss the parameterization choices for the DiFFs, IFFs, and transversity PDFs as well as review the Bayesian methodology employed for the analysis.
In Sec.~\ref{s.data} we compare our theoretical calculations to the data and discuss the quality of the fit.
In Sec.~\ref{s.DiFFs} we show and discuss the extracted DiFFs and IFFs as well as calculate expectation values for $z$ and $M_h$.
In Sec.~\ref{s.TPDFs} we present the extracted transversity PDFs and tensor charges, comparing to findings from other phenomenological analyses and LQCD, and providing additional information beyond Ref.~\cite{Cocuzza:2023oam}. We also elaborate on the importance of high-$x$ measurements in further testing the compatibility between phenomenology and LQCD results for the tensor charges.
Finally, in Sec.~\ref{s.summary} we summarize our results and discuss future directions for this analysis, including the role of new experimental data.

%%%%%%%%%%%%%%%%%%%%%%%%%%%%%%%%%%%%%%%%%%
\section{Experimental Observables Used in the Global Analysis}
\label{s.processes}
%%%%%%%%%%%%%%%%%%%%%%%%%%%%%%%%%%%%%%%%%%

In this analysis we study $\pi^+\pi^-$ dihadron production in $e^+ e^-$ annihilation, SIDIS, and proton-proton collisions with the theoretical formulas given at leading order (LO) in the strong coupling  $\alpha_s$ using leading-twist collinear factorization.  Since we only consider $\pi^+\pi^-$ pairs, we will drop the superscript from our notation for the DiFFs.
In the subsections below we summarize each of the three processes and the relevant equations.  We note that the formulas in this section use a new definition of the DiFFs that has a number density interpretation~\cite{Pitonyak:2023gjx}.

%%%%%%%%%%%%%%%%%%%%%%%%%%%%%%%%%%%%%%%%%%
\subsection{Dihadron Production in Electron-Positron Annihilation}
\label{ss.SIA}
%%%%%%%%%%%%%%%%%%%%%%%%%%%%%%%%%%%%%%%%%%

In the $e^+ e^-$ annihilation process, $e^+ e^- \rightarrow (\pi^+ \pi^-) X$, an electron and positron annihilate to form an inclusive spectrum of $\pi^+ \pi^-$ pairs with invariant mass $M_h$ and fractional energy $z$.  The center of mass (COM) energy is denoted by $\sqrt{s}$.
The cross section for this process is given by~\cite{Pitonyak:2023gjx}
\begin{align}
\frac{\diff \sigma}{\diff z\, \diff M_h} = \frac{4\pi N_c\alpha_{\rm em}^2}{3s}\sum_q \bar{e}_q^2\, D_1^{q}(z,M_h)\,,
\label{e.dsigdzdMh}
\end{align}
where the sum is over quarks and antiquarks, $\alpha_{\rm em}$ is the fine structure constant, and the hard scale of the process is set to be $\mu = \sqrt{s}$.
In addition, one has
\begin{align}
\bar{e}_q^2 = e_q^2 + (1+a_Z^2) (1+a_q^2) \frac{G_F^2 M_Z^4 s^2}{128 \pi^2 \alpha_{\rm em}^2 [(s-M_Z^2) + \Gamma_Z^2 M_Z^2]}  - a_Z e_q a_q \frac{G_F M_Z^2 s (s-M_Z^2)}{4 \sqrt{2} \pi \alpha_{\rm em} [(s-M_Z^2) + \Gamma_Z^2 M_Z^2]} \,,
\end{align}
where $e_q$ is the charge of the quark of flavor $q$ in units of the elementary charge,  $G_F$ is the Fermi constant, and $M_Z$ and $\Gamma_Z$ are the mass and decay width of the $Z$ boson, respectively.
Furthermore, $a_Z = -1 + 4 \sin^2{\theta_W}$, and
$a_q = 1 - 8/3 \sin^2{\theta_W}$ for $q = u, c, \bar{u}, \bar{c}$ or
$a_q = -1 + 4/3 \sin^2{\theta_W}$ for $q = d, s, b, \bar{d}, \bar{s}, \bar{b}$,
where $\theta_W$ is the weak mixing angle.
As we will argue below (see Eq.~(\ref{e.D1symmetry}) and Appendix \ref{a.symmetry}),
there are five independent DiFFs ($D_1^u, D_1^s, D_1^c$, $D_1^b$, and $D_1^g$), and this observable is only capable of constraining one of them (which we choose to be $D_1^u$).  Thus, we  supplement the experimental data from Belle with information from PYTHIA~\cite{Sjostrand:2001yu}, which will be discussed in more detail in Sec.~\ref{ss.PYTHIA}.

We also consider the process $e^+ e^- \rightarrow (\pi^+ \pi^-) (\pi^+ \pi^-) X$ where two dihadron pairs are detected, with the second having invariant mass $\overline{M}_h$ and fractional energy $\bar{z}$.
By measuring an azimuthal correlation of two hadron pairs detected in opposite hemispheres, one obtains an observable that is sensitive to the IFF $H_1^{\sa}(z,M_h)$.
This modulation (known as the Artru-Collins asymmetry \cite{Artru:1995zu} and denoted as $a_{12R}$ by Belle~\cite{Belle:2011cur}) is given by \cite{Boer:2003ya,Bacchetta:2008wb, Courtoy:2012ry,Radici:2015mwa,Matevosyan:2018icf, Artru:1995zu}
\begin{align}
A^{e^+ e^-}(z, M_h, \bar{z}, \overline{M}_h)=\frac{ \sin^2 \theta \sum_q e_q^2 \, H_1^{\sa,q}(z,M_h) H_1^{\sa,\bar{q}}(\bar{z},\overline{M}_h)}{(1+\cos^2\theta)\sum_q e_q^2 \, D_1^{q}(z,M_h) D_1^{\bar{q}}(\bar{z},\overline{M}_h)}\,,
\label{e.artrucollins}
\end{align}
where $\theta$ is defined as the polar angle between the beam axis and the reference axis in the COM system \cite{Boer:2003ya}.
As we will argue later (see Eq.~(\ref{e.H1symmetry}) and Appendix \ref{a.symmetry}), there is only one IFF that is needed for phenomenology,
which we choose to be $H_1^{\sa, u}$.  Thus, only one observable is needed to constrain the IFFs.
However, one can see from Eq.~(\ref{e.artrucollins}) that $A^{e^+ e^-}$ is proportional to $[H_1^{\sa,u}]^2$.  Consequently, the asymmetry cannot uniquely determine the sign of $H_1^{\sa,u}$, and it must be fixed by hand.
Given the sign of the SIDIS asymmetries from experimental measurements (see the following section) and assuming that the transversity up quark PDF must be positive (as is found in all phenomenological, model, and lattice QCD studies) leads to the conclusion that $H_1^{\sa,u}$ must be negative (which is also supported by model calculations for $H_1^{\sa}$~\cite{Matevosyan:2017uls}). In the analysis we therefore choose $H_1^{\sa,u}$ to be negative.

%%%%%%%%%%%%%%%%%%%%%%%%%%%%%%%%%%%%%%%%%%
\subsection{Dihadron Production in SIDIS}
\label{ss.SIDIS}
%%%%%%%%%%%%%%%%%%%%%%%%%%%%%%%%%%%%%%%%%%

The SIDIS process is given by $\ell N^{\uparrow} \to \ell' (\pi^+ \pi^-) X$, where a lepton with 4-momentum $k$ scatters off a transversely polarized proton ($N=p$) or deuteron ($N=D$) with 4-momentum $P$ and is detected with 4-momentum $k'$ along with a $\pi^+ \pi^-$ pair with 4-momentum $P_h$.
The 4-momentum transfer squared is given by $Q^2 \equiv -(k - k')^2$, and we define the Bjorken scaling variable $\xb \equiv Q^2/(2 P \cdot q)$ and the inelasticity $y \equiv P \cdot q/P \cdot k$.
We denote the virtual-photon 4-momentum and relative 4-momentum between the two hadrons (divided by 2) by $q$ and $R$, respectively.
The transverse component $\vec{R}_T$ of $R$ makes an azimuthal angle $\phi_R$ about the virtual-photon direction, and the transverse nucleon spin $\vec{S}_T$ makes an azimuthal angle $\phi_S$.
COMPASS \cite{COMPASS:2023cgk} defines the angle $\phi_{RS}$ as $\phi_{RS} \equiv \phi_R + \phi_S - \pi$, while HERMES \cite{HERMES:2008mcr} defines it as $\phi_{RS} \equiv \phi_R + \phi_S$.

The asymmetry (usually denoted as $A_{UT}^{\sin{(\phi_R + \phi_S)}\sin{\theta}}$ \cite{HERMES:2008mcr} or $A_{UT}^{\sin{(\phi_R + \phi_S)}}$ \cite{COMPASS:2023cgk}) can be written as~\cite{Bacchetta:2002ux, Bacchetta:2003vn, Bacchetta:2011ip, Bacchetta:2012ty, Radici:2015mwa}
\begin{align}
A_{UT}^{\rm SIDIS} = c(y) \frac{\sum_q e_q^2\,h_1^q(x)\,H_1^{\sphericalangle,q}(z,M_h)}{\sum_q e_q^2\,f_1^q(x)\,D_1^{q}(z,M_h)}\,.
\label{e.AUTSIDIS}
\end{align}
The factor $c(y)$ is given by
\begin{align}
c(y) 
= \begin{cases}
    \hspace{1cm} 1 ~~ \hspace{0.75cm}({\rm COMPASS})\,, \notag \\[0.3cm]
 -\dfrac{1-y}{1-y+\frac{y^2}2} ~~ ({\rm HERMES})\,. \notag
\end{cases}
\end{align}
The opposite sign between the asymmetries for COMPASS and HERMES comes from the different definitions for the angle $\phi_{RS}$ mentioned above.
The hard scale is set to be $\mu = Q$, and the unpolarized PDF $f_1$ is taken from Ref.~\cite{Cocuzza:2022jye} (we utilize only the mean value).

Both COMPASS and HERMES measured the asymmetry for proton targets, while COMPASS also used a deuteron target.
For the deuteron target we neglect nuclear corrections and take the (unpolarized and transversity) PDFs in Eq.~(\ref{e.AUTSIDIS}) to be the average of proton and neutron PDFs, using isospin symmetry to relate the neutron PDFs to those of the proton.
Due to the symmetry of the $\pi^+ \pi^-$ IFFs (see Eq.~(\ref{e.H1symmetry})), one has in the numerator for a proton target
\begin{align}
\sum_q e_q^2\,h_1^q\,H_1^{\sa,q} = \frac{1}{9}H_1^{\sa,u} [4 (h_1^u - h_1^{\bar{u}}) - (h_1^d - h_1^{\bar{d}})] = \frac{1}{9}H_1^{\sa,u} [4 h_1^{u_v} - h_1^{d_v}] \approx \frac{4}{9}H_1^{\sa,u}  h_1^{u_v}, 
\label{e.SIDISproton}
\end{align}
while for the deuteron one has
\begin{align}
\sum_q e_q^2\,h_1^{q/D}\,H_1^{\sa,q} = \frac{1}{6}H_1^{\sa,u} [(h_1^u - h_1^{\bar{u}}) + (h_1^d - h_1^{\bar{d}})] = \frac{1}{6}H_1^{\sa,u} [h_1^{u_v} + h_1^{d_v}]\,,
\label{e.SIDISdeuteron}
\end{align}
where $h_1^{q/D}$ are the deuteron transversity PDFs.
From this one immediately sees that the SIDIS data is sensitive only to the valence  transversity  PDFs~\cite{Bacchetta:2012ty,Radici:2015mwa,Benel:2019mcq}, with the proton asymmetry primarily sensitive to $u_v$ and the deuteron asymmetry equally sensitive to $u_v$ and $d_v$.

%%%%%%%%%%%%%%%%%%%%%%%%%%%%%%%%%%%%%%%%%%
\subsection{Dihadron Production in Proton-Proton Collisions}
\label{ss.pp_asym}
%%%%%%%%%%%%%%%%%%%%%%%%%%%%%%%%%%%%%%%%%%

In the process $p^{\uparrow} p \to (\pi^+ \pi^-) X$, a transversely polarized proton with 4-momentum $P_A$ collides with an unpolarized proton with 4-momentum $P_B$ producing an outgoing dihadron pair with 4-momentum $P_h$ (with $P_{hT}$ the magnitude of the transverse component) and pseudo-rapidity $\eta$.
The measured asymmetry (denoted as $A_{UT}$ by STAR \cite{STAR:2015jkc}) is given by~\cite{Radici:2016lam}
\begin{equation}
A_{UT}^{pp} = \frac{\mathcal{H}(M_h,P_{hT},\eta)}{\mathcal{D}(M_h,P_{hT},\eta)}\,,
\label{e.AUTpp}
\end{equation}
where the numerator and denominator read \cite{Bacchetta:2004it}
\begin{align}
\mathcal{H}(M_h,P_{hT},\eta) &= 2 P_{hT} \sum_i \sum_{a,b,c,d} \int_{x_{a}^{\rm min}}^1 \diff x_a\int_{x_{b}^{\rm min}}^1 \frac{ \diff x_b}{z}  h_1^a(x_a)\,f_1^b(x_b)\frac{\diff \Delta\hat{\sigma}_{a^\uparrow b\to c^\uparrow d}}{\diff \hat{t}}\,H_1^{\sa,c}(z,M_h)\,, \label{e.ppUT} \\
\mathcal{D}(M_h,P_{hT},\eta) &= 2 P_{hT} \sum_i \sum_{a,b,c,d} \int_{x_{a}^{\rm min}}^1 \diff x_a\int_{x_{b}^{\rm min}}^1 \frac{\diff x_b}{z} 
f_1^a(x_a)\,f_1^b(x_b) \frac{\diff \hat{\sigma}_{ab\to cd}}{\diff \hat{t}} \,D_1^{c}(z,M_h)\,, \label{e.ppUU}
\end{align}
with $x_a$ ($x_b$) the momentum fraction of the parton coming from the proton with momentum $P_A$ ($P_B$). 
We note that $A_{UT}^{pp}$ is sensitive to both the quark and antiquark transversity PDFs.
The sum $\sum_i$ is over all partonic sub-processes, with $\sum_{a,b,c,d}$ summing over all possible parton flavor combinations for a given channel.
The relevant Mandelstam variables are $s = (P_A+P_B)^2$, $t = (P_A-P_h)^2$, and $u = (P_B-P_h)^2$.  The total momentum fraction of the dihadron is fixed at LO to be 
\begin{align}
z=\frac{P_{hT}}{\sqrt{s}}\left(\frac{x_ae^{-\eta}+x_be^\eta}{x_a x_b}\right).
\end{align}
The hard scale is set to be $\mu = P_{hT}$.
The hard (perturbative) partonic cross sections $\diff \hat{\sigma}$ and $\diff \Delta \hat{\sigma}$~\cite{Bacchetta:2004it}
depend on $\hat{s}=x_ax_b s$, $\hat{t} = x_a s/z$, and $\hat{u} = x_b u/z$.
The explicit expressions for $\diff \Delta \hat{\sigma}$ can be found in Appendix \ref{a.pp}.
The limits on the integration are
\vspace{-0.1cm}
\begin{align}
x_{a}^{\rm min} = \frac{P_{hT} e^\eta}{\sqrt{s}-P_{hT}e^{-\eta}}\,, ~~ x_{b}^{\rm min} = \frac{x_a P_{hT} e^{-\eta}}{x_a\sqrt{s}-P_{hT} e^\eta}\,.
\end{align}

%%%%%%%%%%%%%%%%%%%%%%%%%%%%%%%%%%%%%%%%%%
\section{Phenomenological Methodology}
\label{s.pheno}
%%%%%%%%%%%%%%%%%%%%%%%%%%%%%%%%%%%%%%%%%%

In this section we discuss the parameterization of the DiFFs, IFFs, and transversity PDFs. 
All functions are evolved using the DGLAP equations~\cite{Gribov:1972ri, Dokshitzer:1977sg, Altarelli:1977zs} appropriate for the transversity PDFs~\cite{Baldracchini:1980uq, Artru:1989zv, Blumlein:2001ca, Stratmann:2001pt} and DiFFs/IFFs~\cite{Pitonyak:2023gjx} with LO splitting functions. 
We mention that evolution equations were previously derived for ``extended'' DiFFs (at LO) in Ref.~\cite{Ceccopieri:2007ip} (and commented on in Ref.~\cite{Bacchetta:2008wb})
and for collinear DiFFs (functions of $(z_1,z_2)$) in Refs.~\cite{Sukhatme:1981ym, deFlorian:2003cg, Majumder:2004br, Majumder:2004wh} at LO and recently in Refs.~\cite{Chen:2022pdu, Chen:2022muj} at next-to-leading order (NLO).
The strong coupling $\alpha_s$ is evolved at leading-logarithmic accuracy with the boundary condition $\alpha_s(M_Z)=0.118$. 
We use the zero mass variable flavor scheme for DGLAP evolution.
The values of the heavy quark mass thresholds for the evolution are taken from the PDG as $m_c=1.28$~GeV and $m_b=4.18$~GeV in the $\overline{\textrm{MS}}$ scheme~\cite{ParticleDataGroup:2018ovx}.

%%%%%%%%%%%%%%%%%%%%%%%%%%%%%%%%%%%%%%%%%%
\subsection{DiFF Parameterization}
\label{ss.DiFF_par}
%%%%%%%%%%%%%%%%%%%%%%%%%%%%%%%%%%%%%%%%%%

The DiFF $D_1$ for $\pi^+ \pi^-$ production satisfies (see Appendix \ref{a.symmetry} and Ref.~\cite{Courtoy:2012ry})
\begin{align}
D_1^u = D_1^d &= D_1^{\bar{u}} = D_1^{\bar{d}}, \notag \\
D_1^s = D_1^{\bar{s}}, ~~~ D_1^c &= D_1^{\bar{c}}, ~~~ D_1^b = D_1^{\bar{b}}\,,
\label{e.D1symmetry}
\end{align}
which means, including the gluon, there are five independent $D_1$ functions to be fitted.
$D_1^u$, $D_1^s$, and $D_1^g$ are parameterized at the input scale $\mu_0 = 1$ GeV, while $D_1^c$ and $D_1^b$ are parameterized at $\mu = m_c$ and $\mu = m_b$, respectively.
The primary challenge with extracting DiFFs is their dependence on both $z$ and $M_h$.  Here we choose to discretize the $M_h$ dependence while using an explicit functional form for the $z$ dependence.  Such a choice has the computational advantage of being convertible to Mellin space.  For $D_1^u$, we choose the following $M_h$ grid:
\begin{eqnarray}
{\mathbf M}_{h}^u &=& [2 m_{\pi}, 0.40, 0.50, 0.70, 0.75, 0.80, 0.90, 1.00, 1.20, 1.30, 1.40, 1.60, 1.80, 2.00] ~{\rm GeV}. \notag 
\label{e.Mugrid}
\end{eqnarray}
We note that the grid is not uniform and is instead chosen in a way to best describe the detailed resonance structure of the $e^+ e^-$ cross section \cite{Belle:2017rwm}.  
For $s,c,b$ and $g$ we choose grids that are less dense:
\begin{eqnarray}
{\mathbf M}_h^s = {\mathbf M}_h^c &=& [2 m_{\pi}, 0.50, 0.75, 1.00, 1.20, 1.60, 2.00] ~{\rm GeV}, \notag \\
{\mathbf M}_h^b &=& [2 m_{\pi}, 0.70, 1.00, 1.40, 2.00] ~{\rm GeV}, \\ \notag
{\mathbf M}_h^g &=& [2 m_{\pi}, 0.70, 1.40, 2.00] ~{\rm GeV}. \notag
\label{e.Mqgrids}
\end{eqnarray}
We find that these grids, and the parameters associated with them (see Eq.~(\ref{e.zparD1})), are necessary and sufficient to describe the Belle and PYTHIA data.
At each value of $M_h$ on the grid, denoted by ${\rm M}_h^{i,j}$, the $z$ dependence is parameterized as
\begin{eqnarray}
D_1^i(z, {\rm M}_h^{i,j}) = \sum_{k=1,2,3} \frac{N_{jk}^i}{{\cal M}_{jk}^i} z^{\alpha_{jk}^i} (1-z)^{\beta_{jk}^i},
\label{e.zparD1}
\end{eqnarray}
where $i=u,s,c,b,g$, $\bm{a} \equiv \{ N_{jk}^i, \alpha_{jk}^i, \beta_{jk}^i\}$ is the set of parameters to be inferred, and 
    ${\cal M}_{jk}^i = {\rm B}[\alpha_{jk}^i+1,\beta_{jk}^i+1]$, 
where $B(a,b)$ is the beta function, normalizes the function to the first moment to largely decorrelate the normalization and shape parameters.
For the up quark, it is necessary to include terms up to $k=3$ in order to describe the Belle cross section data.  
The other functions will be constrained by PYTHIA-generated data (see Sec.~\ref{ss.PYTHIA}) and we find that only the $k=1$ term is necessary to describe them.  This leads to $14 \times 9 = 126$ parameters for $D_1^u$, $7 \times 3 = 21$ parameters each for $D_1^s$ and $D_1^c$, $5 \times 3 = 15$ parameters for $D_1^b$, and $4 \times 3 = 12$ parameters for $D_1^g$, for a total of 195 free parameters.
The parameters are generally restricted within the ranges $0 < N_{jk}^i < 15$, $-2 < \alpha_{jk}^i < 10$, and $0 < \beta_{jk}^i < 10$, but in practice these ranges may be restricted further depending on $i$, $j$, and $k$.  These ranges are chosen to give the fit sufficient but not unnecessary flexibility.

The functional form in Eq.~(\ref{e.zparD1}) is easily converted into Mellin space at each ${\rm M}_h^{i,j}$.
The inverse Mellin transform requires a complex contour integration (see Eq.~(\ref{e.mellinspace})), which is performed by discretizing the contour. To obtain the function at any value of $M_h$, we interpolate both the real and imaginary components of the function in Mellin space using the values calculated at ${\rm M}_h^{i,j}$.  Once interpolated, the real and imaginary components are recombined and the inverse Mellin transform is performed to obtain the function in momentum space at any value of $M_h$.

We contrast our choice of parameterization with that of Ref.~\cite{Courtoy:2012ry}, which used a considerably more complicated functional form for $u$, $s$, and $c$ with different functions for the continuum and three resonant channels.
The bottom quark DiFF was not included in the analysis of Ref.~\cite{Courtoy:2012ry}, and the gluon DiFF was only generated perturbatively.
In Ref.~\cite{Radici:2018iag} three choices for the gluon DiFF at the input scale were used: $D_1^g = 0$, $D_1^g = D_1^u/4$, and $D_1^g = D_1^u$. 
Our results in Fig.~\ref{f.diffs} suggest that none of these scenarios seem to be correct across all $z$ and $M_h$.
Overall, our parameterization has the advantages of requiring fewer assumptions about the functional forms, reducing complicated correlations between parameters, and being convertible to Mellin space.  The latter two features make the analysis far more computationally efficient.

We also enforce the positivity bound \cite{Bacchetta:2002ux}
\begin{eqnarray}
D_{1}^i(z,M_h;\mu) > 0
\label{e.D1pos}
\end{eqnarray}
approximately on each Monte Carlo replica by using a Bayesian prior (see Sec.~\ref{ss.bayes}) that in effect imposes a penalty on the $\chi^2$ function when the bounds are violated~\cite{Ball:2013lla}.
For each replica and at each step of the $\chi^2$ minimization we first calculate $D_1$ at the input scale $\mu_0 = 1$~GeV at 300 points spaced linearly in the $(z, M_h)$ plane, with $0.2 < z < 1$ and $2 m_{\pi} < M_h < 2.0$ GeV, noting that if positivity is enforced at the input scale it will automatically hold at larger scales for any function that evolves through the DGLAP equation \cite{Altarelli:1998gn, Forte:1998kd}.  We repeat this process for $u,s,c,b,$ and $g$.
Any DiFFs that are negative contribute to the overall $\chi^2$ through
%The prior is given by (see Eq.~(\ref{e.prior}))
%
\begin{eqnarray}
\Delta^2_{D_{1} < 0} = w^2 \Big( \sum_{z,M_h} \sum_i 
\Theta \big(\!-\!D_{1}^i(z,M_h;\mu_0)\big)~ \big\vert D_{1}^i(z,M_h;\mu_0) \big\vert \Big)^2 ,
\label{e.chi2D1pos}
\end{eqnarray}
where the sum $\sum_{z,M_h}$ represents summing over the 300 $(z, M_h)$ points, and $\Theta(X)$ is the step function.
If the contribution is non-zero, it is proportional to the size of the violation.
The weight $w$ is chosen to be $3$ so that the initial contribution to the $\chi^2$ is generally $\mathcal{O}(1000)$, although the size of the initial contribution can vary significantly depending on the starting parameters.  This ensures that the contribution is non-negligible but that it also does not dominate the entire $\chi^2$ function which otherwise is also $\mathcal{O}(1000)$.
This choice ensures that the violation of the bound is small.

%%%%%%%%%%%%%%%%%%%%%%%%%%%%%%%%%%%%%%%%%%
\subsection{IFF Parameterization}
\label{ss.IFF_par}
%%%%%%%%%%%%%%%%%%%%%%%%%%%%%%%%%%%%%%%%%%

The IFF $H_1^{\sa}$ for $\pi^+ \pi^-$ production satisfies (see Appendix \ref{a.symmetry} and Ref.~\cite{Courtoy:2012ry})
\begin{align}
H_1^{\sa,u} = -H_1^{\sa,d} &= -H_1^{\sa,\bar{u}} = H_1^{\sa,\bar{d}}, \notag \\
H_1^{\sa,s} = H_1^{\sa,\bar{s}} = H_1^{\sa,c} &= H_1^{\sa,\bar{c}} = H_1^{\sa,b} = H_1^{\sa,\bar{b}} = 0\,.
\label{e.H1symmetry}
\end{align}
Consequently, there is just a single independent IFF (since the transversity cannot couple to the IFF for gluons at LO), which we choose to be $H_1^{\sa,u}$.
The IFF $H_1^{\sa,u}$ is parameterized similarly to $D_1^i$.  Since the relevant data for $H_1^{\sa,u}$ is comparatively sparse and has larger errors, far fewer parameters are needed.  Thus, we are able to choose a less dense $M_h$ grid
\begin{eqnarray}
{\mathbf M}_h^u &=& [2 m_{\pi}, 0.50, 0.70, 0.85, 1.00, 1.20, 1.60, 2.00] ~{\rm GeV}, \notag
\end{eqnarray}
and at each ${\rm M}_h^{u,j}$, the $z$ dependence is given by
\begin{eqnarray}
H_1^{\sa, u}(z, {\rm M}_h^{u,j}) = \sum_{k=1,2} \frac{N_{jk}^u}{{\cal M}_{jk}^u} z^{\alpha_{jk}^u} (1-z)^{\beta_{jk}^u},
\label{e.zparH1}
\end{eqnarray}
at the input scale $\mu_0 = 1$ GeV.  This leads to a total of $8 \times 6 = 48$ free parameters for the IFF.  
The parameters are restricted within the ranges $-1 < N_{jk}^u < 0$ (see discussion below Eq.~(\ref{e.artrucollins})), $-1 < \alpha_{jk}^u < 5$, and $0 < \beta_{jk}^u < 10$ for all $j$ and $k$.
As with $D_1^i$, this form is easily converted into Mellin space.
%which is again in contrast to the more complicated parameterization used in Refs.~\cite{Courtoy:2012ry,Radici:2015mwa}.

We also enforce the positivity bound \cite{Bacchetta:2002ux}
\begin{eqnarray}
\vert H_{1}^{\sa,i}(z,M_h;\mu) \vert < D_{1}^i(z,M_h;\mu),
\label{e.H1bound}
\end{eqnarray}
which, due to the symmetry relations Eqs.~(\ref{e.D1symmetry}) and (\ref{e.H1symmetry}), only needs to be applied to $i=u$.
Since we have fixed $D_1^u$ to be positive and $H_1^{\sa,u}$ to be negative, we only need to check the condition $D_1^u > -H_1^{\sa,u}$.
This is implemented similarly to the positivity constraint on $D_1$ in Eq.~(\ref{e.D1pos}), with a Bayesian prior (see Sec.~\ref{ss.bayes}) that in effect leads to a $\chi^2$ penalty equal to 
\begin{align}
\Delta^2_{\vert H_{1}^{\sa} \vert > D_{1}} = w^2 \Big( \sum_{z,M_h}
\Theta \big(\! -\![H_{1}^{\sa,u}(z,M_h;\mu_0) + D_{1}^u(z,M_h;\mu_0)]   \big)
~\vert H_{1}^{\sa,u}(z,M_h;\mu_0) + D_{1}^u(z,M_h;\mu_0) \vert  \Big)^2,
\label{e.chi2H1bound}
\end{align}
% %
with $w$ again chosen to be $3$ (see discussion below Eq.~(\ref{e.chi2D1pos})).

%%%%%%%%%%%%%%%%%%%%%%%%%%%%%%%%%%%%%%%%%%
\subsection{Transversity PDF Parameterization}
\label{ss.TPDF_par}
%%%%%%%%%%%%%%%%%%%%%%%%%%%%%%%%%%%%%%%%%%

The transversity PDFs are parameterized at the input scale $\mu_0 = 1$ GeV using the form
\begin{align}
h_1^i(x) 
= \frac{N^i}{{\cal M}^i}\, x^{\alpha^i}(1-x)^{\beta^i}(1+\gamma^i \sqrt{x} + \delta^i \, x),
\label{e.template}
\end{align}
normalized to the first moment ${\cal M}^i = {\rm B}[\alpha^i+1,\beta^i+1] + \gamma^i {\rm B}[\alpha^i + \frac32,\beta^i + 1] + \delta^i \, {\rm B}[\alpha^i + 2,\beta^i + 1]$.
We choose to parameterize the valence distributions $h_1^{u_v}$ and $h_1^{d_v}$, as well as the antiquark distributions $h_1^{\bar{u}} = - h_1^{\bar{d}}$.
Since we only have three unique observables to constrain the transversity PDFs (proton SIDIS, deuteron SIDIS, and $pp$ collisions), we choose the relation between the antiquarks based on predictions from the large-$N_c$ limit \cite{Pobylitsa:2003ty}.
We note that only one previous phenomenological analysis has included antiquark transversity PDFs, which was an exploratory study in Ref.~\cite{Gamberg:2022kdb} that found $h_1^{\bar{u}}$ and $h_1^{\bar{d}}$ to be small and consistent with zero (although had set $h_1^{\bar{u}}=h_1^{\bar{d}}$, at variance with the large-$N_c$ limit).
The $N$ parameter is restricted between $0 < N^i < 1$ for $i=u$ (see discussion below Eq.~(\ref{e.artrucollins})) and $-1 < N^i < 1$ for $i = d, \bar{u}$, while the other parameters are always restricted between $0.09 < \alpha^i < 0.26$ (see discussion below surrounding Eq.~(\ref{e.alpha})), $0 < \beta^i < 20$, $-20 < \gamma^i < 20$, and $-20 < \delta^i < 20$.
We have tested the model dependence of our input transversity functions by adding a second shape of the form in~Eq.~(\ref{e.template}) to all three quark flavors.  We find that the results only change marginally, indicating that the experimental data do not call for a more expressive form for these PDFs. 
We caution, however, that this test is not exhaustive of all possible functional forms one could choose for the transversity functions.
We have a total of $3 \times 5 = 15$ parameters for the transversity PDFs.  Combined with the $195$ parameters for $D_1$, the $48$ parameters for $H_1^{\sa}$, and $7$ normalization parameters (see Eq.~(\ref{e.chi2}) below), we end up with a total of $265$ fitted parameters.

We also place a constraint on the small-$x$ behavior of the transversity PDFs (governed by the $\alpha^i$ parameter in Eq.~(\ref{e.template})), where no experimental data is available.
Theoretical calculations have placed limits on this parameter as $x \rightarrow 0$ (ignoring saturation effects)~\cite{Kovchegov:2018zeq}:
\begin{align}
\alpha^i \xrightarrow[]{x \to 0} 1 - 2 \sqrt{\frac{\alpha_s N_c} {2 \pi}}\,.
\label{e.alpha}
\end{align}
We apply this limit to both the valence quarks and antiquarks, and there is a roughly 50\% uncertainty on this value from $1/N_c$ and NLO corrections~\cite{Kovchegov:2023pc}.
At the input scale, we calculate $\alpha^i \rightarrow 0.17$, and so we limit the $\alpha^i$ parameter to the range $0.085 \leq \alpha^i \leq 0.255$ for all quark flavors.
Strictly speaking, this limit only applies as $x \rightarrow 0$, while our approach places a limit on the entire range of $x$.
We find, however, that limiting $\alpha^i$ as such has no impact on the resulting transversity PDFs in the measured region or on our ability to describe the experimental data.
Thus, this simplified approach is sufficient to capture the $x \rightarrow 0$ behavior while not affecting results at moderate or high $x$.

We also enforce the Soffer bound \cite{Soffer:1994ww} on the transversity PDFs, given by
\begin{eqnarray}
\abs{h_{1}^i(x;\mu)} \leq \frac12 \bigg[ f_1^i(x;\mu) + g_1^i(x;\mu) \bigg],
\label{e.SB}
\end{eqnarray}
where $f_1^i$ and $g_1^i$ are taken from Ref.~\cite{Cocuzza:2022jye}.
This is again enforced through a Bayesian prior (see Sec.~\ref{ss.bayes}) that in effect leads to a $\chi^2$ penalty, similar to Eqs.~(\ref{e.chi2D1pos}), (\ref{e.chi2H1bound}):
\begin{eqnarray}
\Delta^2_{{\rm SB}} = w^2 \Big( \sum_{\pm} \sum_{x} \sum_i 
\Theta \big(\! -\!F_{\pm}^i(x;\mu_0) \big)~ \vert F_{\pm}^i(x;\mu_0) \vert \Big)^2,
\label{e.chi2SB}
\end{eqnarray}
where $i = u, d, \bar{u}, \bar{d}$ and $F_{\pm}^i(x;\mu) \equiv \frac12 \big [ f_1^i(x;\mu) + g_1^i(x;\mu) \big] \pm h_{1}^i(x,\mu)$.
The sum over $\pm$ is due to the fact that the positivity bound applies to the absolute value of $h_1^i$.
The sum $\sum_{x}$ represents summing over 100 points in $x$ linearly spaced in the range $0.001 < x < 0.99$. The weight $w$ is chosen to be 10 (see discussion below Eq.~(\ref{e.chi2D1pos})).
In order to account for the errors on $f_1$ and $g_1$, we add their 1$\sigma$ errors in quadrature, add this error to the mean of the Soffer bound, and use these values in the fit.

%%%%%%%%%%%%%%%%%%%%%%%%%%%%%%%%%%%%%%%%%%
\subsection{Bayesian Analysis}
\label{ss.bayes}
%%%%%%%%%%%%%%%%%%%%%%%%%%%%%%%%%%%%%%%%%%

Our Bayesian analysis consists of sampling the posterior distribution given by
\begin{align}
\mathcal{P}(\bm{a}|{\rm data})
\propto \mathcal{L}(\bm{a},{\rm data})\, \pi(\bm{a}),
\end{align}
with a likelihood function of Gaussian form,
\begin{align}
\mathcal{L}(\bm{a},{\rm data}) 
= \exp\!\Big(\! \!-\!\frac12 \chi^2(\bm{a},{\rm data})\! \Big),
\label{e.likelihood}
\end{align}
and a prior function $\pi(\bm{a})$.
The $\chi^2$ function in Eq.~(\ref{e.likelihood}) is defined for each replica as
\begin{align}
\label{e.chi2}
\chi^2(\bm{a}) &= \sum_{e,i} 
\bigg( 
\frac{d_{e,i} - \sum_k r_{e,k} \beta_{e,i}^k - T_{e,i}(\bm{a})/N_e}{\alpha_{e,i}} \bigg)^{\!2},
\end{align}
where $d_{e,i}$ is the data point $i$ from experimental dataset $e$, and $T_{e,i}$ is the corresponding theoretical value.
Note that the PYTHIA and LQCD data are not included in this sum.
All uncorrelated uncertainties are added in quadrature and labeled by $\alpha_{e,i}$, while $\beta_{e,i}^k$ represents the $k^{\rm th}$ source of point-to-point correlated systematic uncertainties for the $i^{\rm th}$ data point weighted by $r_{e,k}$.
The latter are optimized per values of the parameters $\bm{a}$ via $\partial \chi^2/\partial r_{e,k}=-2r_{e,k}$, which can be solved in a closed form.
We include normalization parameters $N_e$ for each dataset $e$ as part of the posterior distribution per data set.

The prior $\pi(\bm{a})$ for each replica is given by
\begin{align}
\pi(\bm{a}) &= \prod_l \Theta \big( (a_l-a_l^{\rm min})(a_l^{\rm max}-a_l) \big)
\prod_{f} \prod_{i} \exp{ -\frac{1}{2} \bigg( \frac{d_{f,i} - T_{f,i}(\bm{a})}{\alpha_{f,i}} \bigg)^{\!2}} \notag \\
&\times \prod_e \prod_k \exp{-\frac{1}{2} r_{e,k}^2} 
\prod_e \exp{-\frac{1}{2} \bigg( \frac{1-N_e}{\delta N_e} \bigg)^{\!2}}
\exp{-\frac{1}{2} \Delta^2_{D_1 < 0}} \exp{-\frac{1}{2} \Delta^2_{\vert H_1^{\sa} \vert > D_1}} \exp{-\frac{1}{2} \Delta^2_{\rm SB}},
\label{e.prior}
\end{align}
where $\prod_l$ is a product over the parameters, and $\prod_f$ is a product over the LQCD and PYTHIA datasets.  The step function $\Theta(X)$ forces the parameters to be within the chosen range between $a_l^{\rm min}$ and $a_l^{\rm max}$, which also automatically enforces the small-$x$ constraint Eq.~(\ref{e.alpha}).
A Gaussian penalty controlled by the experimentally quoted normalization uncertainties $\delta N_e$ is applied when fitting $N_e$.
The final three exponentials utilize Eqs.~(\ref{e.chi2D1pos}), (\ref{e.chi2H1bound}), and (\ref{e.chi2SB}).

The posterior distribution is sampled via data re-sampling, whereby multiple maximum likelihood optimizations of $\mathcal{P}(\bm{a}|{\rm data})$ are carried out by adding Gaussian noise with width $\alpha_{e,i}$ to each data point  across all data sets (including the PYTHIA and LQCD priors).
The resulting ensemble of replicas $\{\bm{a}_k; k=1, \ldots, n\}$ is then used to obtain statistical estimators for a given observable (which are functions of DiFFs/PDFs) ${\cal O}({\bm{a}})$, such as the mean and variance,
\begin{align}\label{e.EandVdef}
{\rm E}[{\cal O}]
= \frac{1}{n} \sum_k {\cal O}(\bm{a}_k),
\hspace{1cm}
{\rm V}[{\cal O}]
= \frac{1}{n} \sum_k \big[ {\cal O}(\bm{a}_k)-{\rm E}[{\cal O}] \big]^2.
\end{align}
The agreement between data and theory is assessed by using the ``reduced" $\chi^2$ which is defined as
\begin{align}
\label{e.chired}
\chired \equiv \frac{1}{N_{\rm dat}} 
\sum_{e,i}
\bigg( 
    \frac{d_{e,i}-{\rm E}\big[ \sum_k r_{e,k} \beta_{e,i}^k + T_{e,i}/N_e \big]}
    {\alpha_{e,i}}
\bigg)^2,
\end{align}
where $N_{\rm dat}$ is the number of data points under consideration and ${\rm E}[\cdots]$ represents the mean theory as defined in Eq.~(\ref{e.EandVdef}). We note that in Eq.~(\ref{e.chired}) we also consider the LQCD and PYTHIA-generated data in order to quantify their compatibility with the experimental data.

%%%%%%%%%%%%%%%%%%%%%%%%%%%%%%%%%%%%%%%%%%
\subsection{Mellin Space Techniques}
\label{ss.mellin}
%%%%%%%%%%%%%%%%%%%%%%%%%%%%%%%%%%%%%%%%%%

We have chosen parameterizations for all the functions in a way that can be converted to Mellin space, which has two advantages. 
First, as is well known, it allows us to solve the DGLAP evolution equations analytically rather than using an iterative solution as is required in momentum space.
We now demonstrate the second advantage of this approach when calculating the $pp$ asymmetry  \cite{Stratmann:2001pb}.
One can relate the DiFFs, IFFs, and transversity PDFs in momentum space and Mellin space through
\begin{align}
h_1(x;\mu^2) &= \frac{1}{2\pi i} \int \diff N x^{-N} \bm{h}_1(N;\mu^2)\,, \notag \\
H_1^{\sa}(z,M_h;\mu^2) &= \frac{1}{2\pi i} \int \diff M z^{-M} \bm{H}_1^{\sa}(M, M_h;\mu^2)\,, \\
D_1(z,M_h;\mu^2) &= \frac{1}{2\pi i} \int \diff M z^{-M} \bm{D}_1(M, M_h;\mu^2)\,, \notag
\label{e.mellinspace}
\end{align}
where $\bm{F}(N;\mu^2)$ is the $N^{\rm th}$ moment of $F(x;\mu^2)$ and $\bm{F}(M, M_h;\mu^2)$ is the $M^{\rm th}$ moment of $F(z, M_h;\mu^2)$.
The integration is over a contour in the complex $N$ plane that intersects the real axis to the right of the rightmost poles of $\bm{F}(N;\mu^2)$.
Then, the numerator Eq.~(\ref{e.ppUT}) and denominator Eq.~(\ref{e.ppUU}) of the $pp$ asymmetry can be written as
\begin{align}
\mathcal{H}(M_h,P_{hT},\eta) &= \frac{2 P_{hT}}{(2 \pi i)^2} \sum_i \sum_{a,b,c} \int \diff N~\bm{h}_1^a(N) \int \diff M~\bm{H}_1^{\sa,c}(M, M_h) \notag \\
&\times \Bigg[ \int_{x_{a}^{\rm min}}^1 \diff x_a\int_{x_{b}^{\rm min}}^1 \frac{ \diff x_b}{z} x_a^{-N} z^{-M} f_1^b(x_b) \frac{\diff \Delta\hat{\sigma}_{a^\uparrow b\to c^\uparrow d}}{\diff \hat{t}} \Bigg],  \\[0.3cm]
\mathcal{D}(M_h,P_{hT},\eta) &= \frac{2 P_{hT}}{2\pi i}  \sum_i \sum_{a,b,c} \int \diff M \bm{D}_1^{\sa,c}(M, M_h) \notag \\
&\times \Bigg[ \int_{x_{a}^{\rm min}}^1 \diff x_a\int_{x_{b}^{\rm min}}^1 \frac{\diff x_b}{z} z^{-M} f_1^a(x_a) f_1^b(x_b) \frac{\diff \hat{\sigma}_{ab\to cd}}{\diff \hat{t}} \Bigg].
\end{align}
Since we do not fit $f_1$ and instead take it from Ref.~\cite{Cocuzza:2022jye}, the quantities in brackets are independent of any fitting parameters and so only need to be calculated once prior to the actual fit.
Thus, the integrals over $x_a$ and $x_b$ need not be calculated at every step of the fitting procedure, massively reducing the computation time, especially  for the double convolution in the numerator.
We take advantage of this computational efficiency, allowing us to perform a simultaneous fit of the DiFFs, IFFs, and transversity PDFs.

%%%%%%%%%%%%%%%%%%%%%%%%%%%%%%%%%%%%%%%%%%
\section{Global Analysis Data and Quality of Fit}
\label{s.data}
%%%%%%%%%%%%%%%%%%%%%%%%%%%%%%%%%%%%%%%%%%

In this section we discuss the experimental and LQCD results that enter the analysis as well as the PYTHIA-generated data.

%%%%%%%%%%%%%%%%%%%%%%%%%%%%%%%%%%%%%%%%%%
\subsection{PYTHIA-Generated Data}
\label{ss.PYTHIA}
%%%%%%%%%%%%%%%%%%%%%%%%%%%%%%%%%%%%%%%%%%

This analysis must be supplemented by PYTHIA~\cite{Sjostrand:2003wg,Bierlich:2022pfr} data due to the fact that there are five independent $D_1$ functions (see Sec.~\ref{ss.DiFF_par}), but only one experimental observable ($e^+e^-$ cross section $\diff\sigma/\diff z\,\diff M_h$) currently available to constrain them. (Although $D_1$ appears in the denominators of the SIDIS and $pp$ asymmetries, those observables are being used to constrain the transversity PDFs and offer no significant sensitivity to $D_1$.)  Thus, we will discuss how we generate the PYTHIA data and how it is used to help constrain $D_1^i(z,M_h)$.

Our goal with the PYTHIA-generated data is to provide reasonable constraints on the $D_1$ functions for the strange, charm, bottom, and gluon in the absence of experimental data beyond that from Belle.
To do this, we first generate data at the Belle energy $\sqrt{s} = 10.58$ GeV for the ratio $\sigma^q/\sigma^{\rm tot}$ with $q = s,c,b$, where $\sigma^{\rm tot} \equiv \sum_q \sigma^q$ with the sum over all quark flavors and $\sigma^q$ is the flavor-dependent cross section.
This data provides priors that constrain the strange, charm, and bottom DiFFs.  In order to provide some constraint on $D_1^g$, we repeat this process at five energy scales, evenly spaced between Belle and LEP energies,
\begin{eqnarray}
\sqrt{s} &=& [10.58, 30.73, 50.88, 71.04, 91.19]~{\rm GeV} \notag.
\end{eqnarray}
This strategy allows us to gain sensitivity to $D_1^g$ scaling violations.
We do not include the bottom quark contribution at the lowest Belle energy $\sqrt{s} = 10.58$ GeV but include it at all energies above this, due to the fact that our massless quark formalism would be highly inadequate at such low energies comparable to $2 m_b$.

This leads to a total of 14 generated datasets from PYTHIA, with each one covering the same $(z,M_h)$ region as the Belle data. 
The same cuts used on the real Belle data are used on the PYTHIA data.  The cut in Eq.~(\ref{e.bellecut}) below is used with $\sqrt{s} = 10.58$ GeV regardless of the value of $\sqrt{s}$, as there is no benefit within our analysis of going beyond the kinematic region of the Belle data.

The PYTHIA~\cite{Sjostrand:2003wg,Bierlich:2022pfr} data can be generated with arbitrarily high statistics, so we neglect the tiny statistical error.  
In order to propagate model uncertainties in PYTHIA, we use a collection of tunes (``PYTHIA 6 default", ``PYTHIA 6 ALEPH", ``PYTHIA 6 LEP/Tevatron", ``PYTHIA 8") to estimate the systematic uncertainties on the flavor-dependent cross sections.
This list of tunes is similar to that used in Ref.~\cite{Belle:2017rwm}. 
The tunes are summarized in Table~\ref{t.tunes} in Appendix~\ref{a.PYTHIA}.
The resulting spread of the flavor-dependent cross sections is used to estimate the errors due to the tunes, with the mean taken as the central value and the minimum and maximum defining the uncorrelated systematic error.
We choose to fit the ratio $\sigma^q/\sigma^{\rm tot}$ as it leads to smaller variance between the different tunes compared to taking the absolute cross sections $\sigma^q$.
Taking such a ratio should also help to cancel NLO and thrust-cut effects.
The $\chired$ for the PYTHIA data is shown in Table \ref{t.chi2_pythia} in Appendix \ref{a.PYTHIA}, along with comparisons between the PYTHIA-generated data and theory calculations.

%%%%%%%%%%%%%%%%%%%%%%%%%%%%%%%%%%%%%%%%%%
\subsection{Experimental Data and Lattice QCD Results}
\label{ss.experiment}
%%%%%%%%%%%%%%%%%%%%%%%%%%%%%%%%%%%%%%%%%

\begin{figure*}[t]
\centering
\includegraphics[width=0.65\textwidth]{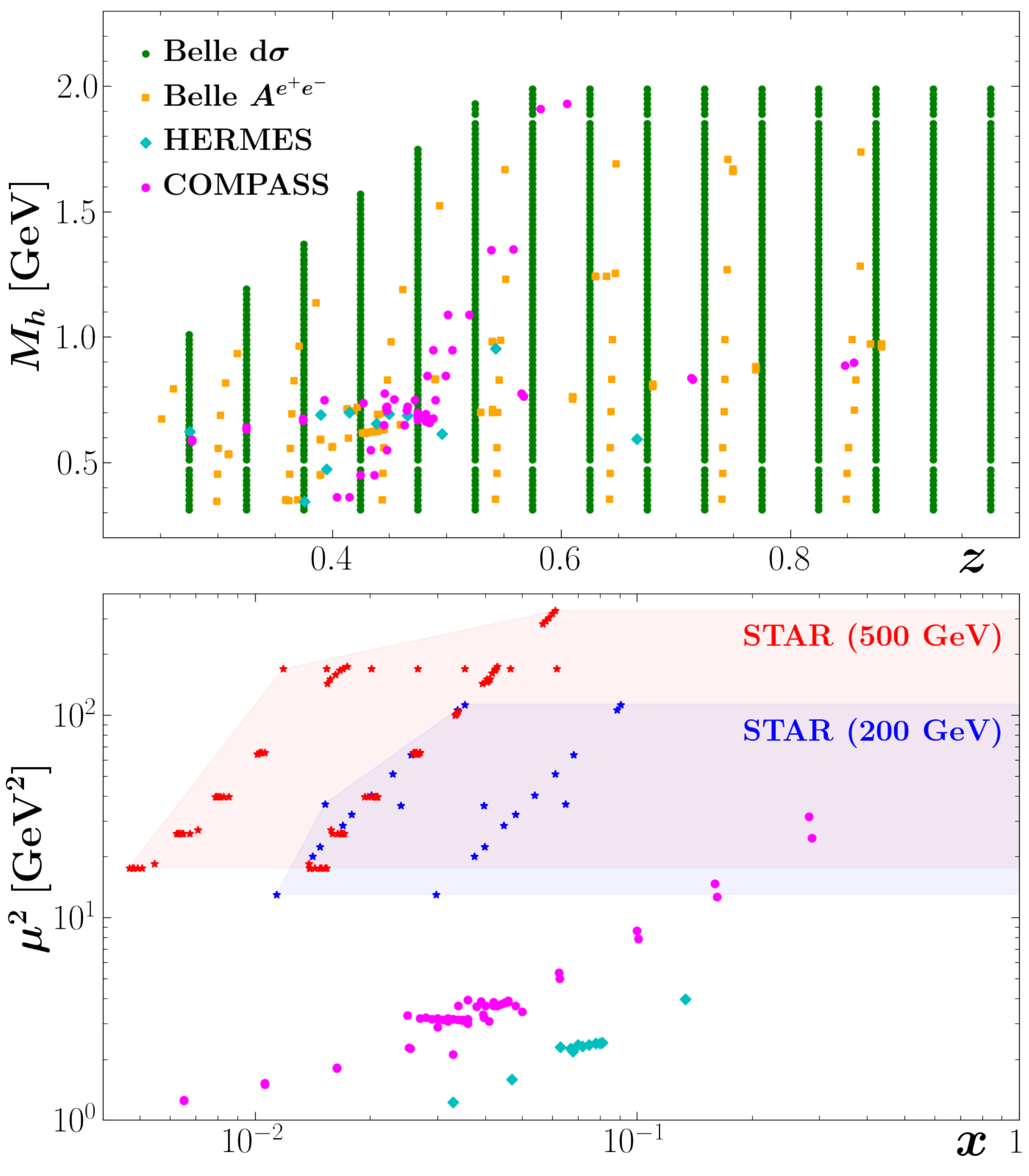}
\caption
{Kinematic coverage of the datasets included in this analysis.  The top panel shows the $e^+ e^-$ annihilation and SIDIS data as a function of $z$ and $M_h$.
The bottom panel shows the SIDIS and $pp$ data as a function of $x$ and $\mu^2$.
The variable $x$ represents $\xb$ for SIDIS and $x_a$ for $pp$ collisions, while the scale $\mu^2$ represents $Q^2$ for SIDIS and $P_{hT}^2$ for $pp$ collisions.
For STAR, the solid points are $x_a^{\rm min}$ and the light shaded region is to indicate that the $x_a$-integration extends up to $x_a=1$.
}
\label{f.kin}
\end{figure*}

A summary of the experimental data and lattice results included in our analysis is given in Table~\ref{t.data_summary}, where we also indicate which fitted non-perturbative functions enter the observables. The kinematic coverage of the measurements is shown in Fig.~\ref{f.kin}.
We use the $\pi^+ \pi^-$ production in $e^+ e^-$ annihilation cross section data from Belle \cite{Belle:2017rwm} at $\sqrt{s} = 10.58$ GeV.
(We note that there is a typo in the Belle publication regarding the units of the cross section.  The units are labeled as $\mu$b/GeV but the cross section is actually shown in units of nb/GeV.)
Since we can only extract information on the IFF up to $M_h = 2.0$ GeV (see below), we place a cut of $M_h < 2.0$ GeV on the cross-section data.
The data are provided in bins of $M_h$, with width $\Delta M_h = 0.02$~GeV, and bins of $z$, with width $\Delta z = 0.05$.  We evaluate the cross section at the mean of each $M_h$ bin $\mean{M_h}$ and $z$ bin $\mean{z}$.
In order to avoid the sharp kaon and $D^0$ resonances, we also cut out all data points with $\mean{M_h} = 0.49$ GeV and $\mean{M_h} = 1.87$ GeV.
The kinematic limit of the Belle data is given by $M_h < \frac{\sqrt{s}}{2} z$, which is reached in the bins of $z$ below $z = 0.45$.  We place a cut of
\begin{align}
\frac{2\mean{M_h}}{\mean{z}\sqrt{s}} < 0.7
\label{e.bellecut}
\end{align}
in order to avoid the region near the kinematic limit.
We also cut out the lowest $z$ bin $0.20 < z < 0.25$ as we find difficulty in describing the data in this bin, which may be an indication of the need to resum small-$z$ logarithms.
In total, with these cuts we include 1,094 of the 1,468 data points provided by Belle, and the kinematic ranges are $0.25 < z < 1.0$ and $0.3 < M_h < 2.0$ GeV.
The data has an overall normalization uncertainty of $1.6 \%$, and the systematic errors are uncorrelated point-to-point~\cite{Belle:2017rwm}.

\begin{table*}
\centering
\resizebox{16cm}{!} {
\begin{tabular}{l c | c | c | c }
\hhline{=====}
\vspace{-0.3cm}  & & & \\
Collaboration & Ref. & Observable & Process & Non-perturbative function(s) \\
\vspace{-0.3cm}  & & & \\ 
\hline \\ [-0.25cm]
Belle      & \cite{Belle:2017rwm}    & $\diff \sigma / \diff z \,\diff M_h$ & $e^+ e^- \!\!\rightarrow (\pi^+ \pi^-)\,X$                  & $D_1$    \\[0.0cm]
Belle      & \cite{Belle:2011cur}    & $A^{e^+ e^-}$                              & $e^+ e^- \!\!\rightarrow (\pi^+ \pi^-)(\pi^+ \pi^-)\,X$   & $D_1, H_1^{\sa}$  \\
HERMES     & \cite{HERMES:2008mcr}   & $A_{UT}^{\rm SIDIS}$                                 & $e\, p^{\uparrow} \rightarrow e' (\pi^+ \pi^-)\,X$        & $D_1, H_1^{\sa}, h_1$ \\
COMPASS     & \cite{COMPASS:2023cgk}   & $A_{UT}^{\rm SIDIS}$                                 & $\mu \{p,D\}^{\uparrow} \rightarrow \mu' (\pi^+ \pi^-)\,X$        & $D_1, H_1^{\sa}, h_1$  \\
STAR       & \cite{STAR:2015jkc, STAR:2017wsi}     & $A_{UT}^{pp}$                    & $p^{\uparrow}p \rightarrow (\pi^+ \pi^-)\,X$         & $D_1, H_1^{\sa}, h_1$  \\
\hline 
ETMC  & \cite{Alexandrou:2021oih} & $\delta u$, $\delta d$ & LQCD & $h_1$ \\
PNDME  & \cite{Gupta:2018lvp} & $\delta u$, $\delta d$ & LQCD & $h_1$ \\
\hhline{=====}
\end{tabular}}
\caption
[Summary of dihadron production data]
{Data from experiment and LQCD used in this analysis for the DiFFs $D_1$ and $H_1^{\sa}$ and the transversity PDF $h_1$.}
\label{t.data_summary}
\end{table*}
%\end{center}

We include the $e^+ e^-$ annihilation Artru-Collins asymmetry data provided by Belle~\cite{Belle:2011cur}.  The same cuts on $M_h$ and $z$ discussed above are applied to both hadron pairs.  
The data is available in three binnings: $(z, M_h)$, $(M_h, \overline{M}_h)$, and $(z, \bar{z})$.
We include all three binnings in our analysis and evaluate the theory formulas using the mean values $\mean{M_h}$, $\mean{z}$, $\mean{\overline{M}_h}$, $\mean{\bar{z}}$.
The systematic errors on the data are uncorrelated point-to-point and there is no overall normalization error~\cite{Belle:2011cur}.
We note that in all Belle measurements in this analysis, a cut is placed on the thrust~$T$~\cite{Brandt:1964sa} of $T > 0.8$.  In our LO formalism, it is not possible to take this thrust cut into account because $T<1$ can only be generated via NLO effects.  We expect that the errors from this omission largely cancel out in the ratio of $H_1^{\sa}/D_1$ which appears in all of the asymmetries, thus reducing the impact on the extraction.  However, since the extraction of $D_1$ relies on cross-section data rather than asymmetries, its extraction could be noticeably influenced by next-to-leading order (NLO) corrections.

The transversity PDFs are constrained by SIDIS asymmetry and $pp$ asymmetry data.
SIDIS data is available from HERMES \cite{HERMES:2008mcr} and COMPASS \cite{COMPASS:2023cgk}.  We use all three available binnings in $\xb$, $z$, and $M_h$, and, as for the $e^+e^-$ annihilation data, apply the cuts $\mean{z} > 0.25$ and $\mean{M_h} < 2.0$ GeV.
We evaluate the theory formulas using the mean values $\mean{\xb}$, $\mean{z}$, $\mean{M_h}$, $\mean{y}$.
The HERMES data has an overall normalization uncertainty of $8.1 \%$, while the COMPASS data has $2.2 \%$ for proton and $5.4 \%$ for deuterium.  
The systematic errors on the HERMES data are correlated point-to-point~\cite{HERMES:2008mcr}, while those on the COMPASS data are uncorrelated~\cite{COMPASS:2023cgk}.

The $pp$ collision data is provided by STAR, at both $\sqrt{s} = 200$ GeV \cite{STAR:2015jkc} and $\sqrt{s} = 500$ GeV \cite{STAR:2017wsi}.  
We apply the cut $\mean{M_h} < 2.0$ GeV to all of the data.
The $\sqrt{s} = 200$ GeV data is provided with three different upper cuts (0.2, 0.3, 0.4) on the opening angle $R$ of the pion pair, with 0.3 treated as the default.  
This cut is used to filter out pion pairs that do not originate from a single parton.
We use the data corresponding to $R < 0.3$ and have verified that the changes to the resulting functions by instead using data from the other cuts are negligible compared to the statistical uncertainties. 
The $\sqrt{s} = 500$ GeV data is provided with an opening angle of $R < 0.7$.  A larger opening angle cut is acceptable here, as the increased energy means that gluon radiation is occurring at wider angles, allowing the dihadron pair to still be considered as originating from a single parton even with a larger $R$-cut value. 
The data is provided binned in $P_{hT}$, $M_h$, and $\eta$, with the results (often) provided for both $\eta > 0$ and $\eta < 0$ when binned in $P_{hT}$ or $M_h$.  We include all binnings and take the central values $\mean{P_{hT}}$, $\mean{M_h}$, and $\mean{\eta}$ when evaluating our theory formulas.
All systematic errors are uncorrelated, and the 200 GeV and 500 GeV data have normalization uncertainties of 4.8\% and 4.5\%, respectively~\cite{STAR:2015jkc, STAR:2017wsi}.

We also consider the inclusion of LQCD data on tensor charges in the fit and treat them as Bayesian priors (see Sec.~\ref{ss.bayes}).  We restrict ourselves to results at the physical pion mass with $2 + 1 + 1$ flavors, where calculations are available from ETMC \cite{Alexandrou:2021oih} and PNDME \cite{Gupta:2018lvp}  on $\delta u$, $\delta d$, and $g_T$.
We choose to fit $\delta u$ and $\delta d$ rather than $g_T$ in order to provide flavor separation.
The reported uncertainties are treated as uncorrelated.

%%%%%%%%%%%%%%%%%%%%%%%%%%%%%%%%%%%%%%%%%%
\subsection{Data vs.~Theory}
\label{ss.datavstheory}
%%%%%%%%%%%%%%%%%%%%%%%%%%%%%%%%%%%%%%%%%%

We now discuss  the results of our global analysis, where we consider three different scenarios.
The first includes all experimental and LQCD data discussed above and will be referred to as ``JAMDIFF (w/ LQCD)." 
The ``JAMDIFF (no LQCD)" fit removes the LQCD data. 
The ``JAMDiFF (SIDIS only)" fit excludes all of the STAR $pp$ and LQCD data, and also sets the antiquark transversity PDFs to zero as they cannot be constrained in this scenario.
The $\chired$ for the three scenarios is summarized in Table \ref{t.chi2}. In the plots that follow comparing our theory results with the experimental measurements, we show the JAMDiFF (w/ LQCD) fit.
We will reserve a discussion about our calculated tensor charges and comparison to those from LQCD computations (and other values from phenomenology) to Sec.~\ref{s.TPDFs}.
We re-emphasize that we have performed a simultaneous global analysis of DiFFs/IFFs and transversity PDFs, where, unlike previous work~\cite{Bacchetta:2012ty, Radici:2015mwa, Radici:2018iag, Benel:2019mcq}, the parameters for the DiFFs are not fixed (from a fit of only $e^+ e^-$ annihilation) but allowed to be free along with the transversity PDF parameters.
We have also studied an exhaustive set of available data on dihadron observables, which includes, for the first time, the Belle cross section~\cite{Belle:2017rwm}, the $\sqrt{s} = 500 \, \textrm{GeV}$ measurements from STAR~\cite{STAR:2017wsi}, and all kinematic variable binnings for the relevant processes under consideration, amassing 1471 experimental data points.
We are able to describe all of the experimental data very well.

\begin{table*}
%\scriptsize
\centering
\resizebox{14cm}{!} {
\begin{tabular}{l  | c | c | c | c | c}
\hhline{======}
\vspace{-0.3cm} & &   \\
& & & \multicolumn{3}{c}{$\chired$} \\
\vspace{-0.3cm} & &  \\
& & & \multicolumn{3}{c}{JAMDiFF} \\
Experiment  & Binning & $N_{\rm dat}$ & (w/ LQCD) & (no LQCD) & (SIDIS only) \\
\vspace{-0.3cm} & & & & \\
\hline
\vspace{-0.3cm} & & & & \\
Belle (cross section)\cite{Belle:2017rwm}    & $z,M_h$      &  1094 & 1.01 & 1.01 & 1.01 \\
\vspace{-0.3cm} & & & & \\
\hline
                                              & $z,M_h$      &  55   & 1.27  & 1.24 & 1.28  \\
Belle (Artru-Collins) \cite{Belle:2011cur}    & $M_h,\Mhb$   &  64   & 0.60  & 0.60 & 0.60  \\
                                              & $z,\bar{z}$  &  64   & 0.42  & 0.42 & 0.41  \\
\hline
                               & $\xb$        &  4    & 1.77  & 1.70 & 1.67 \\
HERMES \cite{HERMES:2008mcr}   & $M_h$        &  4    & 0.41  & 0.42 & 0.47 \\
                               & $z$          &  4    & 1.20  & 1.17 & 1.13 \\
\hline
                                        & $\xb$        &  9    & 1.98 & 0.65 & 0.59  \\
COMPASS ($p$) \cite{COMPASS:2023cgk}    & $M_h$        &  10   & 0.92 & 0.94 & 0.93  \\
                                        & $z$          &  7    & 0.77 & 0.60 & 0.63  \\
\hline
                                        & $\xb$       &  9    & 1.37 & 1.42 & 1.22  \\
COMPASS ($D$) \cite{COMPASS:2023cgk}    & $M_h$       &  10   & 0.45 & 0.37 & 0.38  \\
                                        & $z$         &  7    & 0.50 & 0.46 & 0.46  \\
\hline
                                & $M_h, \eta < 0$      &  5    & 2.57 & 2.56 & --- \\
STAR \cite{STAR:2015jkc}        & $M_h, \eta > 0$      &  5    & 1.34 & 1.55 & --- \\
$\sqrt{s}=200$ GeV              & $P_{hT}, \eta < 0$   &  5    & 0.98 & 1.00 & --- \\
$R < 0.3$                       & $P_{hT}, \eta > 0$   &  5    & 1.73 & 1.74 & --- \\
                                & $\eta$               &  4    & 0.52 & 1.46 & --- \\
\hline
                                 & $M_h, \eta < 0$    &  32   & 1.30 & 1.10 &  --- \\
STAR \cite{STAR:2017wsi}         & $M_h, \eta > 0$    &  32   & 0.81 & 0.78 &  --- \\
$\sqrt{s}=500$ GeV               & $P_{hT}, \eta > 0$ &  35   & 1.09 & 1.07 &  --- \\
$R < 0.7$                        & $\eta$             &  7    & 2.97 & 1.83 &  --- \\
\hline
ETMC $\delta u$ \cite{Alexandrou:2021oih}    & --- &  1       & 0.71 & ---  & --- \\
ETMC $\delta d$ \cite{Alexandrou:2021oih}    & --- &  1       & 1.02 & ---  & --- \\
PNDME $\delta u$ \cite{Gupta:2018lvp}        & --- &  1       & 8.68 & ---  & --- \\
PNDME $\delta d$ \cite{Gupta:2018lvp}        & --- &  1       & 0.04 & ---  & --- \\
\hline
\vspace{-0.3cm} & & & & \\
\textbf{Total} $\boldsymbol{\chired}$ ($N_{\rm dat}$) & &   & {\bf 1.01} (1475) & {\bf 0.98} (1471) & 0.96 (1341) \\
\hhline{======}
\end{tabular}}
\caption
[$\chi^2$ table: Dihadron production data]
{Summary of $\chired$ values for the different fit configurations defined in Sec. \ref{ss.datavstheory}.}
\label{t.chi2}
\end{table*}

\begin{figure}[h]
\includegraphics[width=0.98\textwidth]{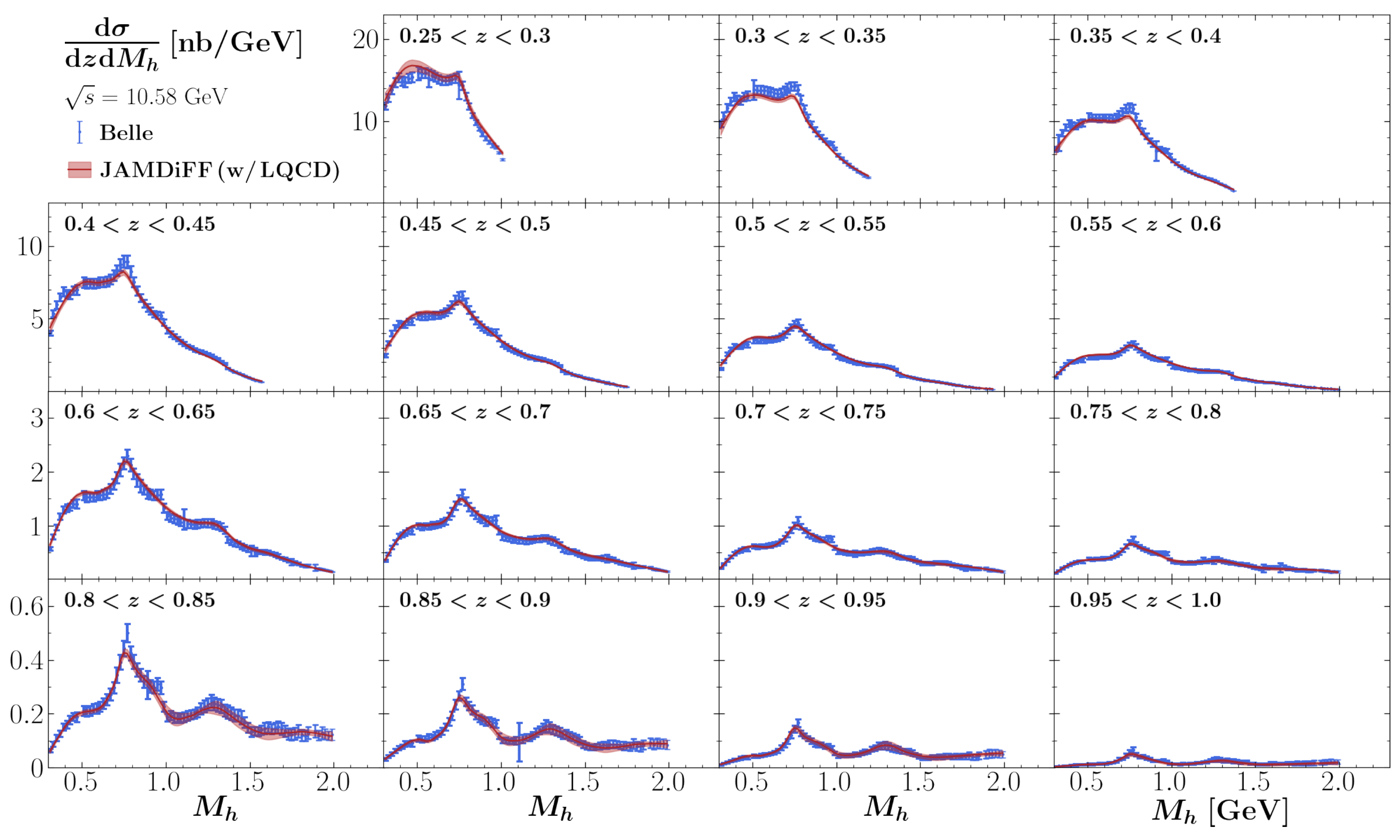}
\caption
{$\diff\sigma/\diff z\,\diff M_h$ cross section data from Belle~\cite{Belle:2017rwm} at $\sqrt{s} = 10.58$ GeV (blue circles) plotted as a function of $M_h$ against the mean JAMDiFF (w/ LQCD) result (using Eq.~(\ref{e.dsigdzdMh})) with 1$\sigma$ uncertainty (red lines with bands).  Each panel is a different bin of~$z$.}
\label{f.belle-unpolarized}
\end{figure}

\begin{figure}[h]
\includegraphics[width=0.89\textwidth]{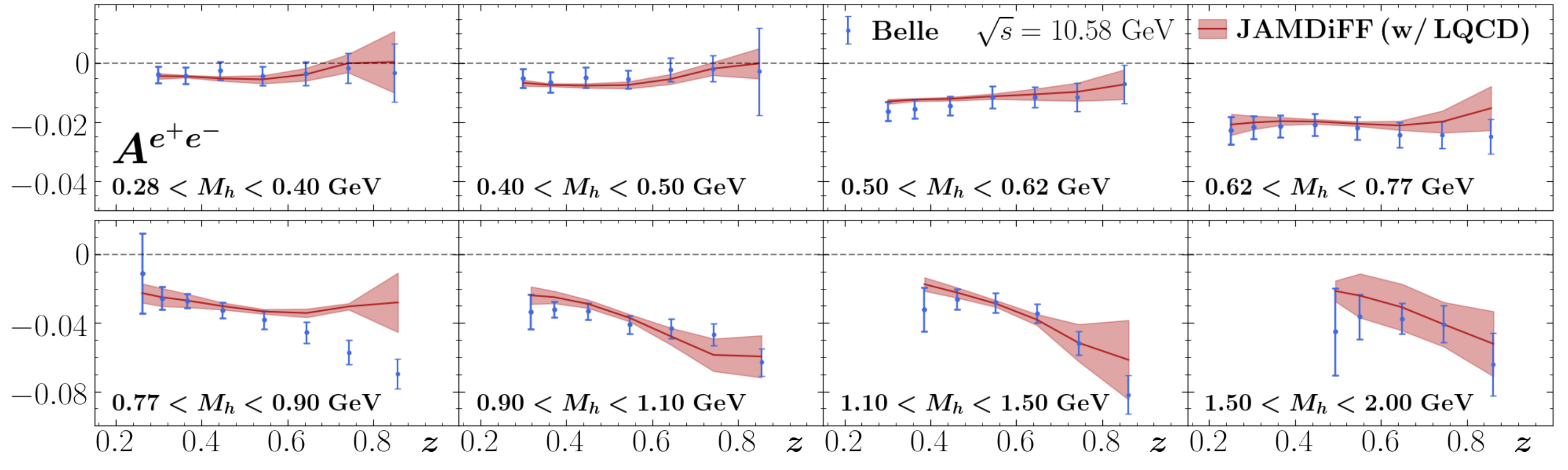}
\caption
{Artru-Collins asymmetry data from Belle~\cite{Belle:2011cur} binned in ($z,M_h$) at $\sqrt{s} = 10.58$ GeV (blue circles) plotted as a function of $z$ against the mean JAMDiFF (w/ LQCD) result (using Eq.~(\ref{e.artrucollins}))  with 1$\sigma$ uncertainty (red lines with bands).  The different panels show different bins of $M_h$.
}
\label{f.belle-a12R-z1-M1}
\end{figure}

\begin{figure}[h]
\includegraphics[width=0.89\textwidth]{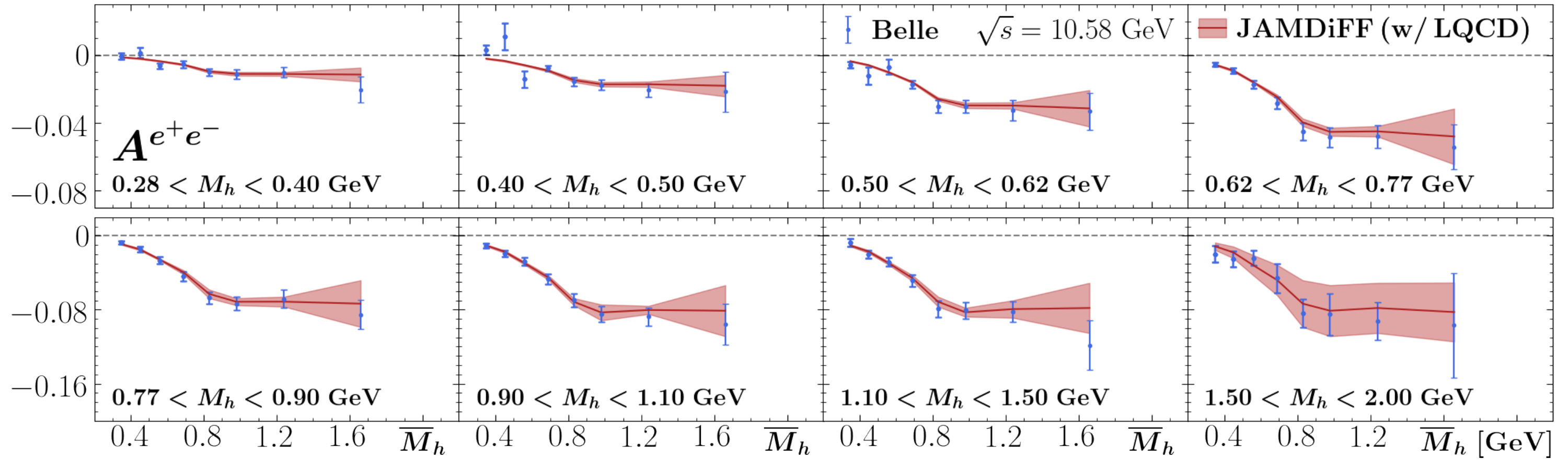}
\caption
{Artru-Collins asymmetry data from Belle~\cite{Belle:2011cur} binned in ($M_h,\overline{M}_h$) at $\sqrt{s} = 10.58$ GeV (blue circles) plotted as a function of $\overline{M}_h$ against the mean JAMDiFF (w/ LQCD) result (using Eq.~(\ref{e.artrucollins})) with 1$\sigma$ uncertainty (red lines with bands).  The different panels show different bins of $M_h$.
}
\label{f.belle-a12R-M1-M2}
\end{figure}

\begin{figure}[h]
\includegraphics[width=0.89\textwidth]{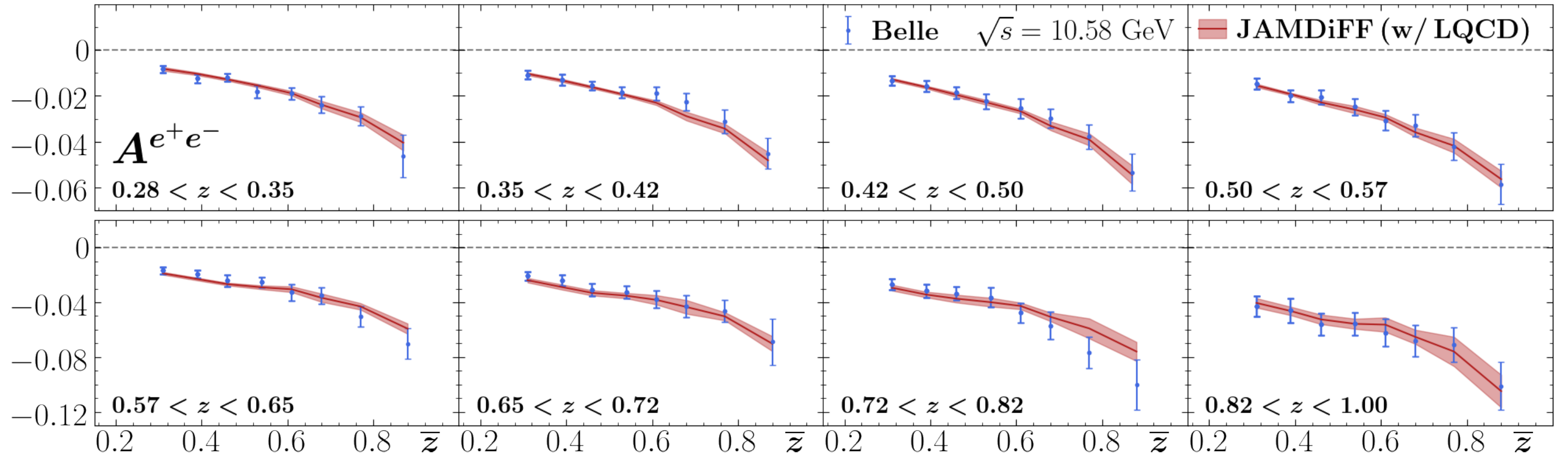}
\caption
{Artru-Collins asymmetry data from Belle~\cite{Belle:2011cur} binned in ($z,\bar{z}$) at $\sqrt{s} = 10.58$ GeV (blue circles) plotted as a function of $\bar{z}$ against the mean JAMDiFF (w/ LQCD) result (using Eq.~(\ref{e.artrucollins})) with 1$\sigma$ uncertainty (red lines with bands).  The different panels show different bins of $z$.
}
\label{f.belle-a12R-z1-z2}
\end{figure}

The results for the Belle cross section  are shown in Fig.~\ref{f.belle-unpolarized}.  We find that our parameterization discussed in Sec.~\ref{ss.DiFF_par} is able to account for the resonance structure in the data and the general behavior of the measurement across the kinematic range in $z$ and $M_h$, with $\chired=1.01$. There are some slight discrepancies in the lowest $z$ bin at small $M_h$.
The fitted normalization $N_e$ (see Eq.~(\ref{e.chi2})) for this dataset is $1.00(1)$.
In Figs.~\ref{f.belle-a12R-z1-M1}--\ref{f.belle-a12R-z1-z2} we display the Belle Artru-Collins asymmetry results.  Again our theory curves are able to describe the data very well across all kinematics, with $\chired=1.27, 0.60, 0.42$ for the $(z,M_h), (M_h,\bar{M}_h),(z,\bar{z})$ binnings, respectively.  The only discrepancy is in the $0.77<M_h<0.90\,{\rm GeV}$ bin at high $z$, which is what causes the $\chired$ for the $(z,M_h)$ binning to be much larger than the other two.  There is no fitted normalization for this dataset.  As seen in Table~\ref{t.chi2}, the $\chired$ for the Belle data (both the cross section and Artru-Collins asymmetry) is nearly identical across the three fit configurations.

\begin{figure}[h]
\includegraphics[width=0.825\textwidth]{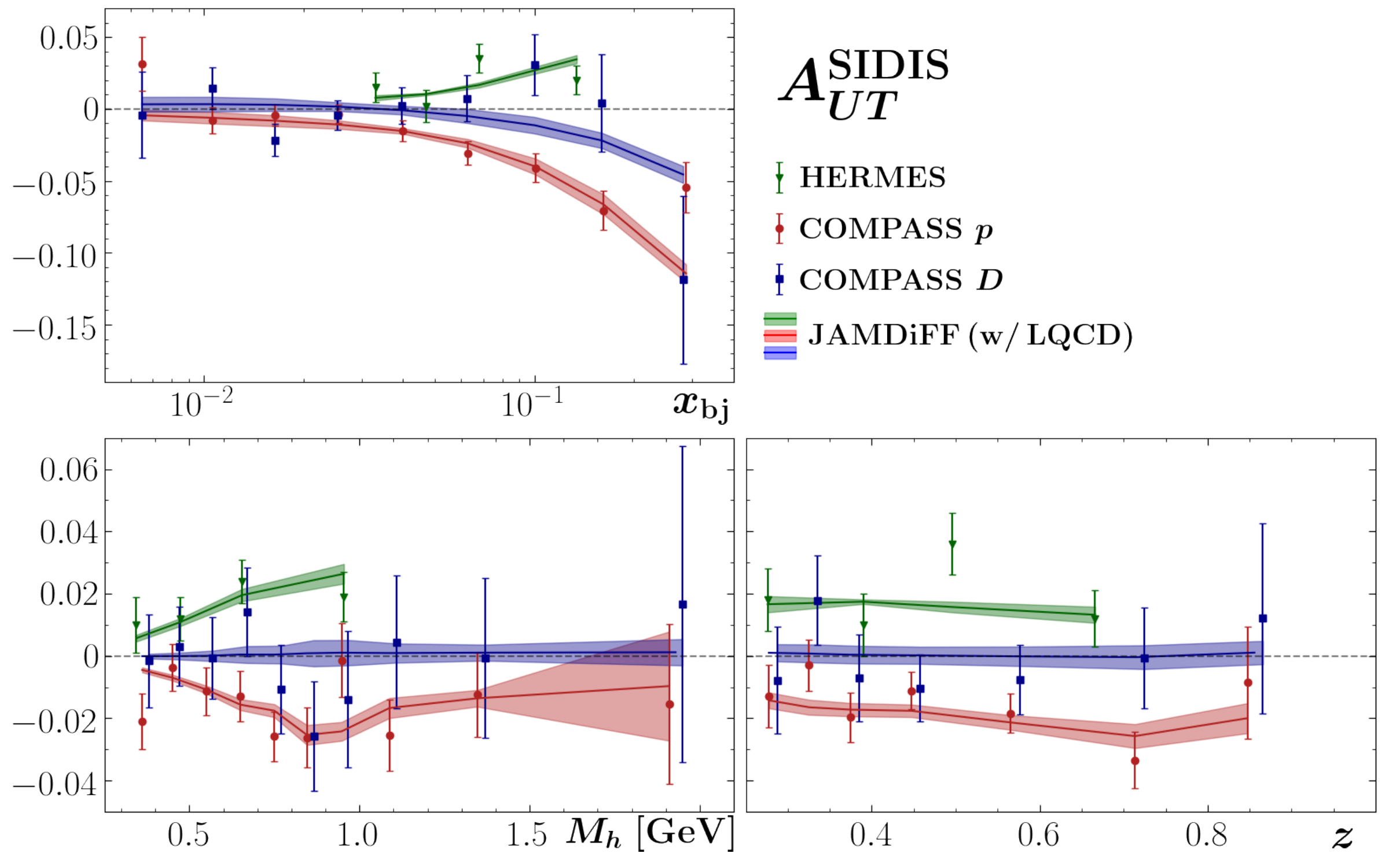}
\caption
{SIDIS asymmetry data from HERMES \cite{HERMES:2008mcr} (green triangles) and COMPASS \cite{COMPASS:2023cgk} (red circles for proton, blue squares for deuteron) plotted against the mean JAMDiFF (w/ LQCD) result  (using Eq.~(\ref{e.AUTSIDIS})) with 1$\sigma$ uncertainty (colored lines with bands).
The data is binned in $\xb$ (top left), $M_h$ (bottom left), and $z$ (bottom right).
We note that the asymmetries from HERMES and COMPASS are defined such that they have opposite signs (see Eq.~(\ref{e.AUTSIDIS}) and the surrounding discussion).
}
\label{f.SIDIS}
\end{figure}

The results for the  SIDIS asymmetry are shown in Fig.~\ref{f.SIDIS}.  The fitted normalizations $N_e$ (see Eq.~(\ref{e.chi2})) are $0.98(4)$ for HERMES,  $1.014(6)$ for COMPASS proton, and $1.007(8)$ for COMPASS deuteron.  The theory calculations are generally in agreement with the data for all three projections ($x_{\rm bj},M_h,z$). 
However, we focus the reader's attention on the highest $x_{\rm bj}$ COMPASS proton point (data/curve in red in the left panel of Fig.~\ref{f.SIDIS}).  The trend is for  JAMDiFF (w/ LQCD) to increase at larger $x_{\rm bj}$ more rapidly than the COMPASS proton data.  The JAMDiFF (no LQCD) fit follows the trend of the data much better (see Fig.~\ref{f.highx} in Sec.~\ref{ss.compatiblity}), which is apparent in the reduction of $\chired$ for the COMPASS proton $\xb$ binning from $1.98$ to $0.65$ when the lattice data is removed.  
In Sec.~\ref{ss.compatiblity} we will elaborate on the importance of high-$x$ experimental measurements in testing the compatibility between phenomenology and LQCD results for the tensor charges.  We lastly remark that the SIDIS only fit has $\chired$ values similar to the analyses that included $pp$ data, demonstrating clear compatibility between the two reactions.

\begin{figure}[h]
\includegraphics[width=0.9\textwidth]{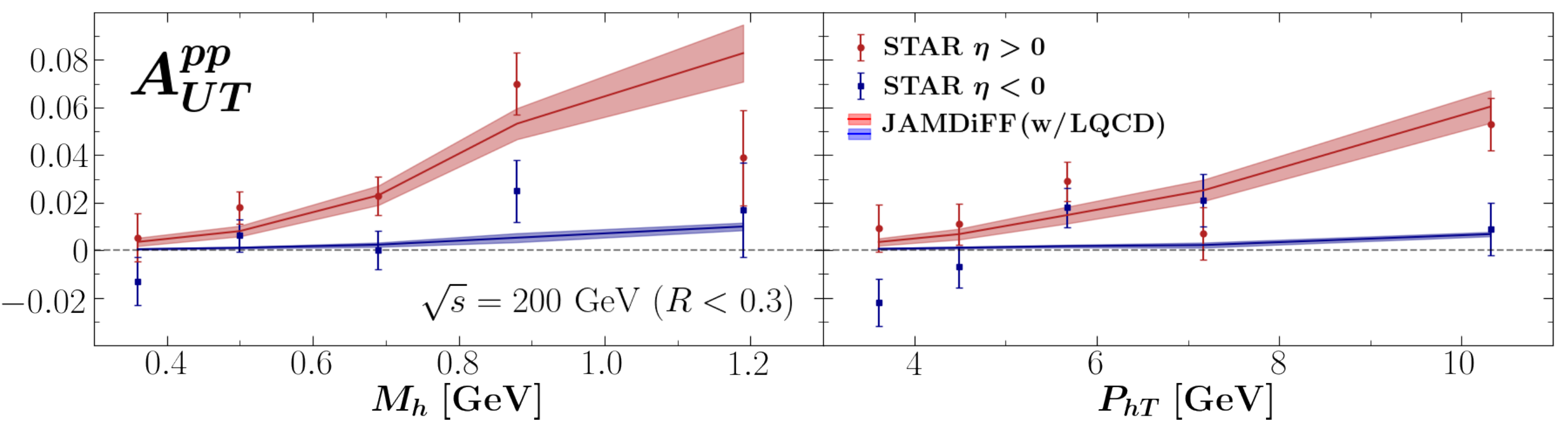}
\caption
{Proton-proton asymmetry data from STAR at $\sqrt{s}=200$ GeV \cite{STAR:2015jkc} with opening angle cut $R < 0.3$ plotted against the mean JAMDiFF (w/ LQCD) result (using Eq.~(\ref{e.AUTpp})) with 1$\sigma$ uncertainty (colored lines with bands).
The left panel shows the results binned in $M_h$ while the right panel shows them binned in $P_{hT}$, for both $\eta > 0$ (red circles) and $\eta < 0$ (blue squares).}
\label{f.star-RS200pub-ang0.3-M-PhT}
\end{figure}

\begin{figure}[h]
\includegraphics[width=0.98\textwidth]{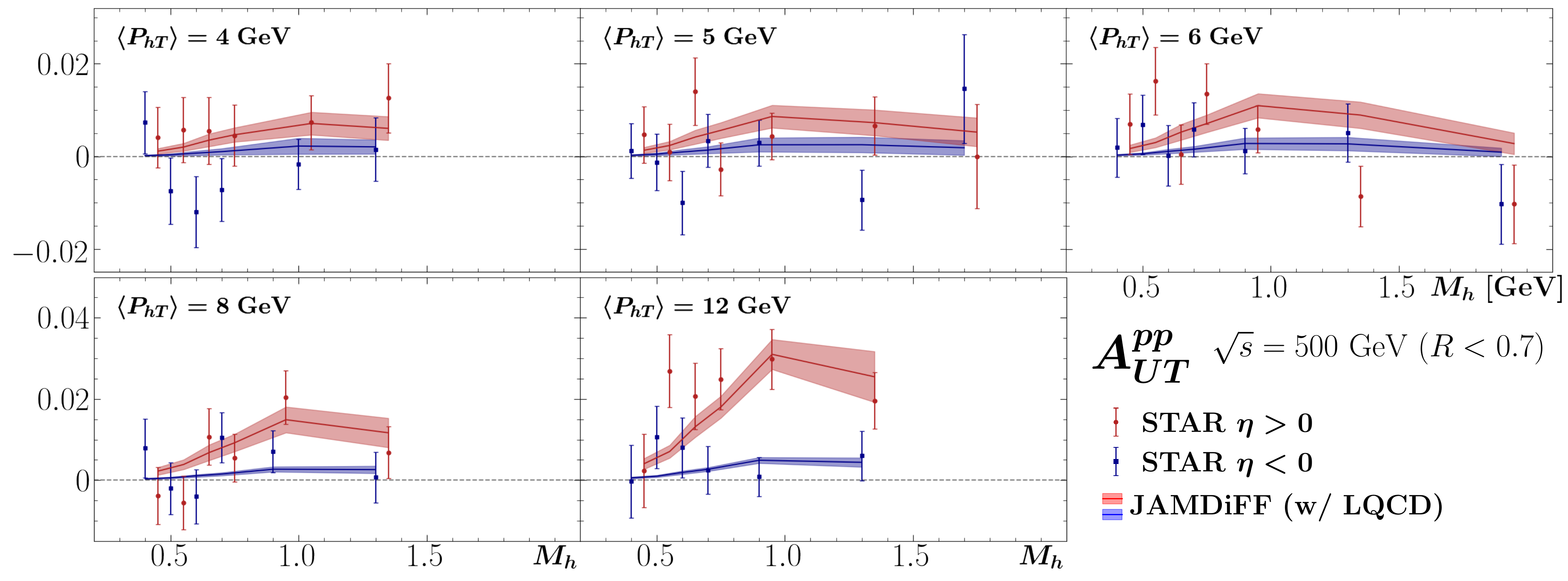}
\caption
{Proton-proton asymmetry data from STAR at $\sqrt{s}=500$ GeV \cite{STAR:2017wsi} with opening angle cut $R<0.7$ (red circles for $\eta > 0$, blue squares for $\eta < 0$) plotted as a function of $M_h$ against the mean JAMDiFF (w/ LQCD) result (using Eq.~(\ref{e.AUTpp})) with 1$\sigma$ uncertainty (colored lines with bands).
The different panels show different bins of $P_{hT}$.}
\label{f.star-RS500-M}
\end{figure}

\begin{figure}[h]
\includegraphics[width=0.98\textwidth]{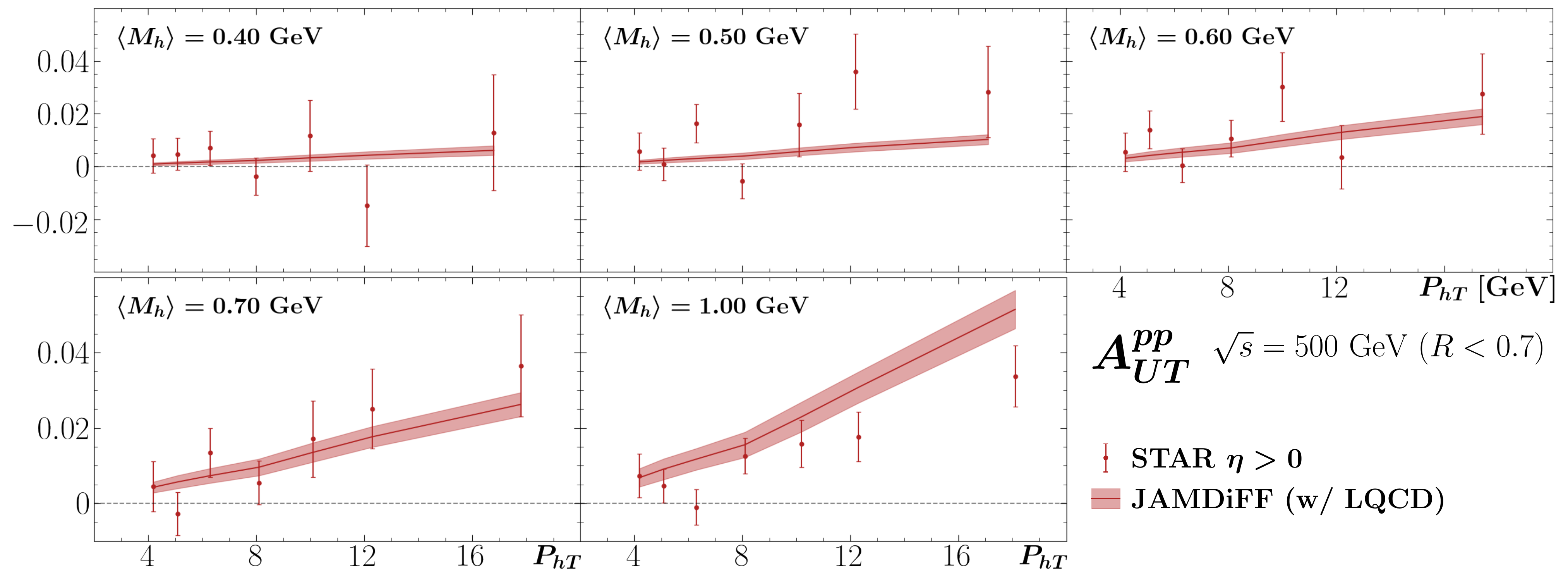}
\caption
{Proton-proton asymmetry data from STAR at $\sqrt{s}=500$ GeV \cite{STAR:2017wsi} with opening angle cut $R<0.7$ and $\eta > 0$ (red circles) plotted as a function of $P_{hT}$ against the mean JAMDiFF (w/ LQCD) result (using Eq.~(\ref{e.AUTpp}))  with 1$\sigma$ uncertainty (red lines with bands).
The different panels show different bins of $M_h$.}
\label{f.star-RS500-PhT}
\end{figure}

\begin{figure}[h]
\includegraphics[width=0.80\textwidth]{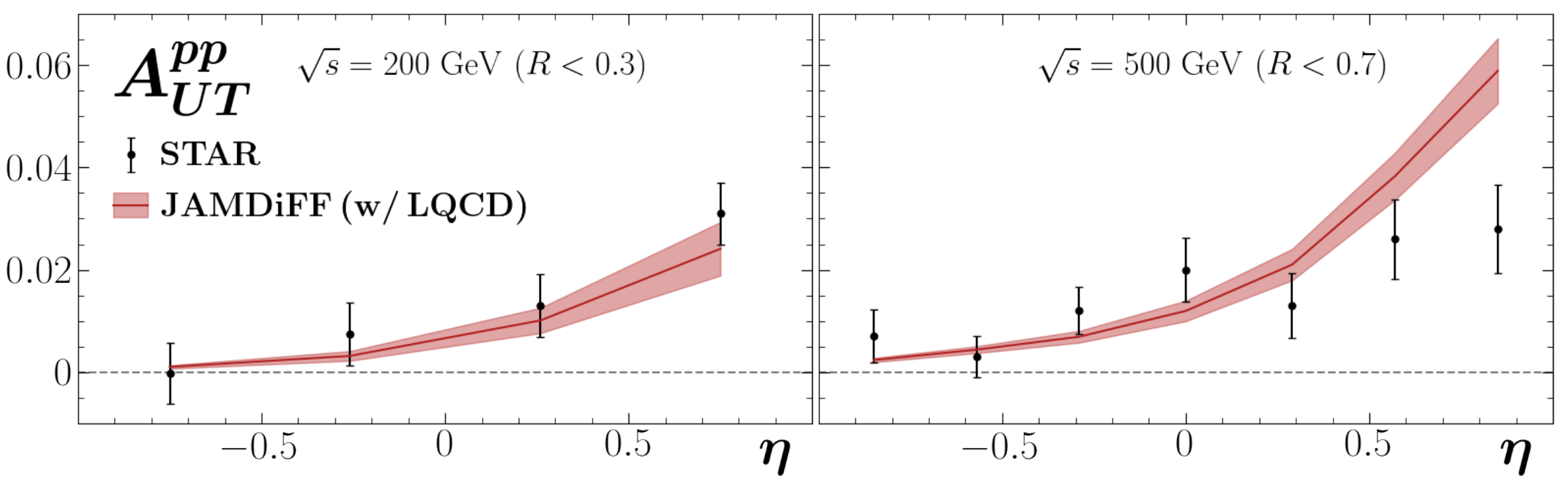}
\caption
{Proton-proton asymmetry data from STAR (black circles) plotted as a function of $\eta$ against the mean JAMDiFF (w/ LQCD) result (using Eq.~(\ref{e.AUTpp})) with 1$\sigma$ uncertainty (red lines with bands).
The left panel shows the data at $\sqrt{s}=200$ GeV~\cite{STAR:2015jkc} with opening angle cut $R<0.3$, $\mean{P_{hT}} = 6.0$ GeV, $\mean{M_h} = 0.6$ GeV.
The right panel shows the data at $\sqrt{s}=500$ GeV \cite{STAR:2017wsi} with opening angle cut $R<0.7$, $\mean{P_{hT}} = 13$ GeV, and $\mean{M_h} = 1$ GeV.}
\label{f.star-eta}
\end{figure}

The results for the $pp$ asymmetry are shown in Figs. \ref{f.star-RS200pub-ang0.3-M-PhT} and \ref{f.star-eta} for the STAR $\sqrt{s} = 200$ GeV data and in Figs. \ref{f.star-RS500-M}--\ref{f.star-eta} for the $\sqrt{s} = 500$ GeV data.  
The fitted normalizations $N_e$ (see Eq.~(\ref{e.chi2})) are $1.00(3)$ for the 200~GeV STAR data and $1.14(6)$ for the 500~GeV STAR data. 
As with the $e^+e^-$ and SIDIS observables, we again find generally good agreement with the $pp$ reaction across all kinematic binnings.  The one discrepancy of note is with the $\eta$ binning for the 500~GeV STAR  data at the highest $\eta$ value (see Fig.~\ref{f.star-eta}) that has more sensitivity to larger $x$.  The JAMDiFF (w/ LQCD) fit overshoots that point and has a trend of continuing to increase with increasing $\eta$.
Data at larger $\eta$ are needed in order to test this behavior.
The JAMDiFF (no LQCD) fit still increases with $\eta$ but falls below the JAMDiFF (w/ LQCD) curve (see Fig.~\ref{f.highx}).  This accounts for the improvement in the $\chired$ for the 500~GeV STAR data in the $\eta$ binning, going from $2.97$ to $1.83$, once the LQCD data is removed.  Interestingly, the 200~GeV STAR  data for the $\eta$ binning improves when LQCD data is included, going from 1.46 to 0.52. 
So it seems that, within our analysis, there are competing trends at larger $\eta$ in $pp$ that measurements at more forward rapidities may be able to resolve.

\clearpage

\clearpage

%%%%%%%%%%%%%%%%%%%%%%%%%%%%%%%%%%%%%%%%%%
\section{Extracted Dihadron Fragmentation Functions}
\label{s.DiFFs}
%%%%%%%%%%%%%%%%%%%%%%%%%%%%%%%%%%%%%%%%%%

\begin{figure}[b]
\includegraphics[width=0.95\textwidth]{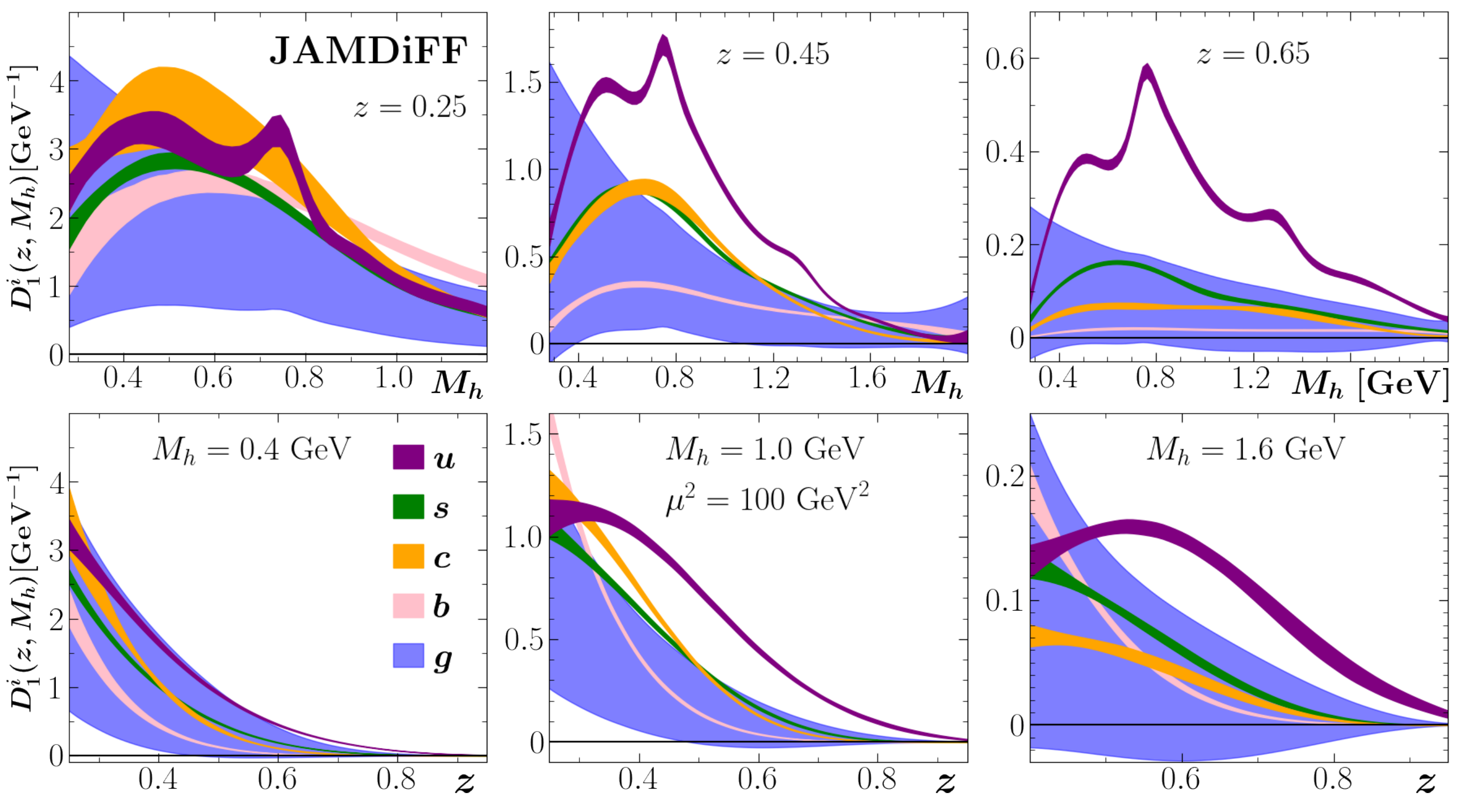}
\caption
{DiFFs $D_1^i(z,M_h)$ from the JAMDiFF (w/ LQCD) fit plotted as a function of $M_h$ with $z = 0.25, 0.45, 0.65$ (top row) and as a function of $z$ with $M_h = 0.4, 1.0, 1.6$~GeV (bottom row) at the scale $\mu^2 = 100$ GeV$^2$.  The up, strange, charm, bottom, and gluon are shown in purple, green, orange, pink, and blue, respectively, with $1\sigma$ uncertainty.}
\label{f.diffs}
\end{figure}

\begin{figure}[t]
\includegraphics[width=0.78\textwidth]{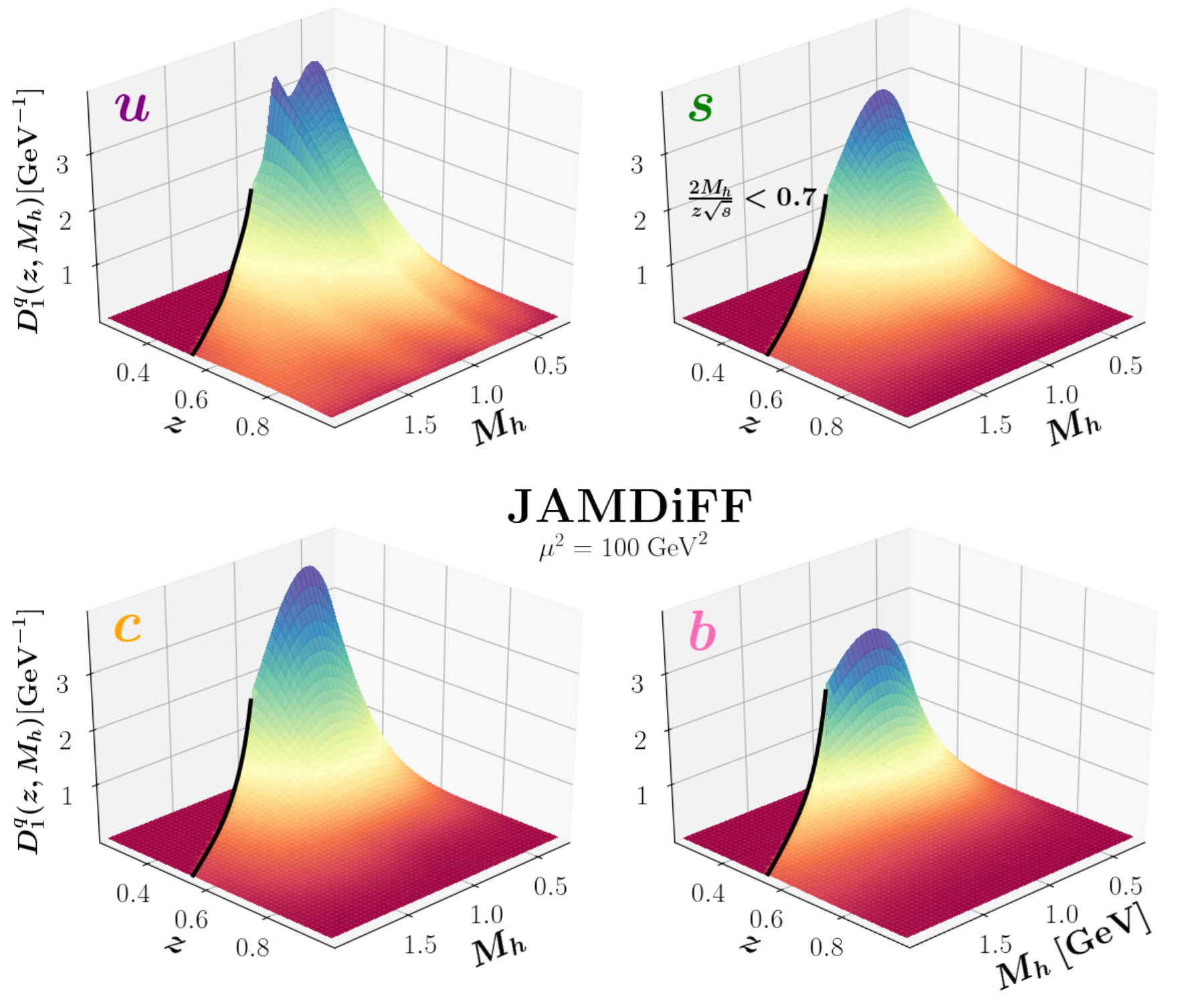}
\caption
{DiFFs $D_1^q$ from the JAMDiFF (w/ LQCD) fit plotted as a function of $M_h$ and $z$ at the scale $\mu^2 = 100$~GeV$^2$.  Only the mean values are shown.  The black line represents the cut Eq. (\ref{e.bellecut}), beyond which we do not show the DiFFs.}
\label{f.diffs-3D}
\end{figure}

In this subsection we discuss our extracted DiFFs.
The results shown below for the no LQCD and w/ LQCD fits are produced from over 900 replicas each.
Due to our new definition~\cite{Pitonyak:2023gjx} we do not make direct comparisons to the results of Refs.~\cite{Courtoy:2012ry,Radici:2015mwa}. We start with $D_1^i(z,M_h)$, shown in Figs.~\ref{f.diffs} and \ref{f.diffs-3D} at the scale $\mu^2 = 100$ GeV$^2$ (approximately the scale of the Belle experiment). We generally find strong constraints on all of the quark DiFFs, but very weak constraints on the gluon DiFF. 
This is to be expected, as we have no observables sensitive to the gluon DiFF at LO and instead depend entirely on evolution to constrain it. (As alluded to previously, the $pp$ asymmetry cannot help, as it is mainly sensitive to the transversity PDFs and $H_1^\sa$ in the numerator.)
In the near future, data on the unpolarized $pp$ cross section from STAR will be available~\cite{Babu:2023pc}, providing stronger constraints on $D_1^g$.
In the Radici, Bacchetta analysis~\cite{Radici:2018iag} (which we will refer to as RB18), they considered three scenarios for the gluon DiFF at the input scale: $D_1^g = 0$, $D_1^g = D_1^u/4$, and $D_1^g = D_1^u$, but it is clear from Fig.~\ref{f.diffs}  that none of these scenarios holds across all $z$ and $M_h$.
The extraction and error quantification for $D_1^g$ shown here should provide a more realistic uncertainty propagation on the extracted transversity PDFs. 
For $D_1^i(z,M_h)$, one sees the resonance structure in the $e^+e^-$ cross section data causes similar behavior in the up quark DiFF as a function of $M_h$. The quark DiFFs generally are larger at smaller $z$ and $M_h$ and decrease as either of those variables increases.  We expect such a behavior since the phase space to produce two hadrons from the same parton-initiated jet becomes smaller at larger $z$ and/or $M_h$. 
Note that the different definition we use for the DiFFs does not allow for a direct quantitative comparison to be made to the extraction of $D_1^i(z,M_h)$ in Ref.~\cite{Courtoy:2012ry}.  Nevertheless, the qualitative features of the DiFFs as a function of $z$ and $M_h$ can be compared, and we find similarities to the behavior seen in Figs.~2, 3 of Ref.~\cite{Courtoy:2012ry} (with the caveat that the functions are plotted at different energy scales).
In Fig.~\ref{f.diffs-3D} we present the $z$ and $M_h$ dependence of $D_1^q$ collectively in a 3-dimensional plot to emphasize, given our new definition of  DiFFs~\cite{Pitonyak:2023gjx}, it is a joint number density in those variables (only quarks are shown since the gluon has a significant uncertainty).

\begin{figure}[t]
\includegraphics[width=0.95\textwidth]{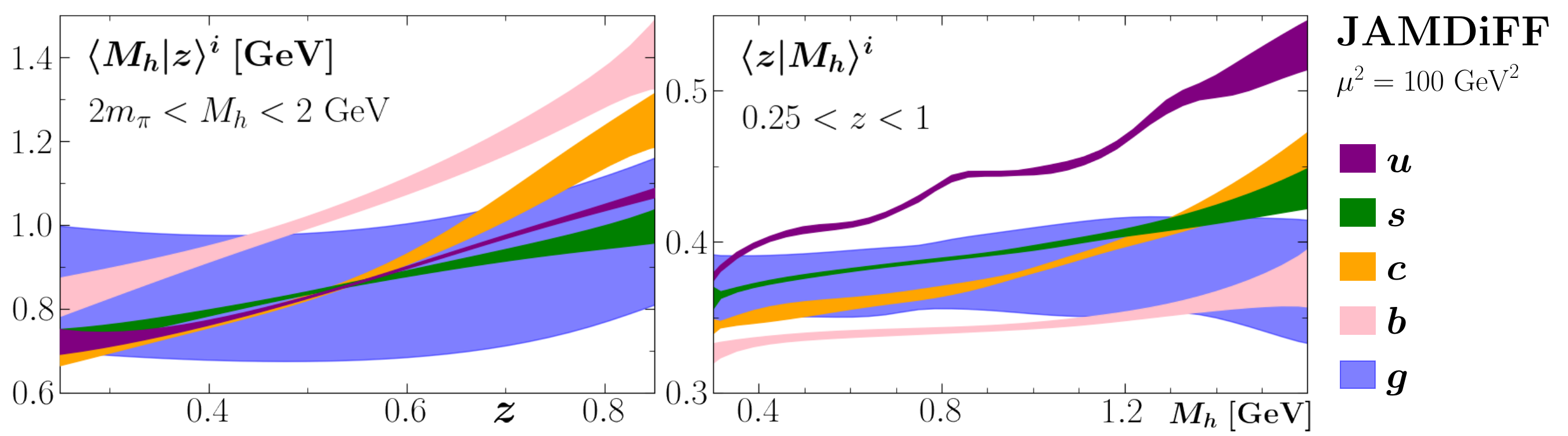}
\caption
{Left panel: Average value $\langle M_h | z \rangle^i$ with $1\sigma$ uncertainty for the ensemble of all $\pi^+\pi^-$ pairs formed from the fragmentation of a parton $i$, computed using $D_1^i(z,M_h)$ at $\mu^2 = 100$ GeV$^2$  (see Eq.~(\ref{e.Mh_given_z})), with $M_h$ integrated from $2m_{\pi}$ to 2 GeV.  Right panel: Similar to the left panel but for $\langle z | M_h \rangle^i$ (see Eq.~(\ref{e.z_given_Mh})), with $z$ integrated from 0.25 to 1.}
\label{f.diffs-expectations}
\end{figure}

\begin{figure}[t]
\includegraphics[width=0.95\textwidth]{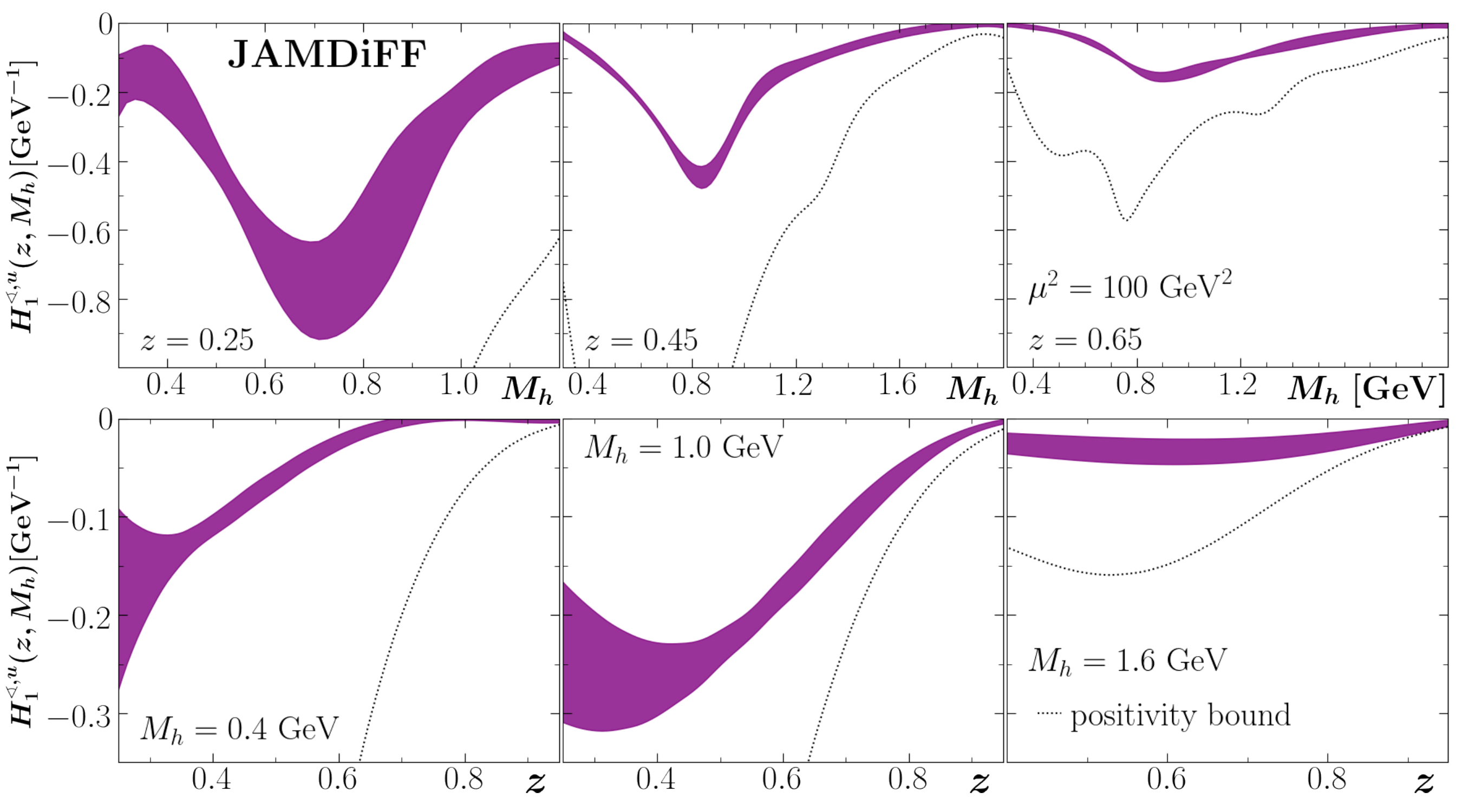}
\caption
{IFF $H_1^{\sa,u}$ from the JAMDiFF (w/ LQCD) fit plotted at $1\sigma$ uncertainty as a function of $M_h$ with $z = 0.25, 0.45, 0.65$ (top row) and as a function of $z$ with $M_h = 0.4, 1.0, 1.6$ GeV (bottom row) at the scale $\mu^2 = 100$~GeV$^2$.  The dashed black lines represents the positivity bound $\vert H_1^{\sa,u} \vert < D_1^u$.}
\label{f.tdiffs}
\end{figure}

\begin{figure}[h!]
\includegraphics[width=0.65\textwidth]{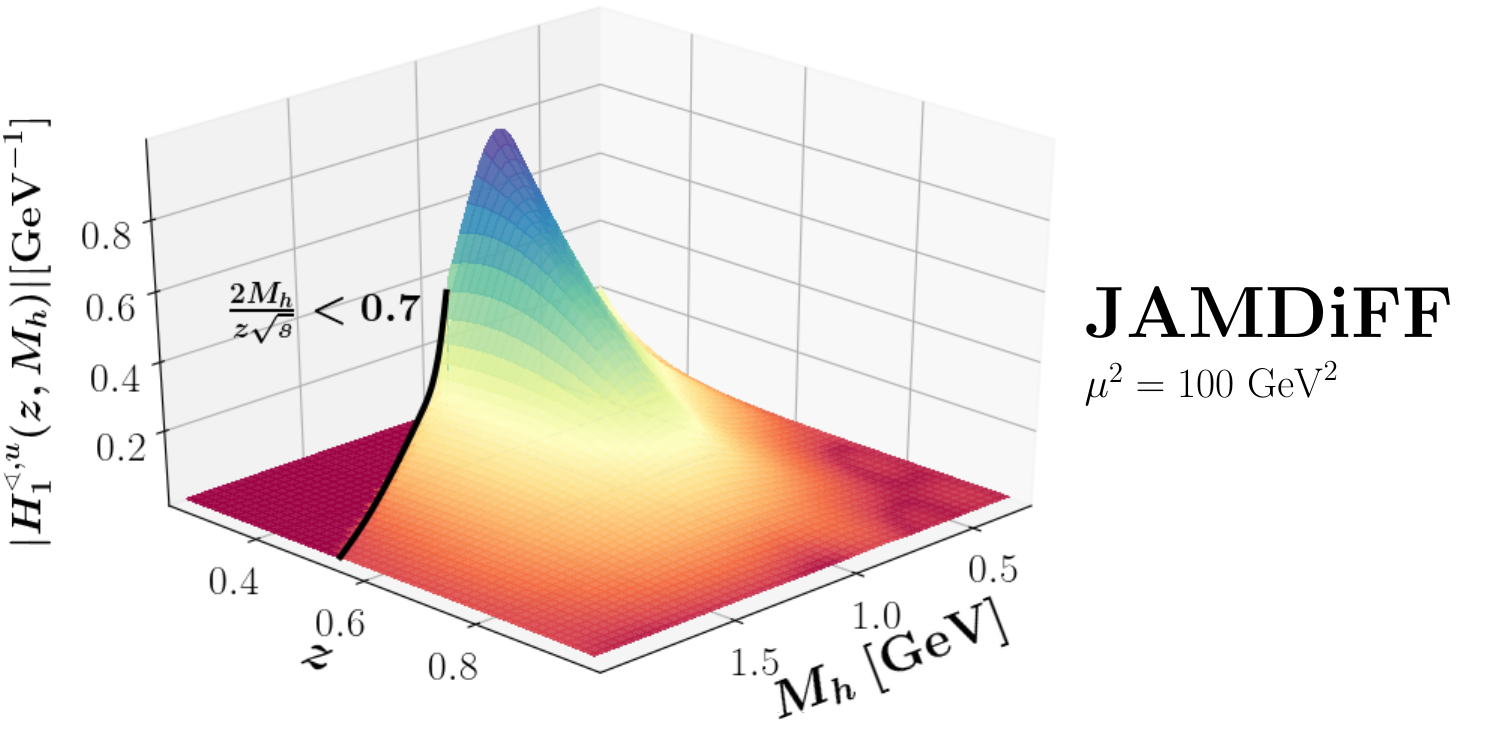}
\caption
{IFF $|H_1^{\sa, u}|$ from the JAMDiFF (w/ LQCD) fit plotted as a function of $M_h$ and $z$ at the scale $\mu^2 = 100$~GeV$^2$.  Only the mean values are shown.  The black line represents the cut Eq.~(\ref{e.bellecut}), beyond which we do not show the IFF.}
\label{f.tdiffs-3D}
\end{figure}
 
To that point, we are now able to calculate expectation values associated with dihadron fragmentation.  
For example, the average value of $M_h$ for a single $\pi^+\pi^-$ pair with a given $z$ formed from the fragmentation of a parton $i$ is
\begin{align}
\langle M_h | z \rangle^i &= \frac{\int \diff M_h\,  M_h \,D_1^i(z,M_h)}{\int \diff M_h  \,D_1^i(z,M_h)}.
\label{e.Mh_given_z}
\end{align}
Similarly, the average value of $z$ for a single $\pi^+\pi^-$ pair with a given $M_h$ formed from the fragmentation of a parton $i$ is
\begin{align}
\langle z | M_h \rangle^i = \frac{\int \diff z\,  z\, D_1^i(z,M_h)}{\int \diff z\,  D_1^i(z,M_h)}.
\label{e.z_given_Mh}
\end{align}
These quantities are shown in Fig.~\ref{f.diffs-expectations}.  The general feature is that the average $M_h$ $(z)$ increases as $z$ $(M_h)$ becomes larger.  
This can perhaps be understood by the fact that the number of dihadrons produced is greater at smaller $z$ and decreases with increasing $z$ (see Fig.~\ref{f.diffs}).
We notice from the left panel of Fig.~\ref{f.diffs-expectations} that the $u$, $d=u$, $s$, and $c$ quarks typically fragment into $\pi^+\pi^-$ pairs with similar invariant masses when the dihadron carries a small to moderate momentum fraction $z$.  As $z\to 1$, there is a tendency for the $c$ quark to produce a heavier $\pi^+\pi^-$ pair than the $u$, $d$, and $s$ quarks.  When more $\pi^+\pi^-$ pairs are produced (smaller $z$), we expect on average they will have a smaller mass, in which case the mass of the quark flavor becomes less relevant.  As less $\pi^+\pi^-$ pairs are produced as one nears threshold ($z\to 1$), then the mass of the quark more directly correlates to the mass of the dihadron.  The mass of the $b$ quark being much heavier than the others, it separates itself to yield heavier dihadrons even at small $z$.  
The large uncertainties for the gluon expectation values are a reflection of the large uncertainties on $D_1^g$.
The right panel of Fig.~\ref{f.diffs-expectations} displays a clear hierarchy of the average dihadron momentum fractions, with $z$ decreasing from the lightest ($u, d, s$ quarks) to the heaviest ($c, b$ quark).  This pattern continues until around $m_c$, where $\pi^+\pi^-$ pairs produced from the $c$ quark start to carry slightly larger momentum fractions that  are comparable to dihadrons produced by the $s$ quark.  The trend in the plot aligns with the idea that for a given $M_h$, $\pi^+\pi^-$ pairs need to carry smaller momentum fractions if they arise from the fragmentation of a heavier quark.  The gluon produces dihadron pairs with the same $z\; (\approx 0.35 - 0.4)$ independent of the invariant mass.

In Figs.~\ref{f.tdiffs} and \ref{f.tdiffs-3D} we show the IFF at the scale $\mu^2 = 100$ GeV$^2$.
As mentioned in Sec.~\ref{ss.SIA}, the sign of $H_1^{\sa, u}$ cannot be fixed by the experimental data alone, and we have thus forced $H_1^{\sa, u}$ to be negative to be consistent with the expected sign of $h_1^{u_v}$.
We generally find that the IFF is not constrained by the positivity bound $| H_1^{\sa, u} | < D_1^u$, except at large $M_h$ and $z$ where its magnitude begins to be limited by the bound.  The IFF decreases with increasing $z$ and peaks at $M_h\approx 0.8\,{\rm GeV}$.

%%%%%%%%%%%%%%%%%%%%%%%%%%%%%%%%%%%%%%%%%%
\section{Extracted Transversity Distributions and Tensor Charges}
\label{s.TPDFs}
%%%%%%%%%%%%%%%%%%%%%%%%%%%%%%%%%%%%%%%%%%

%%%%%%%%%
\subsection{Main Results}
%%%%%%%%

\begin{figure}[b!]
\includegraphics[width=0.95\textwidth]{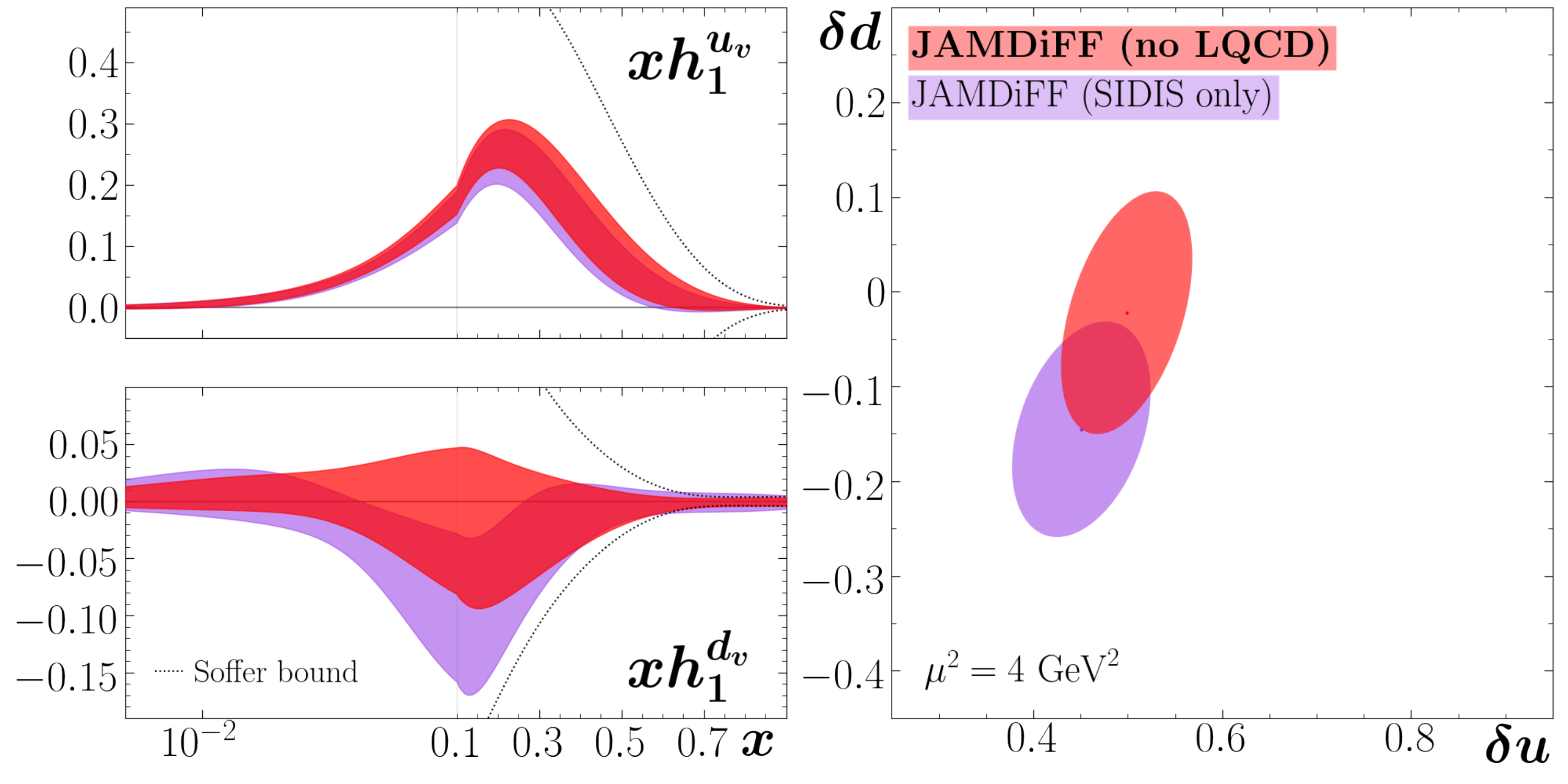}
\caption
{Left panel: Transversity PDFs $xh_1^{u_v}$ and $xh_1^{d_v}$ plotted as a function of $x$ at the scale $\mu^2 = 4$ GeV$^2$. We show results for the JAMDiFF (no LQCD) fit (red) and SIDIS only fit (violet) with 1$\sigma$ uncertainty. The Soffer bound is indicated by the dashed black lines.
Right panel: The tensor charges $\delta u$ and $\delta d$ at the scale $\mu^2 = 4$~GeV$^2$ for the JAMDiFF (no LQCD) and SIDIS only fits with 1$\sigma$ uncertainty.}
\label{f.sidis-only}
\end{figure}

As discussed in Sec.~\ref{ss.experiment}, we perform three different analyses for the transversity PDFs to assess the impact of experimental datasets and LQCD on the results.
In Fig.~\ref{f.sidis-only} we show the JAMDiFF (no LQCD) and SIDIS only results for the transversity PDFs and the tensor charges.  The two fits agree within errors, with the JAMDiFF (no LQCD) fit preferring a slightly larger $u_v$ distribution and the SIDIS only fit preferring a more negative $d_v$.
These results translate directly into the tensor charge values, with the SIDIS only result being slightly smaller for $\delta u$ and more negative for $\delta d$.

From the SIDIS only fit one can easily understand the signs of the extracted transversity PDFs.
Referring to Fig.~\ref{f.SIDIS}, one sees that the proton asymmetry from COMPASS is negative.  
In Eq.~(\ref{e.SIDISproton}), we showed that the proton asymmetry is dominated by $h_1^{u_v}$.  Since $H_1^{\sa,u}$ is (chosen to be) negative (see Fig.~\ref{f.tdiffs}), $h_1^{u_v}$ must be positive to lead to an overall negative asymmetry.
A similar argument applies to the HERMES proton asymmetry, except that the negative sign in Eq.~(\ref{e.AUTSIDIS}) makes the asymmetry positive.
For the deuteron asymmetry from COMPASS one sees that it is largely consistent with zero.  From Eq.~(\ref{e.SIDISdeuteron}) the asymmetry is proportional to the sum of $h_1^{u_v}$ and $h_1^{d_v}$.  Since the proton asymmetry fixes $h_1^{u_v}$ to be positive, the deuteron asymmetry then fixes $h_1^{d_v}$ to be negative so that the sum largely cancels.
This qualitative finding nicely agrees with a model-independent QCD analysis for large $N_c$~\cite{Pobylitsa:2003ty}.
Taking the SIDIS results for $h_1^{u_v}$ and $h_1^{d_v}$ and neglecting antiquarks, one can also understand the sign of the $pp$ asymmetry. The measurements are at mid-rapidity where $\hat{s}\approx -\hat{t}\approx -\hat{u}$.    The $q^\uparrow q(q')\to q^\uparrow q(q')$ and $q^\uparrow g\to q^\uparrow g$ channels give the main contributions and their hard factors in this region are negative (see Appendix~\ref{a.pp}).  The numerator of the asymmetry for these channels involves the combinations (cf.~Eq.~(\ref{e.ppUT})) $H_1^{u,\sa}[f_1^{u(q')}h_1^u-f_1^{d(q')}h_1^d]<0$ and $f_1^gH_1^{u,\sa}[h_1^u-h_1^d]<0$, respectively. 
Thus, when multiplied by the negative hard factors, one ultimately obtains a positive asymmetry.

\begin{figure}[t]
\includegraphics[width=0.95\textwidth]{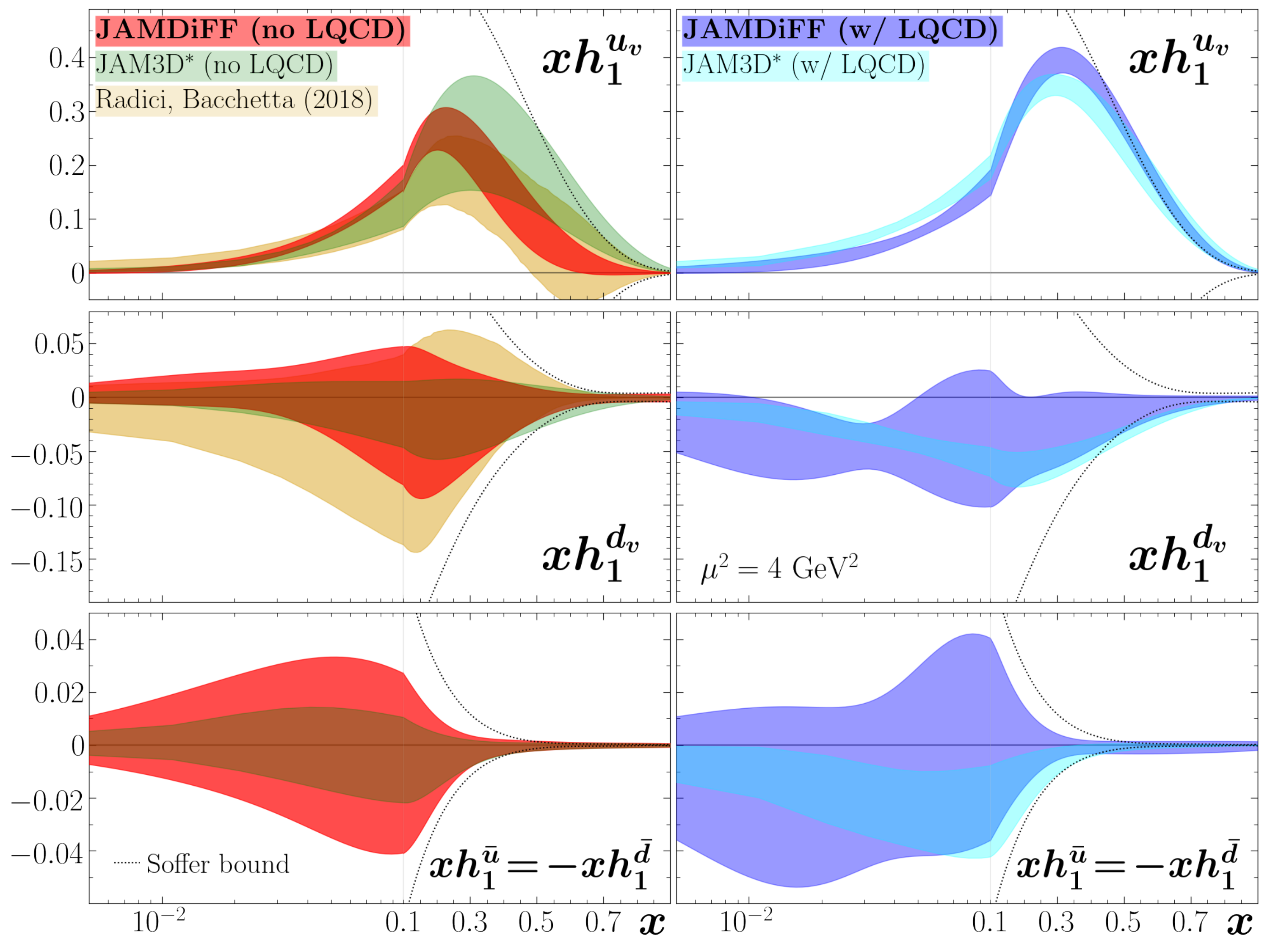}
\caption
{Transversity PDFs $x h_1^{u_v}$ (top row), $x h_1^{d_v}$ (middle row) and $x h_1^{\bar{u}}$ (bottom row) plotted as a function of $x$ at the scale $\mu^2 = 4$ GeV$^2$.
Our results (JAMDiFF) are shown at 1$\sigma$ both with (blue) and without (red) LQCD included in the fit and are compared to those from JAM3D$^*$ \cite{Gamberg:2022kdb,Note1} at 1$\sigma$ with (cyan) and without (green) LQCD and RB18 \cite{Radici:2018iag} (gold, 90\% CL).
The Soffer bound is indicated by the dashed black lines.
Note that for JAMDiFF and JAM3D$^*$ the relation $h_1^{\bar{d}} = - h_1^{\bar{u}}$ from the large-$N_c$ limit is enforced, while for RB18 the antiquarks are not fitted.}
\label{f.tpdfs}
\end{figure}

In Fig.~\ref{f.tpdfs} we compare our results with and without LQCD for the transversity PDFs to those from Radici, Bacchetta~\cite{Radici:2018iag}  (whose analysis did not consider the inclusion of lattice data).
We also compare to a version of JAM3D that has been slightly updated from Ref.~\cite{Gamberg:2022kdb} (see the footnote
\footnote{In order to align with the methodology of JAMDiFF, we show here results from the JAM3D analysis that are slightly updated from Ref.~\cite{Gamberg:2022kdb}:~antiquark transversity PDFs are now included (with $h_1^{\bar{d}} = - h_1^{\bar{u}}$), the small-$x$ constraint Eq.~(\ref{e.alpha}) is imposed, and, for the fit with LQCD, $\delta u$ and $\delta d$ from ETMC and PNDME are used (instead of only the $g_T$ data point from ETMC).})
that we will refer to as JAM3D$^*$.
For the no LQCD results we agree with RB18 within errors, but with a larger $h_1^{u_v}$ in the region $ 0.04 \lesssim x \lesssim 0.3$.
The overall smaller errors on our analysis can be partially attributed to the inclusion of all three SIDIS binnings ($\xb$, $z$, $M_h$), while the RB18 analysis only included the $\xb$ binning.
We note that the inclusion of the small-$x$ constraint (Eq.~(\ref{e.alpha})) and antiquarks in this analysis (neither of which were included in RB18) have no significant impact on the valence transversity distributions in the measured region.  
For details on the comparison to JAM3D$^*$, see Ref.~\cite{Cocuzza:2023oam}.

Within our analysis, the increase in $h_1^{u_v}$ in the $x \gtrsim 0.3$ region when LQCD is included is a consequence of the LQCD results for $\delta u$ being larger than the result of the fit without LQCD (see Fig.~2 of Ref.~\cite{Cocuzza:2023oam}).
The experimental data provides strong constraints in the $0.005 \lesssim x \lesssim 0.2$ region (see Fig.~\ref{f.kin}), while the small-$x$ constraint of Eq.~(\ref{e.alpha}) prevents $h_1^{u_v}$ from becoming significantly larger at $x \lesssim 0.005$.  Similarly, the Soffer bound does not allow the transversity PDF to get large at very high $x$.  Thus, in order to increase $\delta u$, the best option for the fit is to increase $h_1^{u_v}$ in the $x \approx 0.3$ region, although this leads to a slight deterioration in the description of the SIDIS data (see Table \ref{t.chi2}), especially the COMPASS $x_{\rm bj}$ binning.
The lack of overlap within errors between the no LQCD and w/ LQCD results will be discussed in detail in Sec.~\ref{ss.compatiblity}.
For $h_1^{d_v}$, the largest change occurs below $x \approx 0.05$, where, after the inclusion of the LQCD data, it now tends to be negative.  This is a consequence of the LQCD result for $\delta d$ being more negative than the result of the fit without LQCD.

In Fig.~\ref{f.tpdfs} we also show in the bottom row the JAMDiFF (no LQCD) result for the antiquark transversity distributions and compare to JAM3D$^*$, where both analyses assume $h_1^{\bar{u}} = -h_1^{\bar{d}}$ (see Sec.~\ref{ss.TPDF_par}).
The two results are in agreement.
The Soffer bound forces the antiquarks to be very small above $x \gtrsim 0.3$.
Below that region, they still remain small and consistent with zero.
The smaller uncertainties on the JAM3D$^*$ result are partly due to the less flexible parameterization used in that analysis.
The inclusion of the LQCD measurements causes $h_1^{\bar{u}}$
to become slightly negative at small $x$, though the result is still compatible with zero.  For JAM3D$^*$, $h_1^{\bar{u}}$ is negative up to $x \approx 0.2$.  While the observation of nonzero antiquark transversity PDFs is interesting, it would be premature to assign any significance to this result.

%%%%%%%%%%%%%%%%%%%%%%%%%%%%%%%%%%%%%%%%%%
%%%%%%%%%%%%%%%%%%%%%%%%%%%%%%%%%%%%%%%%%%

\begin{figure}[h!]
\includegraphics[width=0.95\textwidth]{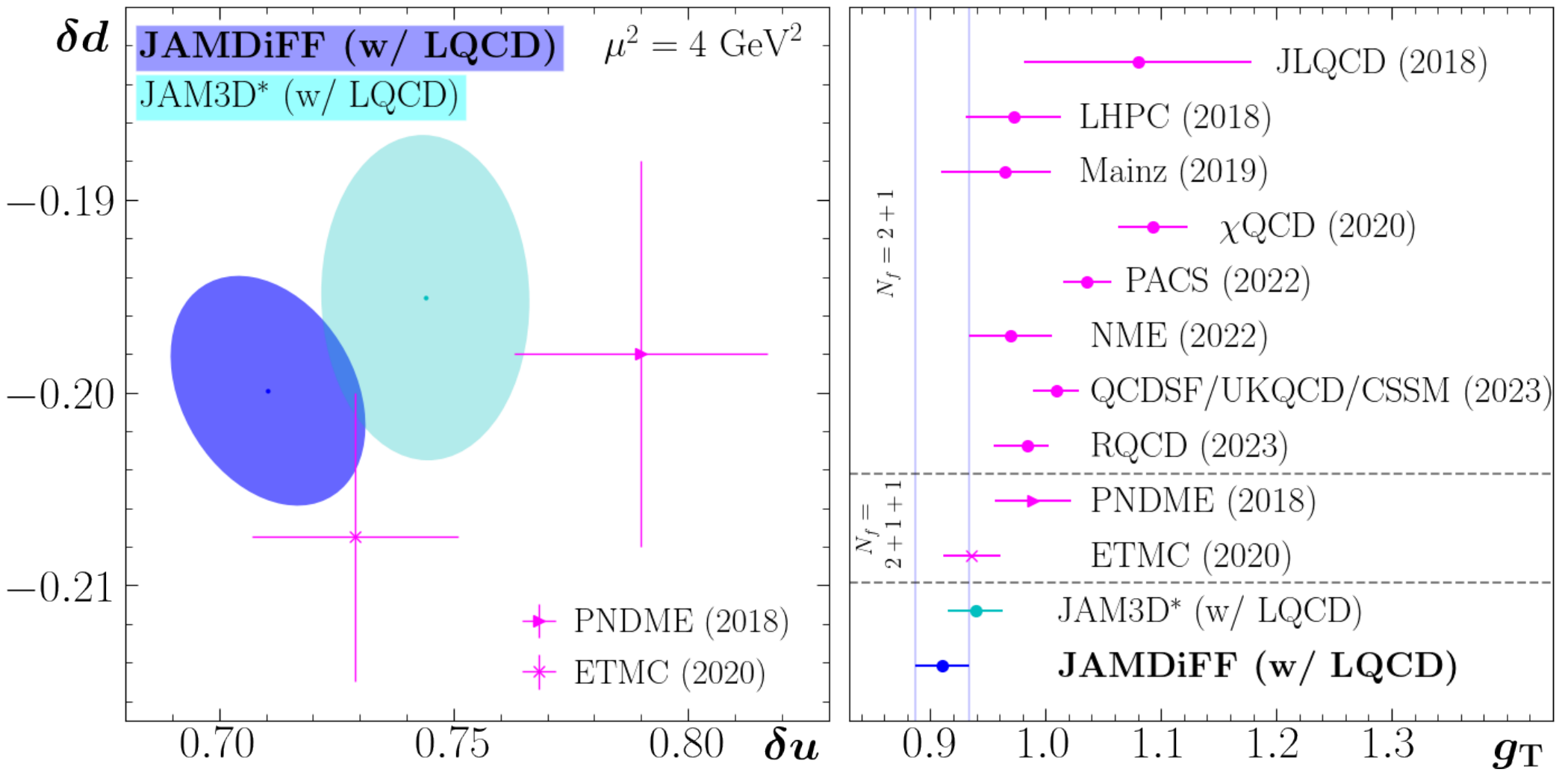}
\caption
{The tensor charges $\delta u$, $\delta d$, and $g_T$ at the scale $\mu^2 = 4$ GeV$^2$.
We show results at 1$\sigma$ for the JAMDiFF (w/ LQCD) fit (blue) and compare to the JAM3D$^*$ \cite{Gamberg:2022kdb, Note1} result with lattice QCD (cyan, 1$\sigma$), and LQCD results~\cite{Gupta:2018qil, Gupta:2018lvp, Alexandrou:2021oih, Yamanaka:2018uud, Horkel:2020hpi, Hasan:2019noy, Harris:2019bih, QCDSFUKQCDCSSM:2023qlx, Smail:2023eyk, Park:2021ypf, Tsuji:2022ric} (magenta points).  The LQCD results are organized by $N_f$ with $N_f = 2+1+1$ or $N_f=2+1$.}
\label{f.tensorcharge-lattice}
\end{figure}

We now move on to the tensor charges.
We refer the reader to Ref.~\cite{Cocuzza:2023oam} for a detailed discussion of the resulting tensor charges from the no LQCD and with LQCD fits.
In Fig.~\ref{f.tensorcharge-lattice} we present a full comparison of the JAMDiFF (w/ LQCD) fit to recent LQCD results \cite{Gupta:2018qil, Gupta:2018lvp, Alexandrou:2021oih, Yamanaka:2018uud, Horkel:2020hpi, Hasan:2019noy, Harris:2019bih, QCDSFUKQCDCSSM:2023qlx, Smail:2023eyk, Park:2021ypf, Tsuji:2022ric}.
The phenomenological results of JAMDiFF (w/ LQCD) and JAM3D$^*$ (w/ LQCD) are generally on the lower end for $g_T$ when considering all LQCD values, but nevertheless are in good agreement with ETMC, NME, Mainz, and LHPC.
They are in reasonable agreement with PNDME, RQCD, and JLQCD, and the worst agreement is found with QCDSF/UKQCD/CSSM, PACS, and $\chi$QCD.
We thus find that our extracted $g_T$ is compatible with most LQCD calculations.
Our tensor charge results for the JAMDiFF (w/ LQCD), JAMDiFF (no LQCD), and JAMDiFF (SIDIS only) fits are summarized in Table~\ref{t.tensorcharges}.

\begin{table}[h]
\resizebox{0.50\textwidth}{!} {
\begin{tabular}{l | c | c | c }
\hhline{====}
\vspace{-0.3cm} & & &  \\
Fit & $\delta u$ & $\delta d$ & $g_T$ \\
\vspace{-0.3cm} & & &  \\
\hline
\textbf{JAMDiFF (w/ LQCD)}    & \textbf{0.71(2)}  & \textbf{-0.200(6)} & \textbf{0.91(2)} \\
\textbf{JAMDiFF (no LQCD)}    & \textbf{0.50(7)}  & \textbf{-0.02(13)} & \textbf{0.52(12)} \\
JAMDiFF (SIDIS only) & 0.45(7)  & -0.15(13) & 0.60(11) \\
\hhline{====}
\end{tabular}}
\caption
{Summary of tensor charge results.  The errors correspond to 1$\sigma$.}
\label{t.tensorcharges}
\end{table}

%%%%%%%%%%%%%%%%%%%%%%%%%%%%%%%%%%%%%%%%%%
\subsection{Further Exploring Compatibility between Lattice QCD and Experimental Data}
\label{ss.compatiblity}
%%%%%%%%%%%%%%%%%%%%%%%%%%%%%%%%%%%%%%%%%%

In this section we explore further the compatibility of the experimental  measurements and LQCD data.  We begin by discussing the resulting $\chired$ values for the LQCD data (see Table \ref{t.chi2}).
While our description of the ETMC and PNDME $\delta d$ are very good, the PNDME $\delta u$ has a large $\chired$ of 8.68 even after its inclusion in the fit.
Following the methodology of NNPDF4.0 \cite{NNPDF:2021njg} (see Sec.~4.2.3), we test the compatibility of the LQCD $\delta u$ data with the experimental data by reweighting its contribution to the total $\chi^2$, so that it is comparable to the $\chi^2$ for the rest of the experimental data, and rerunning the fit.
Specifically, we weight both ETMC and PNDME in a single fit, with weighting factor $w = (1475 + 2978)/2 = 2226.5$ for each. [The numerator is the total number of points in the fit (1475 from experimental + LQCD data, 2978 from PYTHIA data) and the denominator is the number of weighted points (the two $\delta u$ points from ETMC and PNDME).]
Upon doing so, we find a similar $\chired$ of 0.86 for ETMC $\delta u$, and a significantly lower $\chired$ of 2.26 for PNDME $\delta u$.  
Thus, the ability for the fit to describe the PNDME $\delta u$ point is restricted by its tension with the ETMC result combined with the fact that the ETMC point lies closer to the JAMDiFF (no LQCD) result.
The $\chired$ for the STAR data remains nearly identical, while the $\chired$ for SIDIS increases from 1.04 to 1.09.
While the increases in the experimental $\chired$ are small, they are significant enough to prevent a better description of the PNDME $\delta u$ point due to its small weight in the fit. 
We note that in this fit the $u_v$ distribution naturally becomes a bit larger (but still overlaps within errors with the w/ LQCD fit), while the rest of the PDFs and DiFFs remain stable.
We will discuss in detail below the sources of the increases in the $\chired$ of the experimental data when LQCD information is included.

Before doing so, we mention an alternative way of treating the LQCD data, which takes into account the tension between the ETMC and PNDME $\delta u$ measurements in a more conservative manner.
We label this method as ``JAMDiFF (w/ LQCD Flat Sampling)."
Instead of fitting the ETMC and PNDME points directly, we construct a range for $\delta u$ that covers both points including their 1$\sigma$ errors: $0.707 < \delta u < 0.817$.  We then generate two points randomly (flat sampling) within this range with the 1$\sigma$ error taken to be $(0.817-0.707)/2=0.055$.  We repeat this process for $\delta d$.
Upon including these new points in the fit, we find $\delta u = 0.62(5)$, $\delta d = -0.195(11)$, and $g_T = 0.82(5)$.
The result for $\delta d$ unsurprisingly agrees with that of the w/ LQCD fit (see Table~\ref{t.tensorcharges}).
The result for $\delta u$, on the other hand, lies in between the no LQCD and w/ LQCD fits ($1.4\sigma$ larger than the former and $1.7\sigma$  smaller than the latter) in both its central values and in the size of its errors.
This is to be expected given the larger errors on the generated $\delta u$ data, which have less of an impact on the transversity PDFs and cause the extracted result for $\delta u$ to also have larger errors.
Ultimately, we find a 1.9$\sigma$ discrepancy between the extracted $\delta u$ and the constructed data point $\delta u = 0.762(55)$, again demonstrating reasonable agreement between experiment and LQCD.
We note that this method of fitting the LQCD data is more lenient with regard to testing the compatibility between experiment and LQCD. Our ``standard" method of fitting ETMC and PNDME separately is a much more stringent test of their compatibility with experimental data and still successful.

\begin{figure}[t]
\includegraphics[width=0.95\textwidth]{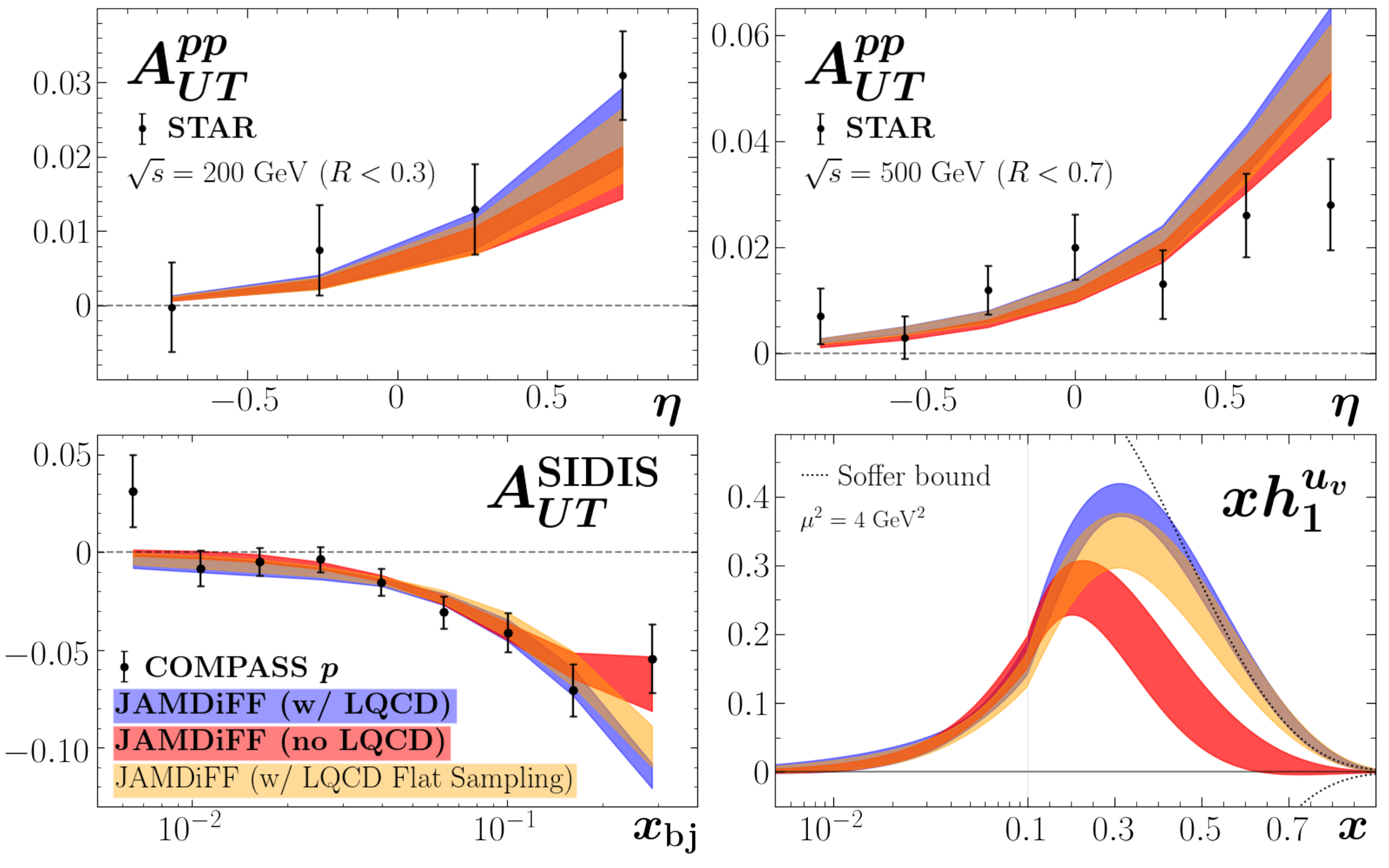}
\caption
{Results for three different fits: JAMDiFF (w/ LQCD) (blue), JAMDiFF (no LQCD) (red), and the JAMDiFF (w/ LQCD Flat Sampling) (orange).
Top left: The $pp$ asymmetry  data from STAR \cite{STAR:2015jkc} at $\sqrt{s} = 200$ GeV binned in $\eta$ (black circles) plotted against the results with 1$\sigma$ uncertainty bands.
Top right: Same as the top left, except for the STAR data at $\sqrt{s} = 500$ GeV \cite{STAR:2017wsi}.
Bottom left: Same as the top left, except for the SIDIS asymmetry data from COMPASS \cite{COMPASS:2023cgk} on a proton target binned in $\xb$.
Bottom right: Transversity PDF $x h_1^{u_v}$ plotted as a function of $x$ at the scale $\mu^2 = 4$ GeV$^2$ with $1\sigma$ uncertainty bands.
The Soffer bound is indicated by the dashed black lines.}
\label{f.highx}
\end{figure}

From Table~\ref{t.chi2} we see that the largest changes in the $\chired$ going from the no LQCD to w/ LQCD fits are seen in the COMPASS proton measurement (increasing from $0.65$ to $1.98$), the STAR $\sqrt{s} = 500$ GeV data binned in $\eta$ (increasing from $1.83$ to $2.97$), and the STAR $\sqrt{s} = 200$ GeV data binned in $\eta$ (decreasing from $1.46$ to $0.52$).
Thus, in Fig.~\ref{f.highx} we show these three datasets, as well as $h_1^{u_v}$, for three different fits: no LQCD, w/ LQCD, and w/ LQCD (flat sampling).
We see that the changes (in absolute terms) between the theory calculations from the fits occur almost entirely at larger $x$ for SIDIS and larger $\eta$ for $pp$ (which likewise corresponds to probing transversity in the higher-$x$ region).
In this regime the observables are determined primarily by the $h_1^{u_v}$ distribution, and so we see a direct correspondence between the magnitude of $h_1^{u_v}$ at higher $x$ and the magnitude of the asymmetries.

With the small-$x$ constraint (Eq.~(\ref{e.alpha})) and strong constraints from experimental data in the region $0.005 \lesssim x \lesssim 0.2$, the fit has no choice but to increase the size of $h_1^{u_v}$ in the large $x > 0.2$ region in its attempt to accommodate the $\delta u$ data from LQCD.
Within our framework, this puts it in tension with a few experimental data points, most notably the highest-$x$ COMPASS proton point and the highest-$\eta$ STAR $\sqrt{s} = 500$~GeV point (but in better agreement with the highest-$\eta$ STAR $\sqrt{s}=200~{\rm GeV}$ point).
Therefore, in order to further test the compatibility between LQCD and experimental data, it is of vital importance to have more measurements at larger $x$ (for SIDIS) and more forward rapidity (for $pp$).
We see that, in the no LQCD fit, $h_1^{u_v}$ has a maximum and then begins to decrease around where the $x$ coverage of the experimental data ends ($x \approx 0.3$) due to the highest-$x$ COMPASS data point.
Within our parameterization, the PDFs fall off smoothly and monotonically as $x \to 1$, and this drives the behavior (and uncertainty) of $h_1^{u_v}$ in the unmeasured ($x>0.3$) region (i.e., $h_1^{u_v}$ cannot suddenly increase at larger $x>0.3$ once it begins its downturn).  The fit with LQCD included has additional constraints at larger $x$ due to the fact that one integrates from $x\in [0,1]$ to calculate the tensor charges.  As mentioned previously, this causes $h_1^{u_v}$ to now peak at slightly higher $x\approx 0.35$ in order to accommodate both LQCD and experimental data.  The Soffer bound forces the with LQCD $h_1^{u_v}$ to then decrease shortly after $x= 0.35$ and the PDF again falls off smoothly and monotonically as $x \to 1$.  
The LQCD data influences the behavior and uncertainty (making it smaller) in the large-$x$ region in order for $h_1^{u_v}$ to give the appropriate value for $\delta u$.  This in part helps to explain why, in our analysis, there is not overlap in the $h_1^{u_v}$ and $\delta u$ error bands between the JAMDiFF (w/ LQCD) and JAMDiFF (no LQCD) fits.

In addition, we have seen in Fig.~\ref{f.highx} that the LQCD data and STAR $\sqrt{s} = 200$~GeV data have a preference for a larger $h_1^{u_v}$ at large $x$, while the COMPASS proton data and STAR $\sqrt{s} = 500$~GeV data prefer a smaller $h_1^{u_v}$.
In such a situation where there are competing preferences, and we compare analyses containing different subsets of the data, the choice of likelihood function $\mathcal{L}$ and prior $\pi$ do not guarantee that the fits overlap within statistical uncertainties (see, e.g., Ref.~\cite{Sivia1996DealingWD}).
Before drawing a conclusion about the compatibility between LQCD tensor charges and experimental data, one needs first to include both in the analysis.
One should only be concerned if the description of the lattice data remains poor even after its inclusion and/or if the description of the experimental data suffers significantly.
Indeed, we have performed several comprehensive tests that show we are able to describe the LQCD results while not significantly sacrificing the description of the experimental data, showing their compatibility.

%%%%%%%%%%%%%%%%%%%%%%%%%%%%%%%%%%%%%%%%%%
\section{Summary and Outlook}
\label{s.summary}
%%%%%%%%%%%%%%%%%%%%%%%%%%%%%%%%%%%%%%%%%%

We have presented the most comprehensive study of dihadron observables (JAMDiFF) to date by performing, for the first time, a simultaneous global QCD analysis of the $\pi^+ \pi^-$ DiFFs/IFFs and transversity PDFs from $e^+ e^-$ annihilation, SIDIS, and $pp$ collisions while including the Belle cross section data~\cite{Belle:2017rwm}, the latest $pp$ dihadron measurements from STAR~\cite{STAR:2017wsi}, and all kinematic variable binnings for the relevant processes under consideration.
Our extracted transversity PDFs and DiFFs at any kinematics can be found in a {\tt github} library~\cite{github-repo} and a {\tt google colab} notebook~\cite{google-colab} that provide a platform to independently access our results.
We utilized a parameterization for $D_1(z,M_h)$ (and $H_1^{\sa}(z,M_h)$) which allowed an accurate description of the $\diff\sigma/\diff z\,\diff M_h$ data from Belle and events generated by PYTHIA while remaining computationally efficient.
Studying multiple center-of-mass energies with the latter allowed for the first extraction of the gluon $D_1(z,M_h)$.  From the extracted quark and gluon $D_1(z,M_h)$, where we use a definition that now has a number density interpretation~\cite{Pitonyak:2023gjx}, we made the first calculations of expectation values for the $\pi^+\pi^-$ dihadron invariant mass (as a function of $z$) and momentum fraction (as a function of $M_h$), studying also how these vary with parton flavor.

Furthermore, we have tested the compatibility of different techniques used for determining the tensor charges of the nucleon.  By incorporating theory constraints on the transversity PDFs at small $x$ and large $x$ we were able to meaningfully include LQCD results for $\delta u$ and $\delta d$ from ETMC~\cite{Alexandrou:2019brg} and PNDME~\cite{Gupta:2018lvp} into our analysis. We found compatibility with this data while maintaining a very good description of the experimental measurements.  In addition, our results for $\delta u$, $\delta d$ and the $x$ dependence of transversity match closely to those from the single-hadron TMD/collinear twist-3 analysis of JAM3D~\cite{Gamberg:2022kdb,Note1}.  We have thus demonstrated, for the first time, the universal nature of all available information on the transversity PDFs and tensor charges of the nucleon.
We also note that we extracted information about the antiquark transversity PDFs, which had not been done previously for dihadron production.

There are several future directions for extending this analysis.  More data is expected from STAR on not only the dihadron azimuthal asymmetry but also the unpolarized cross section.  The latter will be important for providing the first experimental constraint on the gluon $D_1(z,M_h)$. We also anticipate measurements of dihadron multiplicities in SIDIS to offer further experimental quark flavor separation of $D_1(z,M_h)$. 
Furthermore, it is mandatory to extend the theoretical framework to next-to-leading order (NLO).  Generally, NLO corrections to cross sections can be large, and they will certainly modify our quantitative results.  However, we expect that our main qualitative findings about the transversity PDFs and the tensor charges will not be affected, as the relevant observables are asymmetries for which typically significant cancellations of higher-order corrections occur.  Nevertheless, a NLO analysis is needed to provide a definitive answer to that question.
Overall, future measurements at STAR, Jefferson Lab, and the Electron-Ion Collider can provide crucial information to help reduce uncertainties on the transversity PDFs and tensor charges~\cite{Gamberg:2021lgx, AbdulKhalek:2021gbh}.  In particular, we identified the large-$x$ (in the case of SIDIS) and forward rapidity (in the case of $pp$) regions as key to further test the compatibility between LQCD results for the tensor charges and experiment.  Measurements in the small-$x$ region will also help test the small-$x$ asymptotic behavior of transversity imposed in our analysis and give further insight on the antiquark transversity PDFs.  Moreover, dihadron, single-hadron TMD/collinear twist-3 observables, and LQCD (tensor charge and $x$-dependent transversity computations~\cite{HadStruc:2021qdf,Alexandrou:2021oih}) should eventually be fit simultaneously in a ``universal'' analysis.

%%%%%%%%%%%%%%%%%%%%%%%%%%%%%%%%%%%%%%%%%%
\begin{acknowledgments}
We thank Wally Melnitchouk for helpful discussions, 
Yuri Kovchegov for useful exchanges on the small-$x$ behavior of transversity,
and Anna Martin and Gunar Schnell for clarification on the COMPASS and HERMES data, respectively.
This work was supported by the National Science Foundation under Grants No.~PHY-2110472 (C.C.~and A.M.), No.~PHY-2011763 and No.~PHY-2308567 (D.P.), and No.~PHY-2012002, No.~PHY-2310031, No.~PHY-2335114 (A.P.), and the U.S. Department of Energy contract No.~DE-AC05-06OR23177, under which Jefferson Science Associates, LLC operates Jefferson Lab (A.P.~and N.S.). 
The work of N.S. was supported by the DOE, Office of Science, Office of Nuclear Physics in the Early Career Program.
The work of C.C., A.M., and A.P.~was supported by the U.S. Department of Energy, Office of Science, Office of Nuclear Physics, within the framework of the TMD Topical Collaboration, and by Temple University (C.C. and A.P.).
\end{acknowledgments}
%%%%%%%%%%%%%%%%%%%%%%%%%%%%%%%%%%%%%%%%%%

\appendix

\section{Partonic Cross Sections in Proton-Proton Collisions}
\label{a.pp}

Here we list the partonic cross sections that enter Eq.~(\ref{e.ppUT}).
The unpolarized partonic cross sections match exactly those from the appendix of Ref.~\cite{Bacchetta:2004it}.
For the transversely polarized partonic cross sections, we use results that are similar to those of Ref.~\cite{Bacchetta:2004it}, except with an opposite overall sign on the four channels involving only quarks.
Furthermore, because we define proton $A$ to be the transversely polarized proton, rather than proton $B$, we interchange $\hat{u} \leftrightarrow \hat{t}$ relative to Ref.~\cite{Bacchetta:2004it}.
This leads to the following for the partonic cross sections involving quark flavors $q,q'$:
\begin{align}
\frac{\diff \Delta \hat{\sigma}_{q^{\uparrow} q \to q^{\uparrow}q}}{\diff \hat{t}} &= \frac{8 \pi \alpha_s^2}{27 \hat{s}^2} \frac{\hat{s} (3\hat{u} - \hat{t})}{\hat{t}^2}, \label{e:qq_channel}\\
\frac{\diff \Delta \hat{\sigma}_{q^{\uparrow} q' \to q^{\uparrow}q'}}{\diff \hat{t}} &= \frac{\diff \Delta \hat{\sigma}_{q^{\uparrow} \bar{q}' \to q^{\uparrow}\bar{q}'}}{\diff \hat{t}} = \frac{8 \pi \alpha_s^2}{9 \hat{s}^2} \frac{\hat{u} \hat{s}}{\hat{t}^2} ,\\
\frac{\diff \Delta \hat{\sigma}_{q^{\uparrow} \bar{q} \to q^{\uparrow} \bar{q}}}{\diff \hat{t}} &= \frac{8 \pi \alpha_s^2}{27 \hat{s}^2} \frac{\hat{u} (3\hat{s} - \hat{t})}{\hat{t}^2},  \\
\frac{\diff \Delta \hat{\sigma}_{\bar{q}^{\uparrow} q \to q^{\uparrow}\bar{q}}}{\diff \hat{t}} &=\frac{8 \pi \alpha_s^2}{27 \hat{s}^2},  \\
\frac{\diff \Delta \hat{\sigma}_{q^{\uparrow} g \to q^{\uparrow} g}}{\diff \hat{t}} &= -\frac{8 \pi \alpha_s^2}{9 \hat{s}^2} \Big(1 - \frac{9}{4} \frac{\hat{u} \hat{s}}{\hat{t}^2}  \Big).\label{e:qg_channel}
\end{align}
The hard factors involving antiquark fragmentation are identical to the hard factors of the corresponding charge-conjugated partonic processes. We note these are also consistent with the partonic cross sections associated with the ``derivative term'' of the  fragmentation part of $A_N$ in $p^\uparrow p\to h\,X$~\cite{Metz:2012ct}.

\section{Symmetry of $\pi^+ \pi^-$ DiFFs}
\label{a.symmetry}

We provide details on deriving the symmetry relations for $\pi^+ \pi^-$ pairs that allow us to reduce the number of DiFFs to be fitted, which is an important aspect of the analysis.
For this discussion, it is necessary to introduce the notation $D_1^{h_1 h_2}$ and $H_1^{\sa,h_1 h_2}$, where $h_1 h_2$ indicates the dihadron pair.
First, under the interchange of $h_1$ and $h_2$ one has the relations
\begin{align}
D_1^{\pi^+ \pi^- / q}      &=   D_1^{\pi^- \pi^+ / q}\,, \notag \\
H_1^{\sa, \pi^+ \pi^- / q} &= - H_1^{\sa, \pi^- \pi^+ / q}\,,
\label{e.h1h2symmetry}
\end{align}
for all quarks $q$.
The sign change for $H_1^{\sa}$ occurs due to the fact that interchanging the hadrons switches the sign of $\vec{R}_T$, which appears in the prefactor of $H_1^{\sa}$ in its correlator definition, but the correlator itself is independent of the ordering of $h_1$ and $h_2$ \cite{Bianconi:1999cd, Pitonyak:2023gjx}.
From charge-conjugation symmetry, one also has the relations
\begin{align}
D_1^{\pi^+ \pi^- / q}      &= D_1^{\pi^- \pi^+ / \bar{q}} \,, \notag \\
H_1^{\sa, \pi^+ \pi^- / q} &= H_1^{\sa, \pi^- \pi^+ / \bar{q}} \,,
\label{e.chargeconjugation}
\end{align}
for all quarks $q$.
Combining Eqs. (\ref{e.h1h2symmetry}) and (\ref{e.chargeconjugation}) to eliminate the $\pi^- \pi^+$ DiFFs, one has
\begin{align}
D_1^{\pi^+ \pi^- / q}      &= D_1^{\pi^+ \pi^- / \bar{q}} \,, \notag \\
H_1^{\sa, \pi^+ \pi^- / q} &= - H_1^{\sa, \pi^+ \pi^- / \bar{q}} \,.
\label{e.h1h2chargeconjugation}
\end{align}

From isospin symmetry (which holds to sufficient accuracy for this discussion), one also has the relations for the light quarks
\begin{align}
D_1^{\pi^+ \pi^- / u} &= D_1^{\pi^- \pi^+ / d} \,, \hspace{1.6cm}
D_1^{\pi^+ \pi^- / \bar{u}} = D_1^{\pi^- \pi^+ / \bar{d}} \,, \notag \\
H_1^{\sa, \pi^+ \pi^- / u} &= H_1^{\sa, \pi^- \pi^+ / d} \,, \hspace{1.0cm}
H_1^{\sa, \pi^+ \pi^- / \bar{u}} = H_1^{\sa, \pi^- \pi^+ / \bar{d}} \,.
\label{e.isospin}
\end{align}
Using Eqs.~(\ref{e.h1h2symmetry}) and (\ref{e.isospin}), one has
\begin{align}
D_1^{\pi^+ \pi^- / u}            &= D_1^{\pi^+ \pi^- / d} \,, \hspace{1.9cm}
D_1^{\pi^+ \pi^- / \bar{u}}       = D_1^{\pi^+ \pi^- / \bar{d}} \,, \notag \\
H_1^{\sa, \pi^+ \pi^- / u}       &= - H_1^{\sa, \pi^+ \pi^- / d} \,, \hspace{1.0cm}
H_1^{\sa, \pi^+ \pi^- / \bar{u}}  = - H_1^{\sa, \pi^+ \pi^- / \bar{d}} \,.
\end{align}
Combining the above with Eq.~(\ref{e.h1h2chargeconjugation}) leads to the final relations for the light quarks
\begin{align}
D_1^{\pi^+ \pi^- / u}             = D_1^{\pi^+ \pi^- / d} &= D_1^{\pi^+ \pi^- / \bar{u}} = D_1^{\pi^+ \pi^- / \bar{d}} \,, \notag \\
H_1^{\sa, \pi^+ \pi^- / u}        = - H_1^{\sa, \pi^+ \pi^- / d} &= -H_1^{\sa, \pi^+ \pi^- / \bar{u}} = H_1^{\sa, \pi^+ \pi^- / \bar{d}}
\,.
\end{align}
For the heavier quarks ($s$, $c$, and $b$), which are not valence quarks of $\pi^+ \pi^-$, one can assume
\begin{align}
H_1^{\sa, \pi^+ \pi^- / q} &= H_1^{\sa, \pi^+ \pi^- / \bar{q}} \,, \hspace{1.0cm} (q = s, c, b)
\end{align}
from which it follows, using Eq.~(\ref{e.h1h2chargeconjugation}), that
\begin{align}
H_1^{\sa, \pi^+ \pi^- / q} = -H_1^{\sa, \pi^+ \pi^- / q} = 0  \,. \hspace{1.0cm} (q = s, c, b, \bar{s}, \bar{c}, \bar{b})
\end{align}

In summary, one has the following symmetry relations (see also Ref.~\cite{Courtoy:2012ry}) for the $\pi^+ \pi^-$ DiFFs
\begin{align}
D_1^u = D_1^d &= D_1^{\bar{u}} = D_1^{\bar{d}}, \notag \\
D_1^s = D_1^{\bar{s}}, ~~~ D_1^c &= D_1^{\bar{c}}, ~~~ D_1^b = D_1^{\bar{b}},
\end{align}
and $\pi^+ \pi^-$ IFFs
\begin{align}
H_1^{\sa,u} = -H_1^{\sa,d} &= -H_1^{\sa,\bar{u}} = H_1^{\sa,\bar{d}}, \notag \\
H_1^{\sa,s} = H_1^{\sa,\bar{s}} = H_1^{\sa,c} &= H_1^{\sa,\bar{c}} = H_1^{\sa,b} = H_1^{\sa,\bar{b}} = 0,
\end{align}
where we have again dropped the $\pi^+ \pi^-$ superscript.

\section{Comparison to PYTHIA-Generated Data}
\label{a.PYTHIA}

Here we provide more information related to the data generated from PYTHIA~\cite{Sjostrand:2003wg,Bierlich:2022pfr} used in the analysis of the DiFF $D_1$ (see Sec.~\ref{ss.PYTHIA}).  
In Table~\ref{t.tunes} we give the details of the different PYTHIA tunes used to estimate the systematic errors on the generated data.
In Table~\ref{t.chi2_pythia} we present the $\chired$ of those data.
Figs.~\ref{f.PYTHIA-RS-10.58}-\ref{f.PYTHIA-RS-91.19} below show the results for the five different energies used.

\begin{table}[h!]
\centering
%\scriptsize
\resizebox{12cm}{!} {
\begin{tabular}{l | c | c | c}
\hhline{====}
\vspace{-0.3cm} & &  \\
Parameter & PYTHIA 6 default & PYTHIA 6 ALEPH & PYTHIA 6 LEP/Tevatron\\
\vspace{-0.3cm} & &  \\
\hline
PARJ(1)   & 0.1    & 0.106   & 0.073  \\
PARJ(2)   & 0.3    & 0.285   & 0.2    \\
PARJ(3)   & 0.4    & 0.71    & 0.94   \\
PARJ(4)   & 0.05   & 0.05    & 0.032  \\
PARJ(11)  & 0.5    & 0.55    & 0.31   \\
PARJ(12)  & 0.6    & 0.47    & 0.4    \\
PARJ(13)  & 0.75   & 0.65    & 0.54   \\
PARJ(14)  & 0      & 0.02    &        \\
PARJ(15)  & 0      & 0.04    &        \\
PARJ(16)  & 0      & 0.02    &        \\
PARJ(17)  & 0      & 0.2     &        \\
PARJ(19)  & 1      & 0.57    &        \\
PARJ(21)  & 0.36   & 0.37    & 0.325  \\
PARJ(25)  & 1      &         & 0.63   \\
PARJ(26)  & 0.4    & 0.27    & 0.12   \\
PARJ(33)  & 0.8    & 0.8     & 0.8    \\
PARJ(41)  & 0.3    & 0.4     & 0.5    \\
PARJ(42)  & 0.58   & 0.796   & 0.6    \\
PARJ(45)  & 0.5    &         &        \\
PARJ(46)  & 1      &         &        \\
PARJ(47)  & 1      &         &        \\
PARJ(54)  & -0.05  & -0.04   & -0.05  \\
PARJ(55)  & -0.005 & -0.0035 & -0.005 \\
PARJ(81)  & 0.29   & 0.292   & 0.29   \\
PARJ(82)  & 1      & 1.57    & 1.65   \\
MSTJ(11)  & 4      & 3       & 5      \\
MSTJ(12)  & 2      & 3       &        \\
MSTJ(26)  & 2      & 2       & 2      \\
MSTJ(45)  & 5      &         &        \\
MSTJ(107) & 0      & 0       & 0      \\
\hhline{====}
\end{tabular}}
\caption
{Summary of the different PYTHIA 6 tunes used to generate the PYTHIA data.  The ``PYTHIA 6 default" tune uses the corresponding default parameters shown above.  The ``PYTHIA 6 ALEPH" and ``PYTHIA 6 LEP/Tevatron" tunes modify those parameters as shown in their respective columns (if the row is blank then the default value is used).  For the ``PYTHIA 8" tune (not shown in the table) we use all default parameters.}
\label{t.tunes}
\end{table}

\begin{table}[h!]
\centering
\resizebox{4cm}{!} {
\begin{tabular}{ c | c | c | c}
\hhline{====}
\vspace{-0.3cm} & &  \\
$\sqrt{s}$ & ~$q$~ & $N_{\rm dat}$ & $\chired$\\
\vspace{-0.3cm} & &  \\
\hline
10.58 & s & 200 &  4.00    \\
10.58 & c & 198 &  1.60    \\
30.73 & s & 229 &  0.37    \\
30.73 & c & 225 &  0.66    \\
30.73 & b & 199 &  0.26    \\
50.88 & s & 223 &  0.57    \\
50.88 & c & 222 &  0.70    \\
50.88 & b & 203 &  0.22    \\
71.04 & s & 223 &  0.58    \\
71.04 & c & 217 &  0.55    \\
71.04 & b & 203 &  0.29    \\
91.19 & s & 223 &  0.59    \\
91.19 & c & 209 &  0.67    \\
91.19 & b & 204 &  0.34    \\
\hline
\textbf{Total} & & 2978 &  0.80 \\
\hhline{====}
\end{tabular}}
\caption
{Summary of $\chired$ values from the JAMDiFF (w/ LQCD) fit for the PYTHIA datasets where the observable is $\sigma^q/\sigma^{\rm tot}$.}
\label{t.chi2_pythia}
\end{table}
%\end{center}

\begin{figure}[h]
\includegraphics[width=0.90\textwidth]{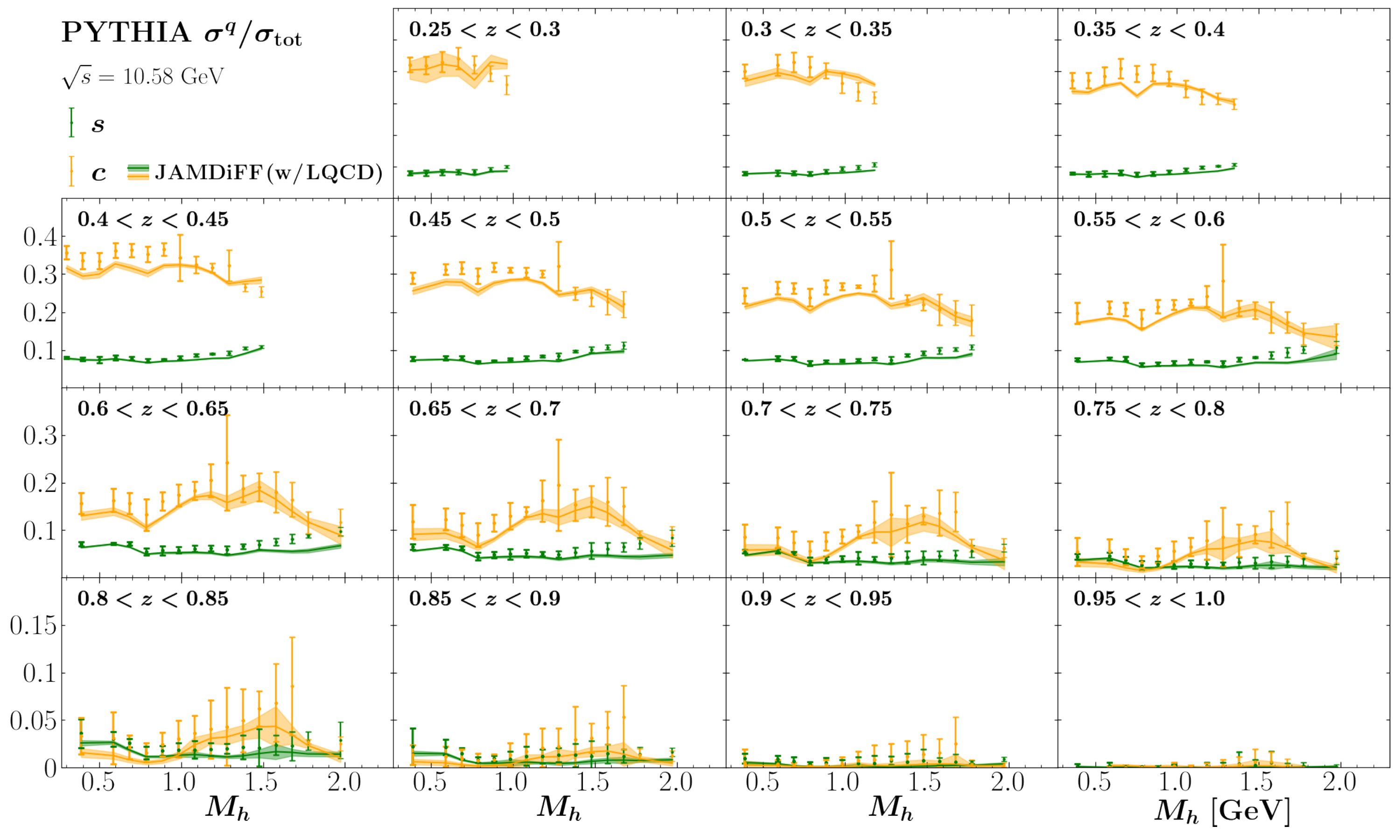}
\caption
{PYTHIA-generated data at $\sqrt{s} = 10.58$ GeV. The strange (green points) and charm (orange points) cross section ratios are plotted as a function of $M_h$ and compared to the mean JAMDiFF (w/ LQCD) results with 1$\sigma$ uncertainty bands (colored bands).  The different panels show different bins of $z$.}
\label{f.PYTHIA-RS-10.58}
\end{figure}

\begin{figure}[h]
\includegraphics[width=0.90\textwidth]{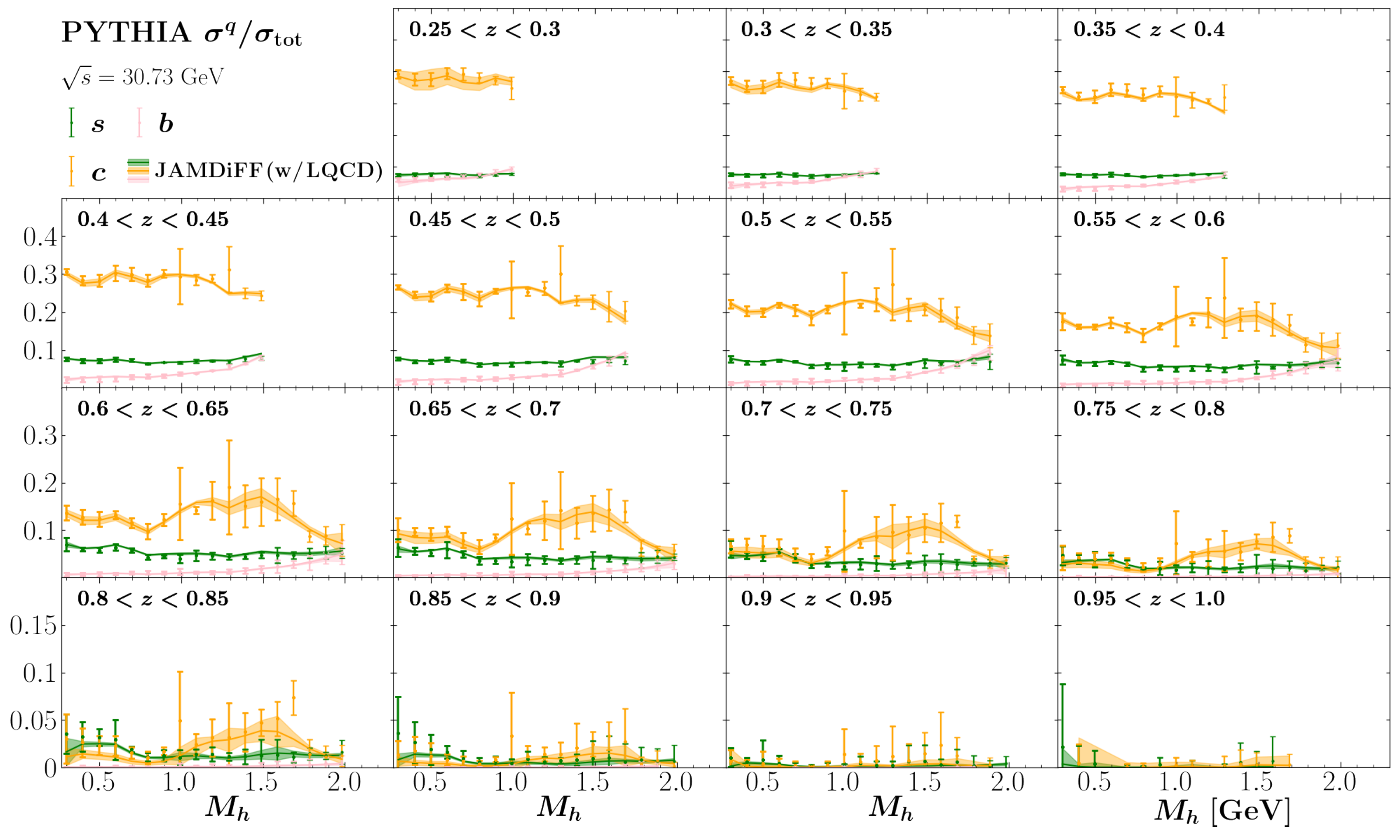}
\caption
{PYTHIA-generated data at $\sqrt{s} = 30.73$ GeV. The strange (green points), charm (orange points), and bottom (pink points) cross section ratios are plotted as a function of $M_h$ and compared to the mean JAMDiFF (w/ LQCD) results with 1$\sigma$ uncertainty bands (colored bands).  The different panels show different bins of $z$.}
\label{f.PYTHIA-RS-30.73}
\end{figure}

\begin{figure}[h]
\includegraphics[width=0.90\textwidth]{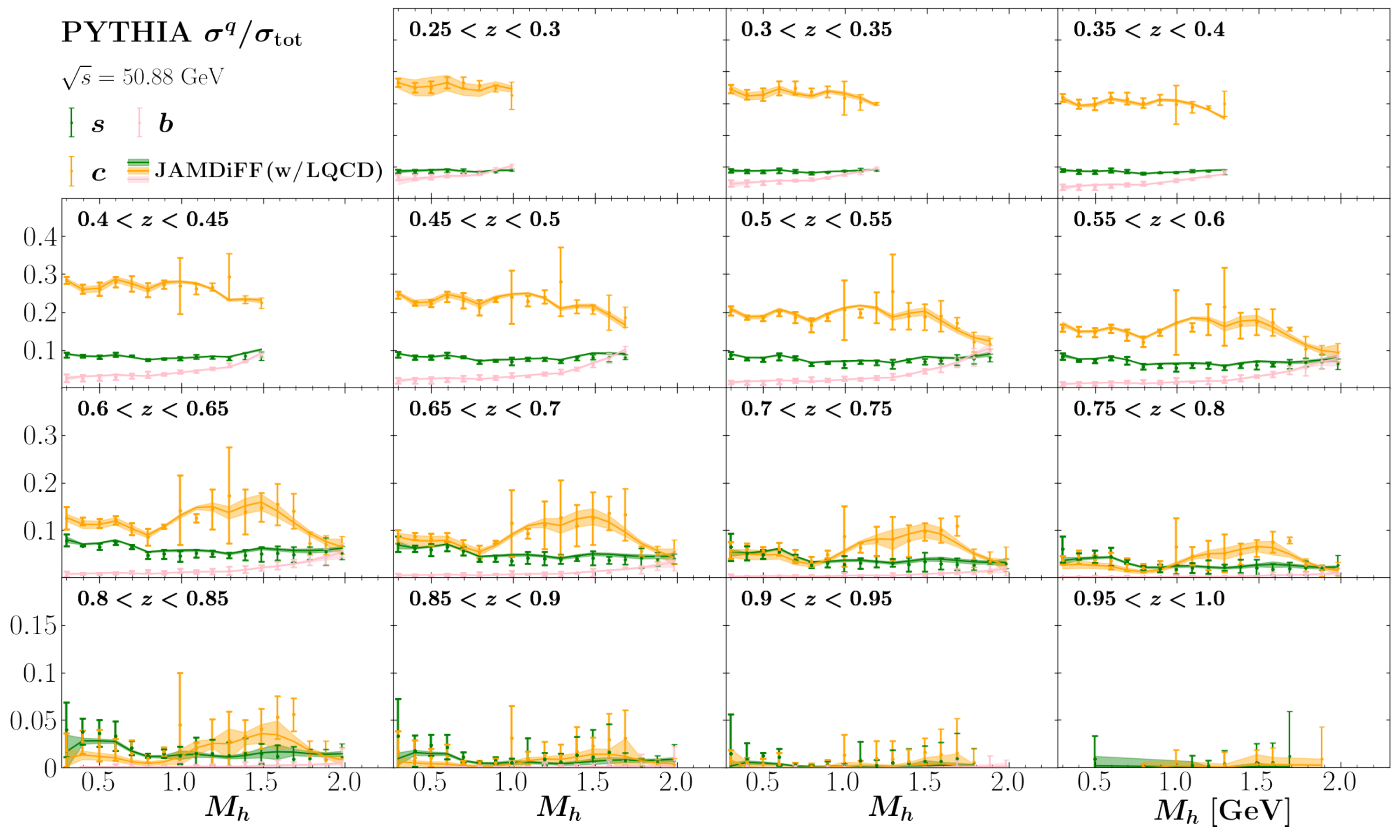}
\caption
{PYTHIA-generated data. Same as Fig.~\ref{f.PYTHIA-RS-30.73} but for $\sqrt{s} = 50.88$ GeV.}
\label{f.PYTHIA-RS-50.88}
\end{figure}

\begin{figure}[h]
\includegraphics[width=0.90\textwidth]{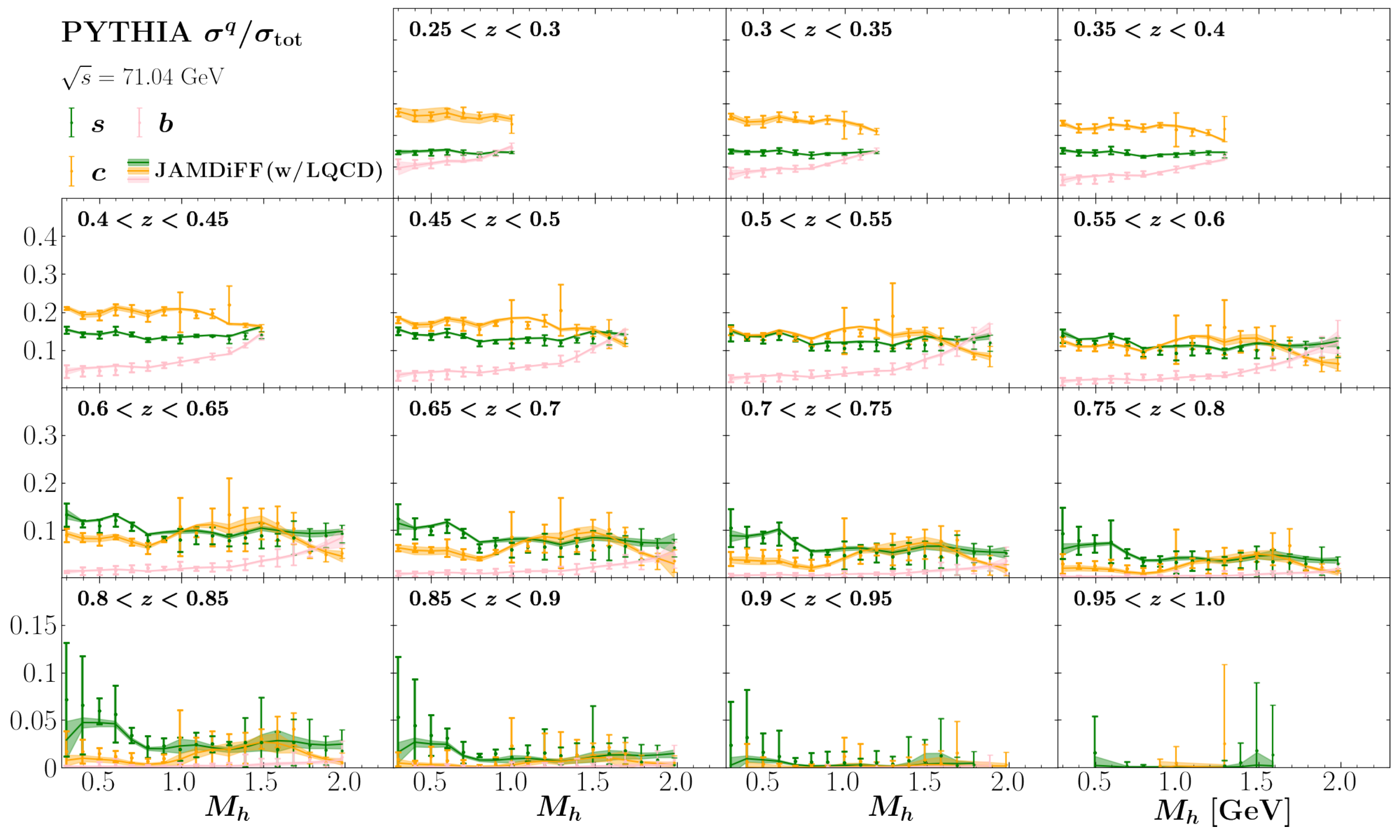}
\caption
{PYTHIA-generated data. Same as Fig.~\ref{f.PYTHIA-RS-30.73} but for $\sqrt{s} = 71.04$ GeV.}
\label{f.PYTHIA-RS-71.04}
\end{figure}

\begin{figure}[h]
\includegraphics[width=0.90\textwidth]{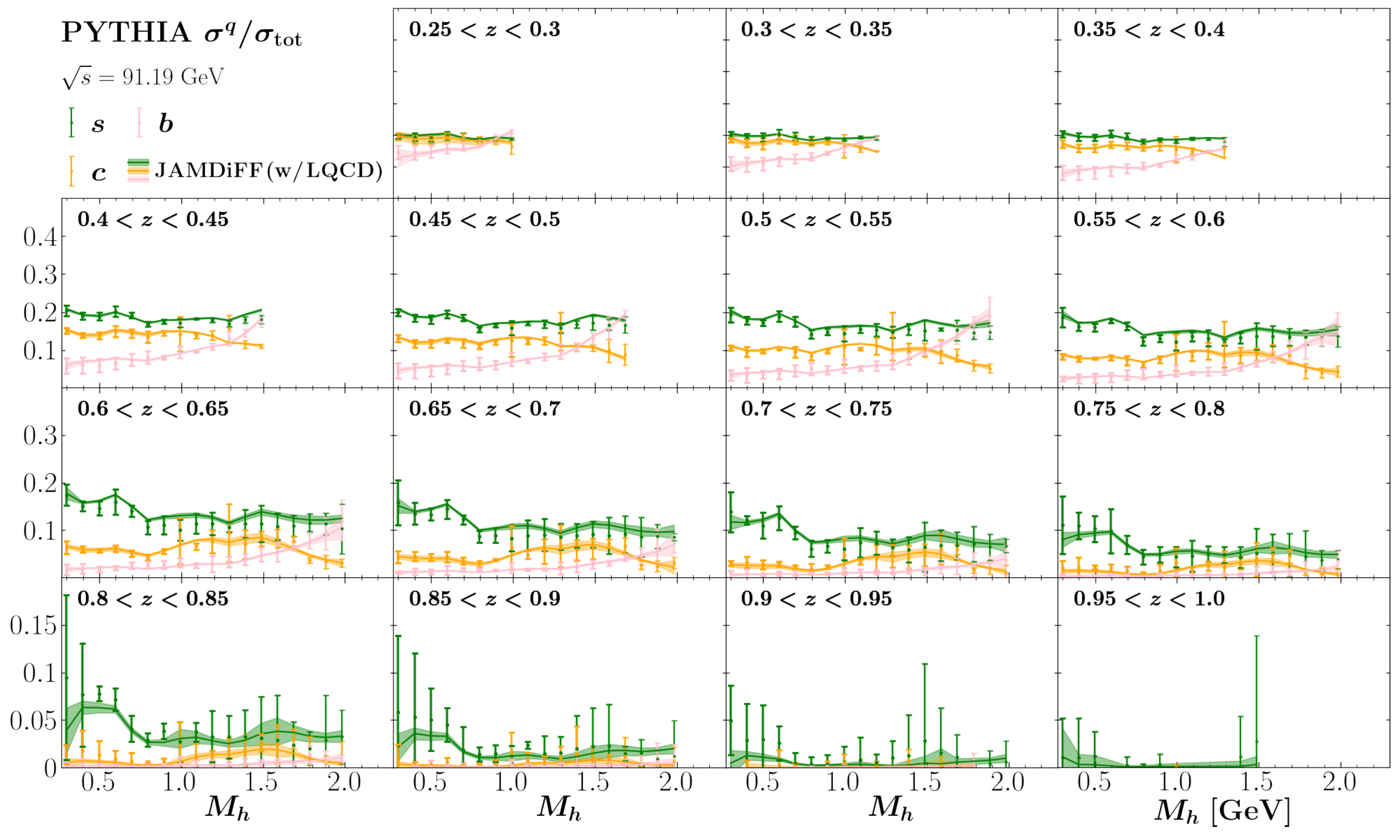}
\caption
{PYTHIA-generated data.  Same as Fig.~\ref{f.PYTHIA-RS-30.73} but for $\sqrt{s} = 91.19$ GeV.}
\label{f.PYTHIA-RS-91.19}
\end{figure}

\clearpage

\bibliography{cc.bib}{}

\end{document}